\documentclass[a4paper]{article}
\usepackage{bm, amsmath,amssymb,amsthm,amsfonts,graphics,graphicx,enumerate, epsfig,rotating,caption,url,natbib,algorithm,algpseudocode}
\usepackage{subfigure}
\usepackage{color}

\newcommand{\noi}{\noindent}
\newcommand{\bbeta}{\bm{\beta}}
\newcommand{\balpha}{\bm{\alpha}}
\newcommand{\blambda}{\bm{\lambda}}

\newcommand{\bxi}{\bm{\xi}}
\newcommand{\btheta}{\bm{\theta}}
\newcommand{\btau}{\bm{\tau}}

\renewcommand{\L}{\mathcal{L}}
\newcommand{\ML}{\text{ML}}
\newcommand{\Y}{\mathbf{Y}}
\newcommand{\X}{\mathbf{X}}
\newcommand{\Z}{\mathbf{Z}}
\newcommand{\tbeta}{\tilde{\beta}}
\newcommand{\argmax}{\text{argmax}}
\newcommand{\expit}{\text{expit}}
\newcommand{\para}{\bigskip\noindent}
\newcommand{\pperp}{\perp\!\!\!\perp}

\newcommand{\tr}{^{\text{T}}}
\newcommand{\diag}{\text{diag}}

\newcommand{\I}{\mathbf{I}}
\newcommand{\T}{\mathbf{T}}

\newcommand{\D}{\mathbf{D}}
\newcommand{\A}{\mathbf{A}}
\newcommand{\ones}{\mathbf{1}}
\newcommand{\zeros}{\mathbf{0}}

\newcommand{\N}{\text{N}}
\newcommand{\GIG}{\text{GIG}}
\newcommand{\IGauss}{\text{IGauss}}
\newcommand{\Un}{\text{U}}

\renewcommand{\(}{\left(}
\renewcommand{\)}{\right)}
\renewcommand{\[}{\left[}
\renewcommand{\]}{\right]}

\title{Learning from a lot: Empirical Bayes in high-dimensional prediction settings}
\date{}
\author{Mark A. van de Wiel$^{1,2}$, Dennis E. te Beest$^{1}$, Magnus M\"unch$^{1,3}$}

\begin{document}
\maketitle

\noindent
1. Dep of Epidemiology \& Biostatistics, VU University Medical Center, Amsterdam, The Netherlands\\
2. Dep of Mathematics, VU University, Amsterdam, The Netherlands\\
3. Mathematical Institute, Faculty of Science, Leiden University, Leiden, The Netherlands


\begin{abstract}
Empirical Bayes is a versatile approach to `learn from a lot' in two ways: first, from a large number of variables and
second, from a potentially large amount of prior information, e.g. stored in public repositories. We review applications of a variety of empirical Bayes
methods to several well-known model-based prediction methods including penalized regression, linear discriminant analysis, and Bayesian models with sparse or dense priors.
We discuss `formal' empirical Bayes methods which maximize the marginal likelihood, but also more informal approaches based on other data summaries. We contrast empirical Bayes to cross-validation and full Bayes, and discuss hybrid approaches. To study the relation between the quality of an empirical Bayes estimator and $p$, the number of variables, we consider
a simple empirical Bayes estimator in a linear model setting.

We argue that empirical Bayes is particularly useful when the prior contains multiple parameters which model a priori information on variables, termed `co-data'.  In particular,
we present two novel examples that allow for co-data. First, a Bayesian spike-and-slab setting that facilitates inclusion of multiple co-data sources and types; second, a hybrid empirical Bayes-full Bayes ridge regression approach for estimation of the posterior predictive interval.
\end{abstract}


\section{Introduction}
High-dimensional data with tens or hundreds of thousands of variables are frequently part of biomedical (or other) studies nowadays. In addition,
a lot of prior information is available in the public domain, for example in genomics data repositories or in data bases containing structural information on the variables, such as genomic pathways. When
one aims to develop a predictor for a new study, one is challenged to learn from this wealth of data. For many high-dimensional prediction methods such learning consists of two phases: 1) learning the tuning parameter(s), e.g. penalty parameters in a frequentist framework or prior parameters in a Bayesian framework; and 2)
learning the predictor as a function of the variables given the tuning parameter(s). Empirical Bayes (EB) is a widely acknowledged approach to complete the first phase.

Broadly speaking, EB is a collection of methods which estimate the tuning parameter(s), often formulated in terms of prior parameters, from the data, thereby borrowing information across variables of the same type. We focus most;y on high-dimensional prediction settings, so $p > n$, with $p$: the number of predictors and $n$: the number of independent samples. For other settings, several excellent contributions exist.  \cite{CarlinLouis} is an extensive introduction to EB. It discusses parametric and nonparametric EB, provides many examples for standard models, presents suggestions on computations (in particular for maximization of the marginal likelihood) and compares performances of EB methods with fully Bayesian and frequentist ones in low-dimensional settings. \cite{EfronBook} has quickly become a standard work for applications of Empirical Bayes to multiple testing, in particular for estimation of the False Discovery Rate and variants thereof. \cite{Houwelingen2014} is a recent, critical review with many data examples on the application of EB to low-dimensional estimation problems, in particular meta-analysis, and to high-dimensional multiple testing problems. Some of the pros and cons of EB mentioned in these references are re-iterated here, but cast in the perspective of high-dimensional prediction.
Note that properties and usefulness of EB estimators may be different in high- and low-dimensional prediction settings. First, high-dimensional data allows for more complex, possibly sparse priors with several hyper-parameters. Moreover, the computational advantage (with respect to full Bayes) is larger in high-dimensional settings. In addition, the large $p$ may lead to better estimation of the prior (see Sections \ref{mmlseb} and \ref{emsepenalized}) and allows for modeling the prior in terms of prior information on the variables (see Section \ref{codata}). Finally, regularization changes the bias-variance trade-off, and hence the properties of the EB estimator (see Section \ref{emsepenalized}).

While the emphasis in this discussion paper is on high-dimensional prediction, we sometimes refer to the `medium-dimensional' setting. The latter is informally defined as a $p<n$ setting, but with $p$ large enough (with respect to $n$) to render estimation ill-behaved and hence desire regularization. We believe that this medium-dimensional prediction setting is becoming increasingly relevant. For example, targeted high-throughput molecular devices have become cheaper, implying that their use for clinical prediction has become realistic. These devices typically measure tens or hundreds of molecular markers, possibly selected from  `whole-genome' screening studies. Examples of such devices are multiplex polymerase chain reaction (PCR) and targeted sequencing platforms.

We review several versions of EB, plus their applications to a variety of prediction methods. We follow \cite{morris1983parametric}: ``Empirical Bayes modeling permits statisticians to incorporate additional information in problems'', and argue that such prior information on the variables, referred to as `co-data' \citep[see also][]{neuenschwander2016use}, is particularly useful in high-dimensional settings, because it may improve prediction and variable selection. Such co-data may be continuous, e.g. $p$-values from a related, but independent study, or nominal, e.g. known sets of variables that share a function. Use of such co-data to accommodate different priors for variables is known as `local adaptation' in full Bayes settings \cite[]{sillanpaa2009review}; we discuss the EB counterpart.

When the unknown hyper-parameter(s) concern tuning parameters in a frequentist setting, cross-validation is a popular alternative for EB. Therefore, we contrast the two approaches and shortly discuss hybrid solutions. We cannot cover the entire scala of high-dimensional prediction methods, and hence focus on model-based prediction. We do discuss a fairly broad spectrum, including penalized regression (e.g. lasso, ridge, elastic net), linear discriminant analysis and Bayesian approaches using sparse or dense priors. Ridge regression is used to illustrate matters on several occasions, in particular to analytically study the expected mean squared error of an empirical Bayes estimator of the prior variance as a function of $p$. Although theory on EB in large $p$ settings is an active field of research, results are still mostly limited to very simple models, as discussed in
Section \ref{theory}.

Finally, we present two novel examples of the usage of EB for high- and medium-dimensional prediction. The two examples both allow to account for co-data when estimating the prior(s). The first example illustrates how EB may be used to inform prior inclusion probabilities in a Bayesian spike-and-slab model that is fit using MCMC sampling. Second, a simulation example demonstrates the benefit of a hybrid Bayes-EB approach for estimation of the posterior predictive interval using group-regularized logistic ridge regression.

\section{Empirical Bayes methodologies}
%

We review several EB methodologies in the context of model-based high-dimensional prediction. Their applicability depends on the prediction method, which we will specify in the sections below. We distinguish:

\begin{enumerate}[1)]
\item MMLU EB: maximize the marginal likelihood product derived from univariate models

\item MMLJ EB: maximize marginal likelihood derived from a joint model
\begin{itemize}
\item Direct EB: maximize an analytical expression for the marginal likelihood
\item Laplace EB: maximize marginal likelihood using Laplace approximation
\item MCMC EB: maximize marginal likelihood using MCMC-sampling
\item VB EB: maximize marginal likelihood using Variational Bayes (VB)
\end{itemize}
\item MoM EB: Method of moments; equate theoretical moments to empirical ones.
\end{enumerate}
Several fundamental similarities and differences across 1) to 3) exist. First, the use of 1) is restricted to prediction methods that combine univariate models into one prediction, such as diagonal linear discriminant analysis.
We show that marginal likelihood-based empirical Bayes, which shrinks the effect sizes, is then very similar to `standard' empirical Bayes in estimation problems.
Methodology 2) applies formal EB to the full multivariate setting, hence to a single joint $p$-dimensional model, like penalized regression. As such it is the most generic methodology. The methodology
is then subclassified by methods that are used to facilitate the maximization, the suitability of which depends on the prediction model used.
Finally, 3) refers to an intuitive, classical use of EB: equating moments. Naturally, this is restricted to predictors for which the moments are known. Below we provide details on 1) to 3).

Throughout this article we denote response by $\Y = (Y_1, \ldots, Y_n)$ and the high-dimensional parameter by $\btheta = (\theta_1, \ldots, \theta_p)$. Variables are denoted by $\X = (\X_1^T,\ldots, \X_n^T)^T, \X_i = (X_{i1}, \ldots, X_{ip})$.

\subsection{Maximum marginal likelihood from univariate models: MMLU EB}\label{mmlseb}
One of the simplest classifiers that may be used in a high-dimensional setting is diagonal linear discriminant analysis (DLDA). It assumes a diagonal covariance matrix $\Sigma$ for the variables. While this is unlikely to be true, the results of DLDA may be better than for ordinary LDA, which requires a (regularized) estimate of $\Sigma$ \cite[]{bickel2004some}. One of the early classifiers introduced for high-dimensional prediction, the shrunken centroid algorithm \cite[]{Tibshirani2002}, may be regarded as a DLDA. DLDA is also discussed here because EB for DLDA turns out to be very similar to EB for estimation problems, allowing a gentle introduction to EB. Here, we follow the notation of \cite{Efron2009Prediction}.  DLDA combines univariate effect-size estimates $\hat{\theta}_j$ in one classification rule by the sign of $S_i$, with
\begin{equation}\label{lda}
S_i = \sum_{j=1}^p \hat{\theta}_j W_{ij},
\end{equation} where $W_{ij}$ is the standardized value of variable $X_{ij}$. For this type of classifier, EB-type shrinkage is based on univariate summaries, as in many multiple testing and estimation settings. Following \cite{Efron2009Prediction}, we compute the $Z$-score $Z_j$, which is the standardized difference in means between the two groups (defined by $\Y$) for variable $j$.
Then, $Z_j$ is expressed as a convolution:
\begin{equation}\label{conv}
Z_j = \theta_j + \epsilon_j,
\end{equation}
 where $\epsilon_j \sim N(0,1)$ and $\theta_j \sim \pi$, with assumptions $\epsilon_j \pperp \theta_j$ and $Z_j \pperp Z_{k}$, for $k \neq j$.
\cite{Efron2009Prediction} then continues by developing a non-parametric estimate of $\pi$ using deconvolution, and this could in fact be regarded as a form of non-parametric EB. The posterior mean,  $\hat{\theta}_j = E(\theta_j|Z_j)$, then provides a shrunken estimate of $\theta_j$. \cite{dicker2016high} use a very similar marginal nonparametric deconvolution approach. Their work is based on the Bayes classifier, and provides theoretical guarantees on the performance if the deconvolution is accurate and the joint densities of the two groups of $\X_i$ variables are far apart in terms of Hellinger distance.

Here, we discuss the parametric counterpart, meaning $\pi = \pi_{\balpha}$ is of a specified parametric form with unknown hyper-parameters $\balpha$. This could be useful when one would desire a sparse DLDA, requiring a sparse prior, e.g. a spike-and-slab prior. In the parametric setting, estimating $\balpha$ then boils down to maximizing the (marginal) likelihood, which factorizes rendering
\begin{equation}\label{mlsimple}
\hat{\balpha} = \text{argmax}_{\balpha} \biggl(\prod_{j=1}^p \int_{\theta_j} \L(Z_j; \theta_j) \pi_{\balpha}(\theta_j) d\theta_j\biggr) = \text{argmax}_{\balpha} \log\biggl(\sum_{j=1}^p \int_{\theta_j} \L(Z_j; \theta_j) \pi_{\balpha}(\theta_j) d\theta_j\biggr),
\end{equation}
where $\L(Z_j; \theta_j)$ is the (Gaussian) likelihood implied by (\ref{conv}).
Maximization of (\ref{mlsimple}) is relatively straightforward, because the integral is one-dimensional.
In case the prior is conjugate, it may be solved analytically; otherwise efficient EM-type algorithms are available, such
as the one in \cite{WielShrinkSeq}, which was proven to converge. Once $\hat{\balpha}$, and hence $\pi_{\hat{\balpha}}$, is known, the computation of the shrunken estimate $\hat{\theta}_j = E(\theta_j|Z_j; \hat{\balpha})$ is straightforward; substitution into (\ref{lda}) then renders the (possibly sparse) DLDA.

Note that convolution (\ref{conv}) and EB estimate (\ref{mlsimple}) are exactly the same as in the well-known normal-means estimation problem. This problem is well-studied, also theoretically \cite[]{johnstone2004needles}.
In the Supplementary Information, we revisit the famous batting averages example \cite[]{Efron1975data}, which is a normal-means problem that is often used as a scholarly example of EB estimation. It concerns data of 18 baseball players.  Using a Gaussian prior, \cite{Houwelingen2014} rightfully criticizes EB in this setting, because it seems to over-shrink the estimate for the best player(s) when using a Gaussian prior. We show that when one would have had additional data of 10,000 players, the over-shrinkage is much less severe, because the EB estimate of the Gaussian prior variance improves a lot. This connects to what we will observe in Section \ref{emsetheory}
for the linear ridge regression model. For the enlarged batting data, the large $p$ also accommodates use of a more complex prior, e.g. a 3-component Gaussian mixture, which slightly further reduces shrinkage for the extremes.

The likelihood product in (\ref{mlsimple}) contrasts the marginal likelihood corresponding to joint prediction models. The latter contains a high-dimensional integral over $\btheta$, and is thus much
 more complex. Given that the vast majority of statistical prediction methods are based on joint models, we now turn our attention to those.

\subsection{Maximum marginal likelihood from a joint model: MMLJ EB}\label{mmleb}
Suppose we wish to use a prediction method based on a joint prediction model that implies likelihood $\L(\Y; \btheta)$. For convenience, variables $\X$  which are usually part of $\L(\Y; \btheta)$ via regression, are not denoted in it. Then, an Empirical Bayes estimate is obtained by maximizing the marginal likelihood:
\begin{equation}\label{marglik}
\hat{\balpha} = \argmax_{\balpha} \ML(\balpha),\ \text{with}\  \ML(\balpha) =  \int_{\btheta} \L(\Y; \btheta) \pi_{\balpha}(\btheta) d\btheta,
\end{equation}
with prior $\pi_{\balpha}(\btheta)$. Often, the prior is assumed to have a product form:
$\pi_{\balpha}(\btheta) = \prod_{j=1}^p \pi_{\balpha}(\theta_j)$.  While marginal likelihood is a Bayesian concept, (\ref{marglik}) may also be used in penalized regression
settings due to the correspondence between $\balpha$ and the penalty parameter(s), say $\blambda$, in the penalized likelihood. A well-known example is the elastic net,
with ridge and lasso as special cases \cite[]{ZouHastie2005}. Below, we discuss several methods to solve (\ref{marglik}).

\subsubsection{Direct EB}
If the prior is conjugate to the likelihood in (\ref{marglik}), computations highly simplify, because this enables direct maximization of the marginal likelihood. E.g. for the linear regression model with a shared Gaussian prior $\theta_j \sim N(0,\tau^2)$, and Gaussian error variance $\sigma^2$, i.e. ridge regression, we have:
$$\ML(\balpha) = \ML((\tau^2,\sigma^2)) = \mathcal{N}(\Y;\boldsymbol{\mu}=\mathbf{0}, \Sigma = \X\X^T\tau^2 + I_{n \times n}\sigma^2),$$ which allows for straightforward likelihood maximization. This directly renders an estimator of the ridge penalty: $\hat{\lambda} = \hat{\sigma}^2/\hat{\tau}^2$, which is computationally more efficient than cross-validation.
Ridge regression is basically a random effects model and hence fits in the setting of mixed models. Such models may include fixed effects as well, useful for accommodating covariates like age or known biomarkers
in a clinical prediction model.
\cite{Jiang2016} discuss the well-known restricted maximum likelihood (REML) estimator of $(\tau^2, \sigma^2)$. They prove consistency of the REML estimator in the high-dimensional setting,
even when the prior is misspecified, in the sense that only a fraction of regression parameters are non-zero in reality.

\cite{karabatsos2017marginal} extends the direct MML estimation to a Bayesian generalized ridge model (using a flat gamma prior on $\sigma^{-2}$), which allows differential penalization of the principal components of $\X$. This setting includes the power ridge as a special case, implying a multivariate Gaussian prior with covariance matrix $\tau^2(\X^T\X)^{\delta}$, where $\delta$ is an additional hyper-parameter and $\tau^2 = \sigma^2/\lambda$. \cite{karabatsos2017marginal} presents a two-stage algorithm to maximize the ML with respect to $\delta$ and $\lambda$.

In many prediction problems, conjugacy is not achieved, either due to the nature of the response $\Y$ (e.g. binary or survival) or due to nature of the (preferred) prior, e.g. a sparse prior.
Then, alternative solutions are needed. Below we present two of these, which both are more generic than MMLU EB and direct EB.

\subsubsection{Laplace EB}
In non-conjugate settings, the high-dimensional integral in (\ref{marglik}) poses a major difficulty, preventing a direct, analytical solution.
Hence, approximations have been developed for $\ML(\balpha)$ for various choices of the likelihood and the prior, in particular for penalized regression
with regression parameters $\btheta$. The integrand of (\ref{marglik}) can often be reformulated in an exponential form, motivating use of Laplace approximations:

\noi
\begin{equation}\label{eq:laplace}
\int_{\btheta} e^{-n h_{\balpha} \left( \btheta \right)} d\btheta \approx e^{-n h_{\balpha} \left( \hat{\btheta} \right)} \left( 2 \pi \right)^{p/2} \operatorname{det} \left( \bm{H}_{\balpha}^{-1} \right)^{1/2} n^{-p/2},
\end{equation}
where $\bm{H}_{\balpha}$ is the Hessian of $h_{\balpha}(\btheta)$, evaluated at $\hat{\btheta}$.
Usually, $\hat{\btheta}=\operatorname{argmax}_{\btheta} \, h_{\balpha}(\btheta)$ is used. This maximum depends on the unknown $\balpha$. For many priors efficient maximizers of the integrand of (\ref{marglik}), and hence $h_{\balpha}(\btheta)$, exist. This suggests numerical optimization or EM-type algorithms alternating between maximization with respect to $\btheta$ given $\balpha$ and Laplace approximation plus maximization in terms of $\balpha$, as in \cite{Heisterkamp1999} for a Poisson model with Gaussian priors.

Concerns have been raised about the accuracy of (\ref{eq:laplace}) in high-dimensional settings. E.g. \cite{shun1995laplace} suggest that when $p > O(n^{1/3})$, the standard Laplace approximation may be unreliable. Sparse priors, which effectuate variable selection, may render approximation (\ref{eq:laplace}) to be accurate, but only when the prior is `sparse enough'. Intuitively, a sparse prior may render the effective dimension of the integral of (\ref{eq:laplace}) much smaller than $p$, because $\hat{\btheta}$ contains many zero's. \cite{Barber2016} consider the Laplace approximation to the marginal likelihood of Bayesian generalized linear models with sparse selection priors of the form

\noi
$$
P_{\nu} (J) \propto {p \choose {|J|}}^{-\nu}  \mathbf{1}_{\{ |J| \leq q \}}, \,\,\, J \subset \{1, \ldots, p\},
$$

\noindent where $J$ is the set of selected variables (i.e. non-zero $\theta_j$'s), $q$ is a maximum of selected variables and $\nu$ is a tuning parameter. Here, $\nu$ determines whether the prior distribution of the models ($\nu=0$), or the prior distribution of the model cardinalities ($\nu=1$) is uniform. They show that with $q$ relatively small (sparse setting) and sample size sufficiently large, the Laplace approximation to the marginal likelihood can be accurate for a potentially large number of models, implying that it may be employed for the estimation of hyper-parameters in strongly sparse settings.

Apart from the accuracy of the Laplace approximation, another issue is that  $h_{\balpha}(\btheta)$ in (\ref{eq:laplace}) may not have a second derivative, rendering the Hessian undefined. An example is regression with a Laplace prior, known as Bayesian lasso. The $L_1$-norm on $\btheta$ is not differentiable at zero with respect to the $\theta_j$ and can therefore not be approximated by the Laplace method without modifications.

%

\subsubsection{MCMC EB}\label{gibbseb}
If Laplace approximation to the integral in the right-hand side of (\ref{marglik}) is not possible or feasible, we may circumvent explicit calculation by an MCMC sampler. Desirable quantities are easily calculated from these samples. \cite{Casella2001empirical} proves that one may employ an EM algorithm to estimate the hyper-parameters from Gibbs samples. The algorithm was extended to general MCMC sampling by \cite{Levine2001}, who also provide an approximation of the Monte Carlo error. The algorithm is an MCEM-type algorithm \citep{wei1990monte}, based on posterior samples of $\bm{\theta}$ instead of point estimates. Here, we shortly describe the method. First, write the marginal likelihood as:


\begin{equation}\label{eq:marginalgibbs}
\ML(\balpha) = \frac{\L(\Y, \bm{\theta}; \balpha)}{p(\bm{\theta}|\Y; \balpha)},
\end{equation}

\noindent where $\L(\Y, \bm{\theta}; \balpha)$ and $p(\bm{\theta}|\Y; \balpha)$ denote the conditional likelihood of $\balpha$ (i.e. the joint distribution of $\Y$ and $\btheta$ given $\balpha$) and posterior distribution of the model parameters, respectively. We take the expectation of both sides with respect to $p (\bm{\theta}| \Y; \balpha')$ and switch to the log-scale to arrive at

\begin{equation}\label{eq:expmarginalgibbs}
E_{\balpha'}[\log \ML(\balpha)] = E_{\balpha'}[\ell (\Y, \bm{\theta}; \balpha)] - E_{\balpha'}[\log p(\bm{\theta}|\Y; \balpha)]
\end{equation}

\noindent for some (current value) $\balpha'$. Expand the last term of (\ref{eq:expmarginalgibbs}):

$$
E_{\balpha'}[\log p(\bm{\theta}|\Y; \balpha)] = \int \log p (\bm{\theta}| \Y; \balpha) p(\btheta|\Y; \balpha') d\btheta
$$

\noindent and note that by Gibbs' inequality this integral is maximized at $\balpha=\balpha'$. Consequently, for every $\balpha \neq \balpha'$,  $ - E_{\balpha'}[\log p(\bm{\theta}|\Y; \balpha')] < - E_{\balpha'}[\log p(\bm{\theta}|\Y; \balpha)]$, such that the sequence which iteratively maximizes the first term in the right-hand side of (\ref{eq:expmarginalgibbs}):

\begin{equation}\label{eq:gibbsseq}
\balpha^{(k + 1)} = \argmax_{\balpha}  E_{\balpha^{(k)}}[\ell (\Y, \bm{\theta}; \balpha)]
\end{equation}

\noindent is non-decreasing and converges.  The expectation in (\ref{eq:gibbsseq}) will generally not be available in closed form. However, one may approximate it by its Monte Carlo estimate:

\begin{equation}\label{eq:mcexpectation}
\argmax_{\balpha}  E_{\balpha^{(k)}}[\ell (\Y, \bm{\theta}; \balpha)] \approx \argmax_{\balpha} \frac{1}{M} \displaystyle\sum_{m=1}^M \ell (\Y, \bm{\theta}^{m,(k)}; \balpha),
\end{equation}

\noindent where $\bm{\theta}^{m,(k)}$ denotes the $m$th MCMC sample from the posterior distribution with hyper-parameters $\balpha^{(k)}$ and $\ell (\Y, \bm{\theta}^{m,(k)}; \balpha)$ is the conditional log-likelihood of $\balpha$ evaluated at the $m$th MCMC sample. Often, $\ell (\Y, \bm{\theta}^{(m)}; \balpha)$ has a fairly simple, tractable form, as exemplified in Section \ref{spikeslab}. Applications of this method are the estimation of the penalty parameter(s) for the Bayesian lasso in \cite{ParkCasella2008} and for the Bayesian elastic net in \cite{Li2010bayesian}. In a penalized logistic regression setting, the efficient Gibbs sampler described in \cite{Polson2013} may be used.

The method above is very generic: it may be applied for hyper-parameter estimation using, in principle, any Bayesian sampling technique. It is computationally costly, though: the EM iterations require multiple MCMC updates, although the number of runs can be reduced by periodically alternating with updates from an importance sampling approximation \cite[]{Casella2001empirical}. To limit Monte Carlo error of the marginal log-likelihood estimate in (\ref{eq:mcexpectation}), the MCMC sample size should be sufficiently large. \cite{Booth1999} propose to start with small sample sizes and increase the sample size as long as the expected likelihood is `swamped' by Monte Carlo error. The small, initial sample size is justified with the EM algorithm's tendency to take large steps towards the optimum in the first few iterations. Any Monte Carlo error in the log-likelihood estimate is relatively small compared to the large increase in log-likelihood during these iterations. Close to convergence, the EM algorithm tends to increment the log-likelihood in smaller steps. Then, the Monte Carlo error is relatively larger, requiring a larger sample size to counteract this. For some models the MCMC sample size may be reduced by introducing stochastic approximation in the E-step \citep{kuhn2004coupling}. In Section \ref{spikeslab},
we illustrate how to apply MCMC EB to high-dimensional spike-and-slab models, and show that it straightforwardly allows to moderate the inclusion prior by use of co-data.

\subsubsection{VB EB}\label{vbeb}
For some models, Variational Bayes (VB) approximations \cite[for a review: see][]{blei2017variational} can be developed as a very efficient alternative to MCMC, also in the EM algorithm above.  VB lends itself well for EB estimation, because the nature of the approximation
often allows expressing the expectation in (\ref{eq:gibbsseq}) analytically in terms of $\balpha$. Let us assume a simple hierarchical model: $\Y \leftarrow \btheta \leftarrow \Z \leftarrow \balpha$. For example, the Bayesian lasso {\cite[]{ParkCasella2008} would have $\btheta$ as $p$-dimensional regression parameter, $\Z$ would be the $p$-dimensional latent mixture parameter in a scale mixture of normals, and $\balpha = \lambda_1$, the lasso penalty parameter. Nuisance parameters, like error variance $\sigma^2$, may be added w.l.o.g.

Let $p(\btheta,\Z|\Y; \balpha)$ denote the full posterior. In the context of our model, VB approximation amounts to determining functions $q_1$ and $q_2$ such that $q_1(\btheta)q_2(\Z;\balpha)$ minimizes the Kullback-Leibner distance $\text{KL}(q_1q_2||p)$. Finding solutions $q_1^*$ and $q_2^*$ requires specific derivations for the model at hand. Several are available in the literature, such as for spike-and-slab regression \citep{carbonetto2012scalable}, the Bayesian ridge model \citep{Leday2017}, and the Bayesian lasso \cite[]{joo2017bayesian}. For example, in the latter model $q_1^*$ is a multivariate Gaussian, whereas $q_2^*$ conveniently factorizes with respect to $Z_1, \ldots, Z_p$ as a product of inverse Gaussians.

The VB analogue of the MCMC EB algorithm above is then straightforward: in the EM algorithm above, replace the Monte Carlo approximation of the posterior, required for the expected joint likelihood (\ref{eq:mcexpectation}), by the VB approximation
$q_1^*(\btheta)q_2^*(\Z;\balpha)$.  In the hierarchical model setting, maximization w.r.t. $\balpha$ then amounts to computing the posterior mean of the log-prior of $\Z$:
\begin{equation}\label{VBalpha}
E_{q_2^*(\Z;\balpha^{(k)})}[\log p(\Z;\balpha)],
\end{equation}
where $q_2^*(\Z;\balpha^{(k)})$ denotes the approximation of $q_2$ given current hyper-parameter(s) $\balpha^{(k)}$. Here, we use that other terms of both the approximate posterior and the conditional log-likelihood
disappear, because they do not contain $\balpha$ (as exemplified for the conditional log-likelihood by (\ref{fact}) for the spike-and-slab model).
Often, (\ref{VBalpha}) can be analytically maximized, as in \cite{joo2017bayesian} for the Bayesian Lasso, implemented in R-package \texttt{BLasso}.

A general concern with VB approximations is the potential underestimation of posterior variances \cite[]{blei2017variational}. However, for EB estimation of the hyper-parameters, the variation across high-dimensional parameters, which is modeled by the prior with parameter(s) $\balpha$, is deemed more relevant than the posterior variances themselves. This suggests combining VB EB with MCMC: use VB for computational efficiency to iteratively estimate $\balpha$, followed by one MCMC run with fixed $\balpha$ to obtain more accurate posteriors. For the latter, the VB posterior mode estimates provide a warm start for the sampling. Due to the connection between VB and Gibbs sampling \cite[]{gelfand1990sampling},  it is usually fairly straightforward to develop a Gibbs sampler once a VB apporximation is available.




\subsection{Moment EB}
An alternative to MML (\ref{marglik}) is moment estimation, which is discussed below.
In case $p$ (univariate) models share a prior (as discussed above), equating theoretical moments to empirical moments is a textbook example on EB. In prediction, however, we often have only one model. Now, assume we have an initial estimate $\hat{\btheta} = \hat{\btheta}(\Y)$. Moreover, $(\theta_j)_{j=1}^p$ share prior $\pi_{\balpha}$, with, say, $\balpha=(\alpha_1, \alpha_2)$. Then, $\alpha_1$ and $\alpha_2$ can be estimated by solving moment equations \emph{if} the conditional moments
$E[\hat{\theta}_j(\Y)|\btheta]$ and $E[\hat{\theta}^2_j(\Y)|\btheta]$ are analytically tractable as functions $f_1$ and $f_2$ of $\btheta$:
\begin{equation}\label{momentest}
\begin{split}
\frac{1}{p}\sum_j \hat{\theta}_j \approx \frac{1}{p}\sum_j E[\hat{\theta}_j(\Y)] &= \frac{1}{p}\sum_j E_{\pi_{\balpha}}\left[E[\hat{\theta}_j(\Y)|\btheta]\right] =\frac{1}{p}\sum_j E_{\pi_{\balpha}}\left[f_1(\btheta)\right] := h_1(\alpha_1,\alpha_2)\\
\frac{1}{p} \sum_j \hat{\theta}_j^2 \approx \frac{1}{p} \sum_j E[\hat{\theta}_j^2(\Y)] &=  \frac{1}{p}\sum_j E_{\pi_{\balpha}}\left[E[\hat{\theta}_j^2(\Y)|\btheta]\right] =
\frac{1}{p}\sum_j E_{\pi_{\balpha}}\left[f_2(\btheta)\right]:= h_2(\alpha_1,\alpha_2),
\end{split}
\end{equation}
where $h_1$ and $h_2$ are known functions. In a group-regularized logistic ridge regression setting, \cite{WielGRridge} use a similar idea. Here, groups of variables are given ($\mathcal{G}_g$; e.g. gene sets), corresponding to priors $\theta_j \sim N(0,\alpha_g)$ if $j \in \mathcal{G}_g$.
They first use a standard ridge estimator for $\hat{\btheta}(\Y)$, and then derive and solve $G$ estimating equations with $G$ unknowns to estimate  $\balpha = (\alpha_g)_{g=1}^{G}$:
\begin{equation}\label{momentest2}
\frac{1}{p} \sum_{j \in \mathcal{G}_g} \hat{\theta}_j^2 \approx \frac{1}{p}\sum_{j \in \mathcal{G}_g} E_{\pi_{\balpha}(\btheta)}\left[E[\hat{\theta}_j^2(\Y)|\btheta]\right] =
\frac{1}{p}\sum_{j \in \mathcal{G}_g} E_{\pi_{\balpha}}\left[f_g(\btheta)\right]:= h_g(\balpha) \quad \forall g=1, \ldots, G.
\end{equation}
\cite{leCessie1992} provide expressions for the mean and variance of the logistic ridge estimator, rendering $f_g(\btheta) = E[\hat{\theta}_j^2(\Y)|\btheta] =  (E[\hat{\theta}_j(\Y)|\btheta])^2 + V[\hat{\theta}_j^2(\Y)|\btheta]$. Due to the bias introduced by penalization, the mean term and hence $f_g$ depends on all $\theta_j$'s (not just those for which $j \in \mathcal{G}_g$), so $h_g$ depends on all $\alpha_g$'s. This leads to a system of $G$ linear equations with $G$ unknowns. For several cancer genomics applications, \cite{WielGRridge} and \cite{novianti2017better} show that using group-penalty parameters that are inverse proportional to solution $\hat{\alpha}_g$ improves predictive performance.

Note that the comparison between likelihood-based (Section \ref{mmleb}) and moment-based estimation is on a somewhat different footing here than for ordinary parameter estimation. In the latter case, likelihood-based estimation is usually preferred, because the estimator has several optimality properties when the likelihood is correctly specified. For many types of data and models, the appropriateness of the likelihood can be verified with a variety of techniques. The latter, however, is much harder for the prior, which contains the hyper-parameters. The moment-estimator depends less on the parametric form of the prior than the marginal likelihood-based one, so it may be more robust against miss-specification of the prior.

\section{EB and cross-validation for multiple hyper-parameters}\label{EBCV}
Cross-validation (CV) is a powerful, alternative principle to obtain hyper-parameters, usually referred to as tuning parameters in this context. A practical asset of CV is that it is easy to implement when
the number of tuning parameters is low. Moreover, it allows to directly optimize the tuning parameter with respect to the out-of-bag predictive performance, thereby matching directly with the main goal of most prediction problems. CV can be computationally unattractive, however, when a) model fitting takes considerable time, like for most MCMC-based solutions or b) multiple tuning parameters are required: the search grid grows exponentially with the number of tuning parameters. In the latter case, sequential tuning approaches could alleviate the computational burden, but due to local optima of the utility function, these may be far from optimal, as shown for the elastic net \cite[]{Waldron2011}.

When hyper-parameters are `competitive', e.g. when they shrink the same parameters, EB approaches may, like CV, struggle to find the optimal ones due to a flat or multi-modal marginal likelihood (\ref{marglik}). Figure \ref{margllEN} shows this for the Bayesian elastic net \citep{Li2010bayesian}. This figure is obtained by estimating the marginal likelihood for varying values of the two hyper-parameters in the elastic net. The model and estimation procedure are given in the Supplementary Information. The data was simulated by first sampling $\X$ with independent entries: $X_{ij} \sim N(0,1), i=1, \ldots, n=200, j=1, \ldots, p=200$. Next, we generated model parameters $\beta_j$ for $j=1, \dots 200$ from the elastic net prior with $\lambda_1,\lambda_2=2$ and set the response $Y_i = \X_i \bbeta + \epsilon_i$, with $\epsilon_i \sim N(0,1)$. A Gibbs sampler was run for every combination of $\lambda_1, \lambda_2 \in \{0.5, 0.8, 1.1, \dots, 3.8 \}$ and the marginal likelihood was calculated for every combination. Figure \ref{margllEN}(a) shows that the marginal likelihood estimation indeed renders a high value for the true $(\lambda_1, \lambda_2)$ combination, $(2,2)$, but many other combinations of one higher and one lower penalty render very similar values. Figure \ref{margllEN}(b) shows that when we extend the simulation to $n=100, p=1000$, the
marginal likelihood is less flat, likely due to the larger $p$. However, while the true value, $(2,2)$, still corresponds to a high marginal likelihood, a bias towards a smaller L1 penalty is observed.

Practical solutions for the competition between hyper-parameters depend on the data, the classifier and the EB approach employed. For prediction, local optima are not necessarily a problem: the corresponding models likely predict almost equally well. If one desires a sparse solution, one could consider a grid for the sparsity parameter, and employ EB to find the other parameter(s) conditional on the sparsity one. Then, one may opt for the smallest model within a pre-specified margin of the best performing model, in terms of marginal likelihood or any other criterion. Alternatively, one fixes the a priori expected (or desired) number of included variables, which is feasible for spike-and-slab models, and uses EB for other parameters.

Sometimes, it may be worthwhile to combine EB with CV. For example, if one wishes to apply different penalties $\lambda_g$ for groups of variables \cite[]{WielGRridge, boulesteix2017ipf}, one may re-parameterize $\lambda_g = \lambda \lambda'_g,$ optimize the global parameter $\lambda$ by CV with respect to predictive performance, while estimating the multipliers $\lambda'_g$ by EB. Alternatively, CV or similar out-of-bag approaches may be used to tune the initial EB estimates to improve predictive performance or to implement parameter thresholding.


\section{Criticisms and theory on EB}\label{theory}
Empirical Bayes comes with assumptions, and hence with criticism. Of course, such criticism should be balanced against potential assets of EB, such as computational efficiency and its ability to account for prior information to improve predictions. We discuss three major criticisms and cast these in the high-dimensional perspective.
First, uncertainty of the hyper-parameter $\balpha$ is not propagated, as it would be for a fully Bayesian approach. In a high-dimensional setting, the prior parameters are estimated from a large number of variables. Hence, depending on the correlation strength, the uncertainty may be relatively small. In fact, in a regression-variable selection context \cite{scott_bayes_2010} argue that uncertainty of the selected model is potentially a larger problem: due to the marginal likelihood maximization EB may lead to a degenerate solution, which may be undesirable when alternative values of the hyper-parameter(s) render marginal likelihoods that are very close to the optimal one. A hybrid Full Bayes-Empirical Bayes approach, as discussed in Section \ref{hybBay}, may provide the best of both worlds.
A second criticism is that EB accommodates the `average ones', not the (possibly more interesting) extremes. In many high-dimensional applications, however, use of more complex priors is feasible, e.g. mixture or heavy-tailed ones. Such priors can better accommodate non-average behavior.

A third criticism is the lack of theoretical guarantees on EB, in particular on how and whether an EB estimator improves when $p$ (instead of $n$) increases. This is likely due to the complex dependency of the hyper-parameters on all variables. In addition, the complex (algorithmic) construction of some EB estimators hampers analytical analysis of their properties. Most theoretical results are available for
the simple normal-means model (so $p=n$ and $\X = I_p$), which allows a factorization of the likelihood as in (\ref{mlsimple}). E.g. \cite{johnstone2004needles} provide asymptotic optimality results for a spike-and-slab-type prior. In addition, \cite{Belitser2015needles} present theoretical evidence that in a sparse spike-and-slab setting EB allows use of a Gaussian slab to obtain good contraction rates of the posteriors, which is a prerequisite for obtaining correct coverage of credibility intervals. Such a Gaussian slab prior is not recommended for the ordinary Bayes setting, because it shows sub-optimal contraction rates as compared to more heavy-tailed slab distributions \cite[]{Castillo2012needles}. For a wider class of models, \cite{rousseau2017asymptotic} recently showed that full (hierarchical) Bayes and MML EB have the same oracle posterior contraction rates $(n \rightarrow \infty$), under weak conditions on the hyper-prior. Below we provide some analytical results for an EB estimator of the prior variance of linear regression parameters, also to show that even for this fairly simple model calculations may be tedious.

\section{Expected MSE for a simple EB estimator}\label{emsetheory}
We study a very simple EB estimator for linear (ridge) regression to gain insight in how the quality of the estimator, as quantified by the expected mean squared error (EMSE), changes with $p$.
We start with the case $p < n$, which allows analytical results. This includes the `medium-dimensional' case with $p$ relatively close to $n$, for which regularization is often desirable. Then, results for $p \geq n$ are obtained by simulation.

\subsection{Setting 1: initial OLS estimator}
Suppose $\beta_j \sim^{\text{iid}} N(0,\tau^2)$. Let $\hat{\bbeta}$ be the OLS estimator of $\bbeta$ in a linear regression model without intercept. For sake of simplicity, we assume the error variance $\sigma^2=1$ to be known. Then,
 $$(\hat{\bbeta} |\X,\bbeta) \sim N(\bbeta, V),$$ with $V=(\X^T\X)^{-1}$ and $v_j = V_{jj}$. \cite{hoerl1975ridge} propose the following simple estimator of $\tau^2$ for $p<n$:
\begin{equation}
(\tau')^2 = \frac{\sum_{j=1}^p \hat{\beta}_j^2}{p}.
\end{equation}
Since $E_{\Y}(\hat{\beta}_j) = \beta_j$ and $V_{\Y}(\hat{\beta}_j) = v_j$, we have
$$E_{\bbeta}(E_{\Y}[(\tau')^2]) = \frac{\sum_{j=1}^{p} \bigl(v_j + E_{\bbeta}(\beta^2_j)\bigr)}{p} = \frac{\sum_{j=1}^{p} v_j}{p} + \tau^2 .$$
Hence, the estimator can be corrected for this expected bias without inflating the variance:
\begin{equation}\label{hattau}
\hat{\tau}^2 = \frac{\sum_{j=1}^p (\hat{\beta}_j^2 - v_j)}{p}.
\end{equation}

\noindent
We wish to study the properties of $\hat{\tau}^2$ in terms of $p$ and $n$. For that we consider the EMSE, where the mean squared error is computed w.r.t. $\Y$, which is then averaged over samples of both $\bbeta$ (drawn from the Gaussian prior) and $\X$. While the latter is often considered as fixed, it is more realistic to assume it random, in particular when $\X$ denotes (e.g. genomic) measurements. This also allows to establish the quality of the estimator across instances of $\X$. We assume that, after standardization,
$X_{i} \sim N(0,\Sigma = \Sigma_{p \times p}), i=1, \ldots, n$, with $\Sigma_{jj}=1$. Then, we study

\begin{equation}\label{emse}
\text{EMSE}(\hat{\tau}^2) = E_{\X}\biggl[E_{\bbeta}\bigl[\text{MSE}_{\Y}(\hat{\tau}^2 | \bbeta, \X)\bigr]\biggr] = E_{\X}\biggl[E_{\bbeta}\bigl[(E_{\Y}(\hat{\tau}^2 | \bbeta, \X)-\tau^2)^2 + V_{\Y}(\hat{\tau}^2 | \bbeta, \X)\bigr]\biggr].
\end{equation}

\noindent
\textbf{Theorem} Let $\text{EMSE}(\hat{\tau}^2)$ be as in (\ref{emse}). Then, we have, with $\Psi = \Sigma^{-1}$, for $p<n-3$:
\begin{equation}\label{emseresult}
\begin{split}
\text{EMSE}(\hat{\tau}^2) &= \frac{2}{(n-p-1)p^2}\biggl[\sum_{j=1}^p \bigl(\frac{2\psi^2_{jj}}{(n-p-1)(n-p-3)} + \frac{\psi^2_{jj}}{(n-p-1)}) + 2\tau^2\sum_{j=1}^p \psi_{jj}\\
&+ \sum_{j,k \neq j}^p\bigl(\frac{(n-p+1)\psi_{jk}^2 + (n-p-1)\psi_{jj}\psi_{kk}}{(n-p)(n-p-1)(n-p-3)} + \frac{\psi^2_{jk}}{(n-p-1)}\bigr)\biggr] + \frac{2\tau^4}{p}.
\end{split}
\end{equation}
\emph{Proof:} See Supplementary Information.

\para
\textbf{Corollary} Let $\text{EMSE}_{\perp}(\hat{\tau}^2)$ be $\text{EMSE}(\hat{\tau}^2)$ for independent $X_i$: $\psi_{jj} = 1$ and $\psi_{jk}=0$. Then, for $p<n-3$:
\begin{equation}\label{emseresultind}
\begin{split}
\text{EMSE}_{\perp}(\hat{\tau}^2) &= \frac{2}{(n-p-1)p}\biggl[\frac{2}{(n-p-1)(n-p-3)} + \frac{1}{n-p-1} + 2\tau^2\\
&+ \frac{p-1}{(n-p)(n-p-3)}\biggr] + \frac{2\tau^4}{p}.
\end{split}
\end{equation}

\noindent
Equations (\ref{emseresult}) and (\ref{emseresultind}) clearly show the balance for increasing $p$, causing $n-p$ to decrease. From (\ref{emseresult}) we observe that the effect of collinearity in $\X$ may
be large when the number of non-zero $\psi_{jk}$'s (i.e. partial correlations) is large, due to the double-summation and the relatively small $\mathcal{O}(n-p)$ denominator of $\psi^2_{jk}$. In addition, we observe that for large $\tau^2$ a large $p$ is relatively more beneficial than for small $\tau^2$.
Figure \ref{emseunpen} shows the root EMSE as function of $p<n$ for $n=1000$ for $\tau^2 = (0.1)^2 = 0.01; \tau^2 = (0.2)^2 = 0.04$, for $\Sigma = I_{p}$ (referred to as `independent $\X$');
$\Sigma = I_{B} \otimes A^{\rho}_{b \times b},$ (block-correlation) with $b$: block size and $B = p/b$: the number of blocks, $A^{\rho}_{jj} = 1, A^{\rho}_{ij} =\rho$, where $\rho$ denotes the correlation between
any two variables $i \neq j$. We show results for $b=10$ and $\rho=0.3, 0.8$; results for $b=50$ were fairly similar. The figures support the conclusions drawn from studying the equations.

\subsection{Setting 2: initial ridge estimator}\label{emsepenalized}
It is not straightforward to extend the formulas above to the penalized, $p>n$ setting, because i) penalization introduces bias in the estimates, so $E_{\Y}(\hat{\beta}^{\lambda}_j) \neq \beta_j$; ii)
unlike $(X^TX)^{-1}$, $(X^TX + \lambda I_{p})^{-1}$ does not follow an inverse-Wishart distribution. Hence, we approximate the
EMSE by simulations. In the penalized setting, estimators of $\tau^2$ more advanced than (\ref{hattau}) are available \citep{cule2013ridge}. We proposed an alternative that accounts for the bias of $\hat{\beta}^{\lambda}_j$ due to penalization \citep{WielGRridge}:
\begin{equation}\label{tau2}
\hat{\tau}_2^2 = \frac{\sum_{j=1}^p ((\hat{\beta}^{\lambda_0}_j)^2/v_j - 1)}{\sum_{j,k=1}^p v_j^{-1} c_{jk}^2},
\end{equation}
where $c_{jk}$ is the known coefficient of the bias: $E_{\Y}(\hat{\beta}^{\lambda}_j) = \sum_{k=1}^p c_{jk}\beta_k$  \citep{WielGRridge}, and $\lambda_0$ is an initial value of $\lambda$. We used $\lambda_0=1$, corresponding to a fairly non-informative initial $N(0,\tau^2_0 = 1)$ prior for $\beta_j$. Results were rather insensitive to the exact value of $\lambda_0$.
Figure \ref{emsepen} shows the root EMSE, estimated from 500 generations of $\X$, $\bbeta$ and $\Y$ per setting, using the settings as above except for $n=100$, where $p \leq 20.000$ and $b=50$.

From Figure \ref{emsepen} we observe that $\hat{\tau}^2$ (\ref{hattau}) and $\hat{\tau}_2^2$ (\ref{tau2}) are competitive for $p \leq n$, but the bias-corrected estimator $\hat{\tau}_2^2$ is far superior for $p \gg n$. In fact, the latter is very well on target for $p \geq 500$, supporting the notion that large $p$ is beneficial for EB. Interestingly, even fairly strong correlation seemed to have little impact on the performance (the dotted lines largely overlap the solid ones). This is possibly due to the de-correlation effect of the initial ridge regression with penalty $\lambda_0=1$. This small penalty (much smaller than the true values $\lambda_{\text{true}} = 1/\tau^2 = 1/0.01 =100; 1/0.04=25$) seems to suffice to initially regularize $X^TX$, which explains why the performance improves after $p=n$.  A striking aspect is that across the range of $p$ the performance of $\hat{\tau}_2^2$ is the worst for $p \approx n = 100$. A smaller simulation for $n=200, 500$ shows the same phenomenon, visible from Figure \ref{emsepen200500}. In addition, the use of an even vaguer Gaussian prior with $\tau^2_0=10 \Rightarrow \lambda_0= 0.1$ leads to a similar, and even somewhat more pronounced pattern
in terms of the peak of root EMSE around $p = n$ (data not shown). An explanation is that for $p < n$ the estimation of $\bbeta$ is stable and well-conditioned, while the fairly weak penalty introduces little bias. So, even though $p$ is small, the information from each $\hat{\beta}^{\lambda_0}_j$ is solid which benefits the estimation of $\tau^2$. For $p \approx n$, the penalty necessarily introduces a larger bias in the estimation of $\beta_j$, while the EB estimator does not yet profit much from a large $p$, as is the case for $p > n$. Others have noted this `peaking around $p=n$ phenomenon' as well, e.g. in the context of test error for (regularized) linear discriminant analysis \cite[]{duin2000classifiers}.

Finally, it is tempting to compare the EB estimates of $\tau^2$ with CV-estimates. We noticed that both 5-fold and 10-fold CV (minimizing cross-validated mean squared prediction error for given $\X$) rendered estimates of $\tau^2$ with a root EMSE substantially larger than that of $\hat{\tau}_2^2$. E.g. for $\tau^2 = 0.01, p=1000, n=100$ and independent $X_i$ (so $\Sigma = I_p$), $\text{root EMSE}(\hat{\tau}^2_{\text{CV10}})= 0.0064$, whereas $\text{root EMSE}(\hat{\tau}^2_2)= 0.0015.$ However, one should bear in mind that CV aims at minimizing prediction error rather than at estimating $\tau^2$. In fact, we noticed that the predictive performances usually differed very little when using either $\lambda_{\text{CV10}} = \hat{\tau}^{-2}_{\text{CV10}}$ or $\lambda_{\text{EB}} = \hat{\tau}^{-2}_2$.
Nevertheless, a practical advantage of the EB estimate is its computational efficiency \cite[]{cule2013ridge}: it requires only one ridge-fit, whereas $k$-fold CV requires $k$ times the number of ridge-fits per fold (which depends on the efficiency of the search and the use of approximations).

\section{Application of EB when using co-data}\label{codata}
 \cite{Tai2007} and \cite{novianti2017better} present specific data examples on how co-data can help to improve prediction and variable selection in high-dimensional setting.
 Here, we present two novel prediction examples, both of which use EB to account for co-data.
\subsection{MCMC EB for spike-and-slab models}\label{spikeslab}
Consider a high-dimensional generalized linear model setting, where response $\Y$ is linked to $\X$ via the linear predictor $\eta = \X\bbeta$. Moreover, $\beta_j$ is endowed with a spike-and-slab prior of the form:
$$(\beta_j|\xi_j=0) \sim F_0, \ \ (\beta_j|\xi_j=1) \sim F_1,\ \ \xi_j \sim \text{Bern}(\nu_j),\ \ \ j=1, \ldots, p,$$
where typically $F_0$ is concentrated around zero, or even $F_0 = \delta_0$, and $F_1$ is more dispersed, e.g. Gaussian \cite[]{newcombe2014weibull} or Laplace \cite[]{rovckova2014emvs}. The alternative mixture prior representation is obtained by marginalization over the latent variables $\xi_j$. The model may contain additional nuisance parameters that do not depend on $\xi_j$ (such as error variance $\sigma$), which we omit in the notation below.
Now, assume that we have (several) additional sources of information on the $p$ variables, coded by a $p \times s$ co-data matrix $C$, with $s \ll p$.  Let us model the prior inclusion probability $\nu_j$ parsimoniously as a function of the co-data:
\begin{equation}\label{prior}
\nu_{j,\balpha} = g^{-1}(C_j\balpha),
\end{equation}
where $C_j$ is the $j$th row of $C$, $\balpha$ is an $s \times 1$ vector of hyper-parameters, and $g$ is a link function, e.g. a logit-link. The EB task is to estimate the hyper-parameters $\balpha$. Suppose we have an MCMC sampler which renders posterior samples for all parameters, including the latent ones, given current hyper-parameters $\balpha^{(k)}$. Then, the conditional log-likelihood in (\ref{eq:mcexpectation}) equals
\begin{equation}\label{fact}
\begin{split}
\ell (\Y, \bm{\theta}^{m,(k)}; \balpha) &= \log \pi(\Y,\bbeta^{m,(k)},\bxi^{m,(k)}; \balpha)\\ &= \log \pi(\Y|\bbeta^{m,(k)}) + \log \pi(\bbeta^{m,(k)}|\bxi^{m,(k)}) + \log \pi(\bxi^{m,(k)};\balpha).
\end{split}
\end{equation}
Hence, only the last term depends on $\balpha$, so (\ref{eq:mcexpectation}) reduces to finding:

\begin{equation*}
\begin{split}
\text{argmax}_{\balpha}\{&\sum_{m=1}^M \sum_{j=1}^p \log[\text{Bern}\bigl(\xi_j^{m,(k)}; \nu_{j,\balpha}\bigr)]\}\\&=
\text{argmax}_{\balpha}\{\sum_{j=1}^p \sum_{m=1}^M \log[\text{Bern}\bigl(\xi_j^{m,(k)}); \nu_{j,\balpha}\bigr)]\}\\
 &= \text{argmax}_{\balpha}\{\sum_{j=1}^p \log[\text{Bin}\bigl(\sum_{m=1}^M \xi_j^{m,(k)}); M, \nu_{j,\balpha}\bigr)]\}.
 \end{split}
\end{equation*}
The latter equality holds, because $\nu_{j,\balpha}$ does not depend on $m=1,\ldots, M$ and the $\text{Bin}(M,q)$ density differs from the product of $M\ \text{Bern}(q)$ densities only by a binomial factor that does not depend on $\balpha$. Hence, estimating $\balpha$ reduces to binomial regression of `observations' $B_j^k = \sum_m \xi^{m,(k)}_j, j=1, \ldots, p,$ on the $s$ columns of design matrix $C$. The previous estimate is then iteratively updated by this one, as in (\ref{eq:gibbsseq}), for a new round of MCMC sampling.

The reduction to simple regression is feasible due to the factorization (\ref{fact}) and the i.i.d. Bernoulli prior for $\xi_j$. Other Bayesian sparse regression models, like the Bayesian elastic net (see Section \ref{EBCV}), can also be represented as (scale) mixtures, but with a remaining dependency of $\bbeta$ on $\balpha$ plus a more complex dependency of the mixture proportions on $\balpha$.
While conceptually simple, the algorithm above is computationally demanding, requiring efficient implementations of spike-and-slab MCMC \cite[such as those by][]{peltola2012finite, newcombe2014weibull}. Variational Bayes approximations may be an alternative \cite[]{carbonetto2012scalable}, in combination with an EM-type maximization \cite[]{bernardo2003variational}.

\subsection{Simulation Example: interval estimation}\label{hybBay}
\subsubsection{Empirical Bayes versus Full Bayes} EB is not `truly' Bayes, because the prior parameters are fixed after estimating these from the data. A disadvantage of many full Bayes settings, however, is computational time: the extra layer of priors may lead to a strong increase, e.g. from seconds to minutes (see the example of \cite{Bar2011} with 2,000 variables) or from minutes to several hours.
For the multivariate low-dimensional setting, \cite{CarlinLouis} show that, despite their lack of error propagation, EB methods can be rather competitive to full Bayes ones in terms of frequentist coverage probabilities of the parameter credibility intervals. Below we compare Bayes, EB and hybrid credibility intervals for predictions in medium-dimensional settings with $p$ of the same order of magnitude as $n$.

\subsubsection{Setting}
As indicated in the Introduction, the medium-dimensional case is likely to become more and more relevant in clinical prediction.
In a clinical setting, the uncertainty of each individual's prediction is of importance. The Bayesian paradigm lends itself well for obtaining interval estimates in (penalized) regression settings, because it allows uncertainty propagation of the tuning parameter(s). In the low-dimensional Bayesian linear regression setting, \cite{morris1983parametric, basu2003empirical} provide theoretical guarantees for the
coverage of an EB interval which accounts for the uncertainty of the prediction and the shrinkage factor. In a Bayesian logistic regression setting, we compare three models for the priors of the coefficients $\beta_j$ in terms of coverage of the posterior predictive intervals. These models differ in the level of error propagation. We assume that the variables are grouped into $G$ groups based on co-data \citep{Tai2007, WielGRridge}.

\subsubsection{Models}
Denote the groups of variables by $\mathcal{G}_g, g= 1, \ldots, G.$ We assume:
\begin{equation}\label{likelihood}
\begin{split}
Y_i &\sim \text{Bernoulli}(\expit(\mathbf{X}_i\bbeta))\\
\beta_j &\sim N(0,\tau^2_g), j \in \mathcal{G}_g,
\end{split}
\end{equation}

\noindent
where $\expit(x) = \exp(x)/(1+\exp(x))$. We consider three models for precisions $\tau^{-2}_g$. First, the \emph{Empirical Bayes (EB)} model:
\begin{equation}\label{EBmodel}
\tau^{-2}_g = \lambda \lambda_g^2,
\end{equation}
where $\lambda$ and $\lambda_g$ are fixed. Second, the (conjugate) \emph{Full Bayes (FB)} model:
\begin{equation}\label{FBmodel}
\tau^{-2}_g \sim \Gamma(\alpha_1,\alpha_2),
\end{equation}
with $\alpha_1$ and $\alpha_2$ such that the prior is rendered uninformative. Third, the \emph{Hybrid} model:
\begin{equation}\label{Hybmodel}
\begin{split}
\tau^{-2}_g &= \tau^{-2} \lambda_g^2\\
\tau^{-2} &\sim \Gamma(\alpha_1,\alpha_2),
\end{split}
\end{equation}
with $\alpha_1$ and $\alpha_2$ such that the prior is rendered uninformative and $\lambda_g$ fixed.

Model (\ref{EBmodel}) is equivalent to the one used in \cite{WielGRridge}. We estimate the global ridge tuning parameter $\lambda$ by cross-validation and the group multipliers $\lambda_g$ by moment-based EB,
as in (\ref{momentest2}). This model generally renders good point predictions, and is computationally very efficient. It may, however, not suffice for interval estimation, because the uncertainty of $\hat{\tau}^{-2}_g$ is not accounted for. Model (\ref{FBmodel}) renders a classical Bayesian random effects model. It may be the preferred model when $G$ is small and the number of features per group is large: the estimation of $\tau^{-2}_g$ will be relatively precise and uncertainty of $\tau$ is propagated. However, this model is computationally cumbersome for large $G$ due the the large number of hyper-priors which need to be integrated out when computing the posterior of $\bbeta$. Moreover, when some groups are small, the imprecise estimation of $\tau^{-2}_g$ may render inferior predictions.
Model (\ref{Hybmodel}) is a compromise: it contains only one random hyper-parameter, $\tau$. So, model (\ref{Hybmodel}) is computationally efficient, while still propagating uncertainty of $\tau$. We assume the group-specific penalty multipliers $\lambda_g$ to be identical to those in Model (\ref{EBmodel}) to ensure comparability.

\subsubsection{A small simulation}
In combination with (\ref{likelihood}), (\ref{EBmodel}) to  (\ref{Hybmodel}) render three Bayesian models that are implemented using the R-package INLA \cite[]{Rue2009},
after substituting the estimated $\lambda$ and $\lambda_g$'s into model (\ref{EBmodel}) and $\lambda_g$ into (\ref{Hybmodel}). We evaluate 95\% posterior intervals for the prediction probabilities on an event, $q_i = \expit(\mathbf{X}_i\bbeta)$. We consider equal-tail intervals and highest-probability density (HPD) intervals \citep{CarlinLouis}. The latter concentrate more around the posterior mode, so may be less vulnerable to shrinkage than the first. The following simulation settings were used for model (\ref{likelihood}):

\begin{itemize}
\item \# groups $G=2,5$. \# variables per group: $p_G$. Total $\#$ variables: $p=p_G * G$.
\item For $G=2$, all 3 models are applied; $n_{\text{train}} = n_{\text{test}} = 100, p_G=10, 20, 30, 40, 50.$
\item For $G=5$, Models (\ref{EBmodel}) and (\ref{Hybmodel}) are applied; $n_{\text{train}} = 200, n_{\text{test}}=100, p_G=10, 20, 40$. Model (\ref{FBmodel}) was not evaluated for this computationally demanding case.
\item For variable $j$ in group $g =1, \ldots, G$, $\beta_j \sim N(0,\tau^2_g)$, where $\tau_g = \tau_0 2^{-(g-1)}$, so prior standard deviations decrease by a factor of 2 for each next group $g$; $\tau_0$ is calibrated such that $\approx 20\%$ of observations render extreme probabilities $q_i$ ($<0.05$ or $>0.95$).
\item Correlation between variables occurs in blocks of 5, with correlation $\rho = 0.1$.
\item Each simulation setting was repeated $n_{\text{rep}} = 50$ times; coverage of 95\% posterior intervals for $q_i, i=1, \ldots, n_{\text{test}},$ is studied.
\end{itemize}

We also considered $\rho = 0.5$ and constant $\beta$'s within each group (hence not obeying the Gaussian prior). Results were very similar and hence not shown.

\subsubsection{Results}
We focus on the intervals here; the results on the point predictions (posterior modes) of $q_i$ are rather similar for models (\ref{EBmodel}) to (\ref{Hybmodel}).
We compute average coverage of the true $q_i$ by the 95\% intervals across all test samples, averaged over $n_{\text{rep}}$ repeats. We then plot $q_i$ versus the moving average coverage on overlapping
windows of 200 predictions. These are displayed in Figures \ref{hpdeq} and \ref{comp} for two simulation settings; see the Supplementary Material for other settings. First, from Figure \ref{hpdeq} it is clear that HPD-intervals outperform their equal-tailed counterparts, in particular for the extreme $q_i$'s. Equal-tailed intervals are more sensitive to the bias introduced by penalization, which is stronger for the extremes. This is in line with the findings of \cite{CarlinLouis}. Possibly more surprising is the somewhat inferior coverage for the FB model (\ref{FBmodel}) in this simulation. It may result from use of the conjugate, but possibly wrong prior in (\ref{FBmodel}), or from the small value of $p_G$. The counterpart, the EB model, performs better, but still renders too low coverage for the extremes. This likely results from too narrow intervals caused by lack-of-propagation of the uncertainty of the global penalization parameter, $\tau^{-2} \propto \lambda$. The hybrid model (\ref{Hybmodel}) seems to correctly balance the empirical Bayes estimation of the group-wise parameters and the full Bayes handling of $\tau^{-2}$. Due to shrinkage, a small bias for the coverage remains for extreme $q_i$'s.

\section{Discussion and extensions}
We showed that Empirical Bayes is a versatile and powerful approach to `learn from a lot' in two ways: first, from the large number of variables and
second, from prior information on the variables, stored as co-data. We reviewed several methods for EB estimation in a broad spectrum
of prediction frameworks. This illustrated that developing EB estimators of penalty, prior or other tuning parameters ranges from simple to challenging, depending on the prediction framework and the ambition in terms of number of hyper-parameters to estimate. While EB can be regarded as a `competitor' for cross-validation and full Bayes in a frequentist or Bayesian setting, respectively, we argued that hybrid solutions may prove useful to exploit the strengths of the approaches.

In the Bayesian framework, maximization of the marginal likelihood is the default EB criterion. This is often computationally intensive. Variational Bayes, which returns a lower bound for the marginal likelihood, in combination with EM-type optimization, can strongly alleviate the computational burden \cite[]{bernardo2003variational}. It requires careful development of the approximations for the model at hand, and verification of accuracy (e.g. by Gibbs sampling) for numerical examples. Alternatively, in a variable selection setting one may settle for a conditional EB approach \cite[]{george2000calibration} by conditioning on the included variables, thereby avoiding integration over the large model space.

This overview is by no means complete. Specific EB methods have been developed, in particular also for model-free predictors.
E.g., for the random forest, \cite{taddy2015bayesian} estimate the trunk of the trees, which may stabilize results and save considerable computing time compared to a fully Bayesian approach. \cite{Beest2017} demonstrate
that co-data may be used to improve random forest predictions by moderating the sampling weights of the variables.


As illustrated, EB allows to account for co-data, but is not the only way. Full Bayes alternatives exist, in particular for the purpose of variable selection \cite[]{Ishwaran2005spike, sillanpaa2009review, quintana2013integrative}. These are generally computationally very demanding for typical high-dimensional settings with a large number of variables. Moreover, frequentist solutions have been proposed, which usually require additional tuning parameter(s) to cross-validate \cite[]{Bergersen2011, jiang2016variable} or a group penalty \cite[]{Meier2008, simon2013sparse}. The latter may perform less well than EB-based regularization per group
when the number of groups is small \cite[]{novianti2017better}. However, a group penalty may be particulary powerful when the number of groups is large given its much more parsimonious representation of the group structure. Combination of the two principles is an interesting future research direction.

``Empirical Bayes is still in its adolescence'' \cite[]{EfronBook}, which is particularly true for high-dimensional prediction and variable selection. More theory on the quality of the estimators as a function of $n$ and $p$  for a variety of prediction models will be welcomed by the community. EB theory for large $p$ settings is an active field of research, which will likely lead to more general results. From a practical perspective, prediction accuracy can always be estimated by (repetitive) training-test splits, which allows evaluation of the EB prediction versus alternatives for the data at hand. Evaluation of variable selection is more difficult. We find it useful to compare indirectly by evaluating the predictive accuracies of models of the same size. This enabled us to show that co-data-based EB may improve the predictive performance of small models \cite[]{novianti2017better}.
New prediction methods with various types of penalties, priors or other tuning parameters are frequently introduced. These may benefit from dedicated EB estimators, in particular when multiple tuning parameters are involved. Extension of EB methods towards estimation of multivariate priors should allow to better accommodate network-type information \cite[]{stingo2011incorporating, rovckova2014emvs}. Finally, priors that are modeled as a function of various sources of co-data are increasingly relevant in this `Big Data era'. Developing EB-estimators of hyper-parameters of such priors will require either a parsimonious representation or regularization on the level of hyper-parameters to avoid over-fitting.

\section{Acknowledgements}
We thank Paul Newcombe for fruitful discussions on spike-and-slab models and Carel Peeters for critical reading of this manuscript.



\pagebreak
\section{Figures}

\begin{figure}[!h]
	\begin{minipage}[c]{\linewidth}
    \centering
  		\subfigure[][p=n=200]{
  			\includegraphics[width=1.00\textwidth]{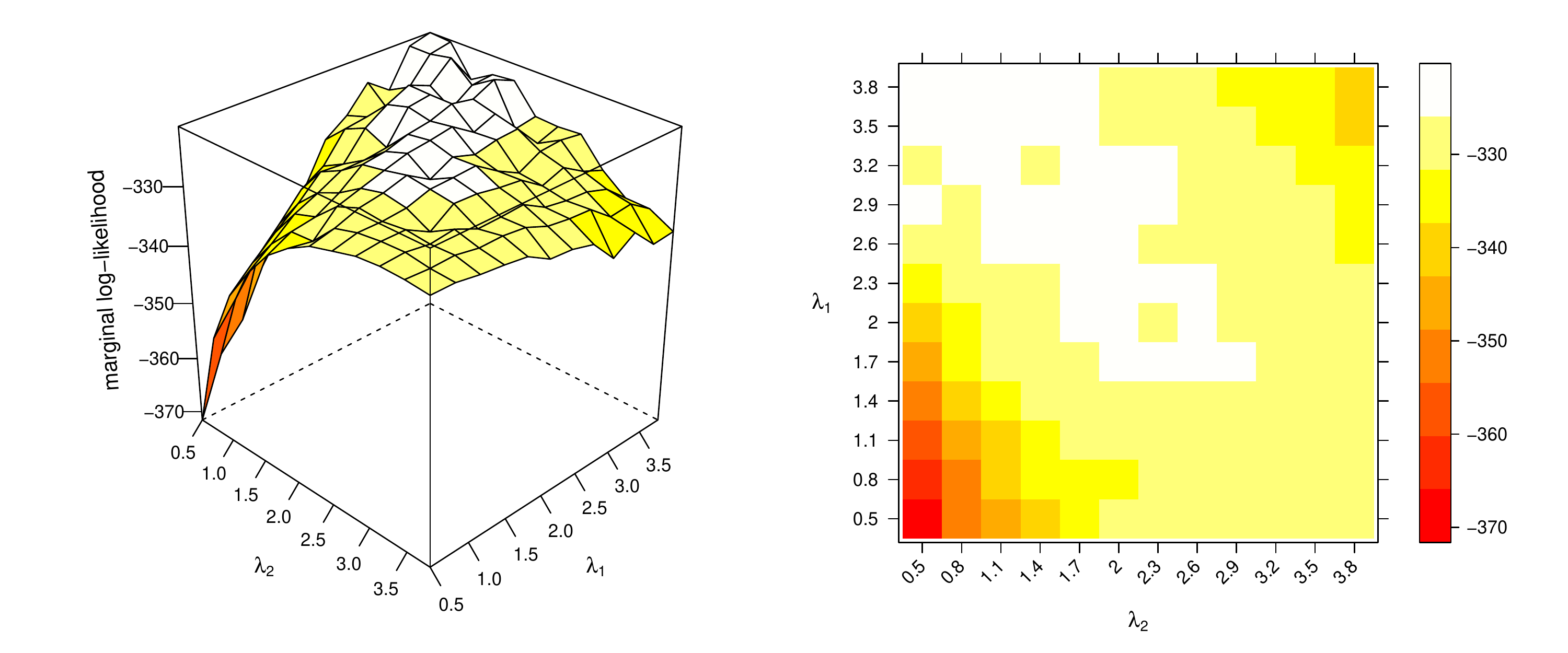}
  		}
  \end{minipage}\vfill
  \begin{minipage}[c]{\linewidth}
    \centering
  		\subfigure[][p=1000, n=100]{
  			\includegraphics[width=1.00\textwidth]{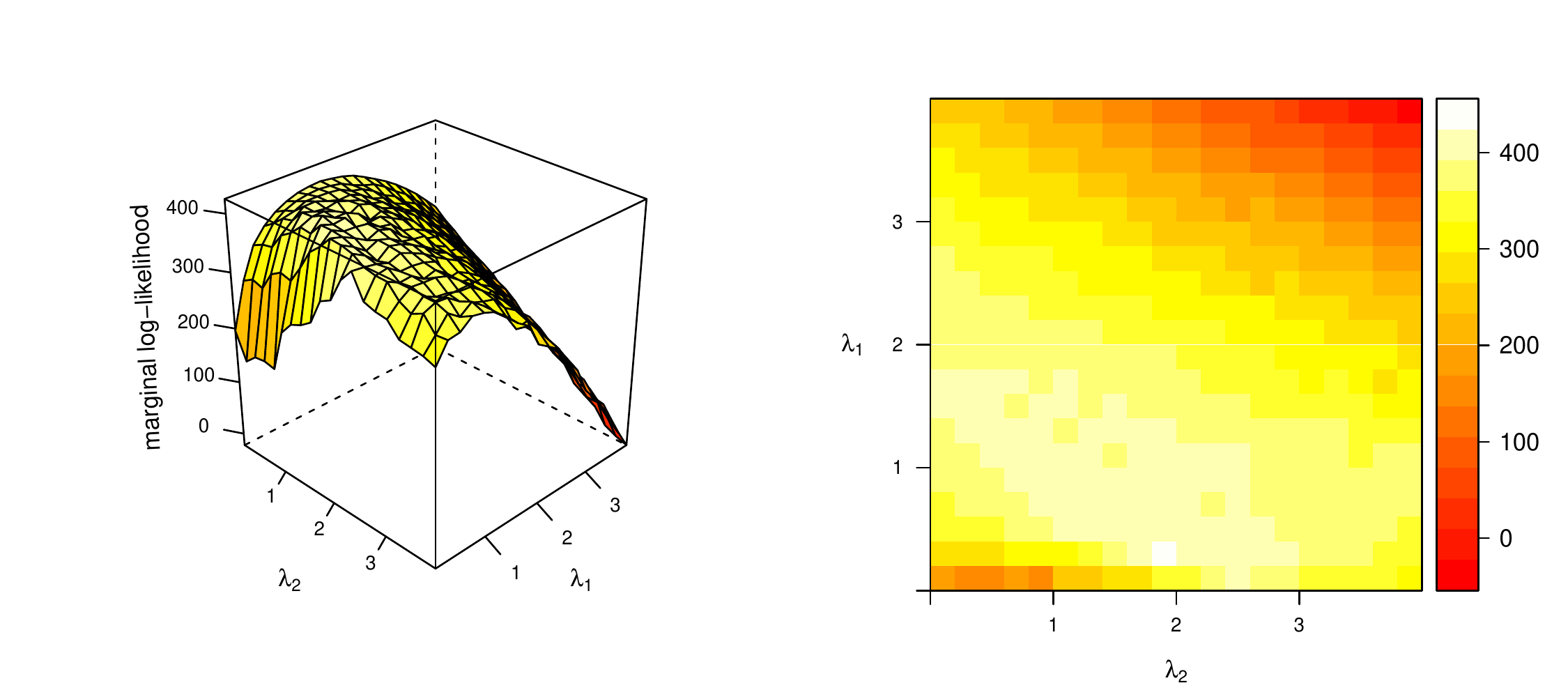}

  		}
  \end{minipage}
  \caption{Marginal likelihood (up to constant) as a function of $\lambda_1$ and $\lambda_2$ in the Bayesian elastic net}\label{margllEN}
  \end{figure}

\begin{figure}[h]
\includegraphics[width=0.95\textwidth]{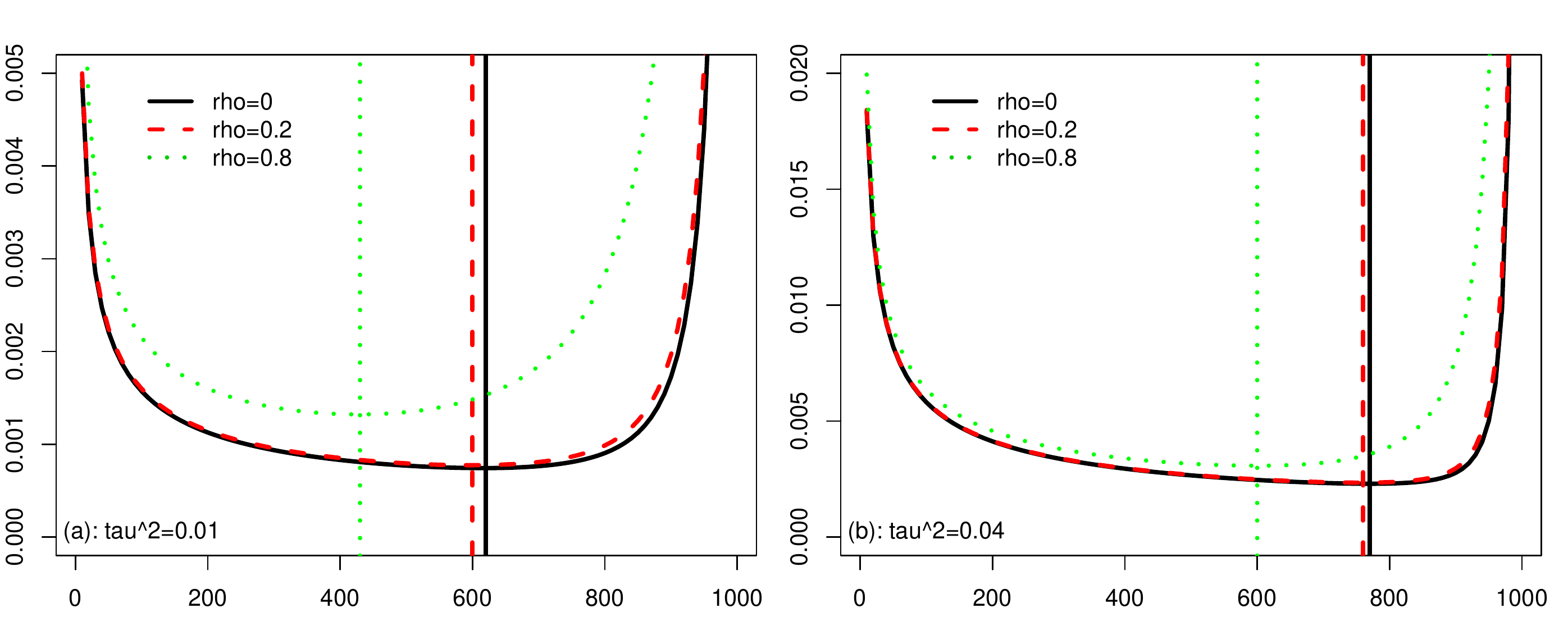}
\caption{Y-axis: root EMSE (\ref{emseresult},\ref{emseresultind}), X-axis: $p$. Settings: $n=1000$; independent $\X$ $(\rho=0)$, $\rho = 0.3, 0.8$; $b=10$; and $\tau^2=0.01$ (a), $\tau^2=0.04$ (b). Vertical line denotes the minimum.
}\label{emseunpen}
\end{figure}

\begin{figure}[h]
\includegraphics[width=0.95\textwidth]{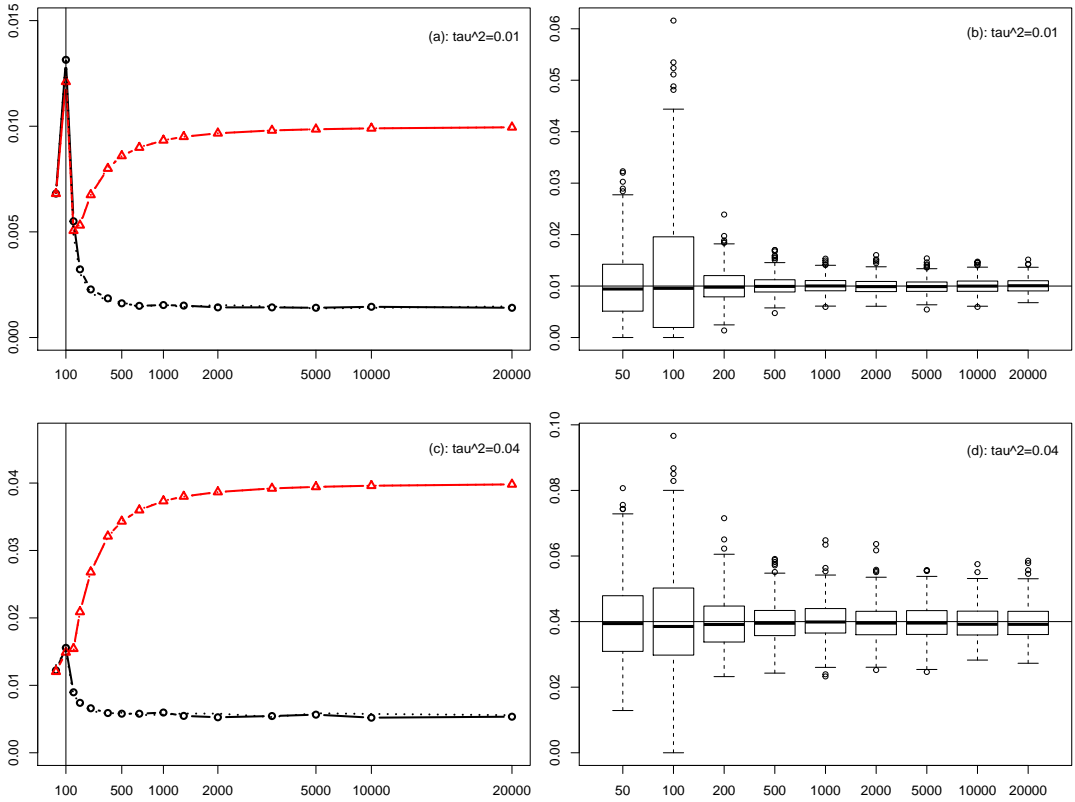}
\caption{(a) and (c): root EMSE (y-axis) versus $p$ (x-axis; square-root scale) for $\tau^2=0.01, 0.04$ and $n=100$. Red: estimator $\hat{\tau}^2$ (\ref{hattau}), black: bias-adjusted estimator $\hat{\tau}_2^2$ (\ref{tau2}).
Solid lines: independent $\X$, dotted line: block-correlation, $b=50, \rho=0.8$. Vertical line denotes $p=n=100$. Sub-figures (b) and (d): Corresponding box-plots of $\hat{\tau}_2^2$ for 500 simulations in the independent $\X$ setting.}\label{emsepen}
\end{figure}

\begin{figure}[h]
\begin{center}
\includegraphics[width=0.75\textwidth]{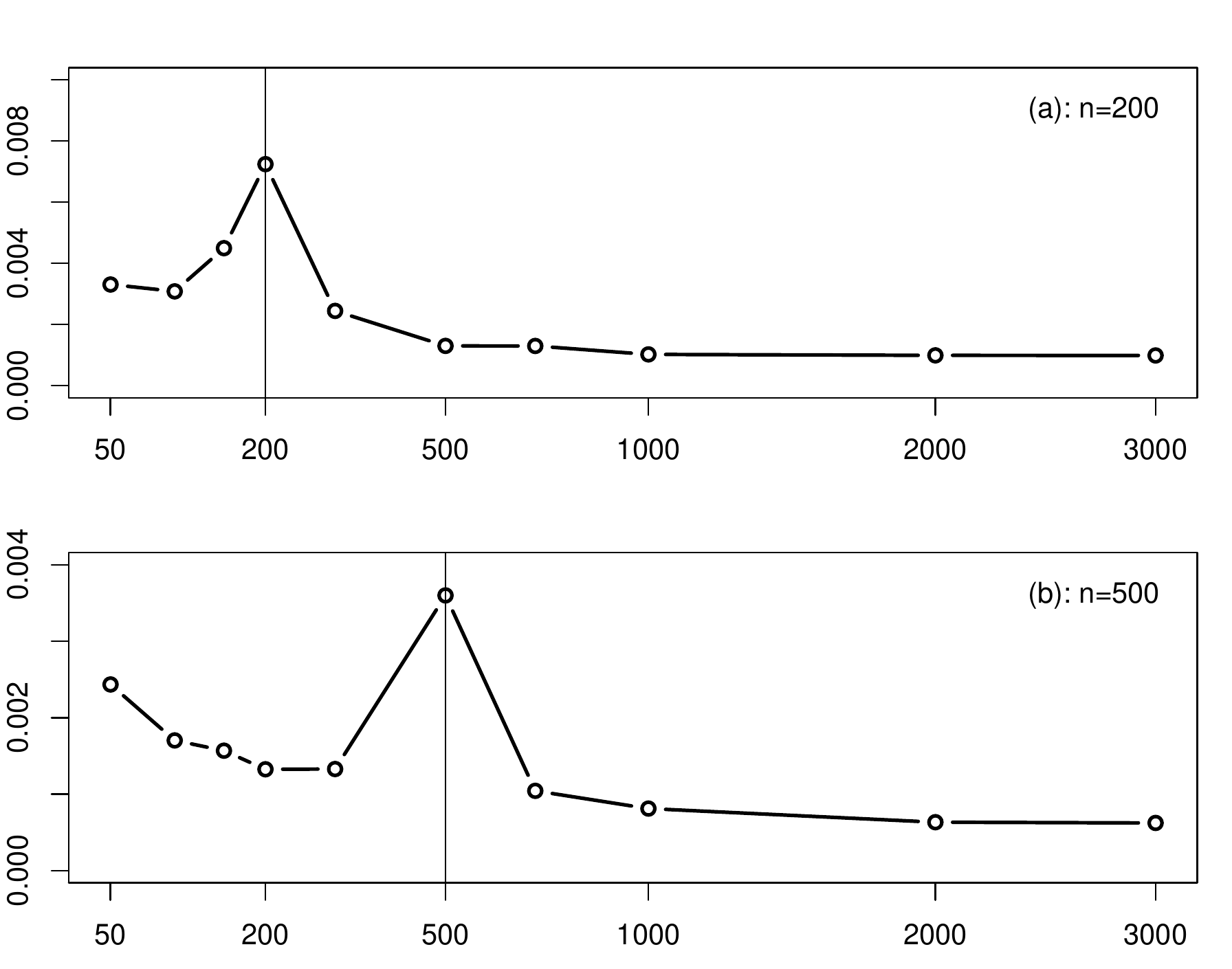}
\caption{Y-axis: root EMSE of $\hat{\tau}_2^2$ (\ref{tau2}); X-axis: $p$ (square-root scale). Settings: $\tau^2=0.01$; independent $\X$; and $n=200$ (a), $n=500$ (b). Vertical line denotes $p=n$. Results based on 200 simulations.}\label{emsepen200500}
\end{center}
\end{figure}

\begin{figure}[h]
	\begin{minipage}[c]{0.50\linewidth}
    \centering
  		\subfigure[][Equal-tailed interval]{
  			\includegraphics[width=1.00\textwidth]{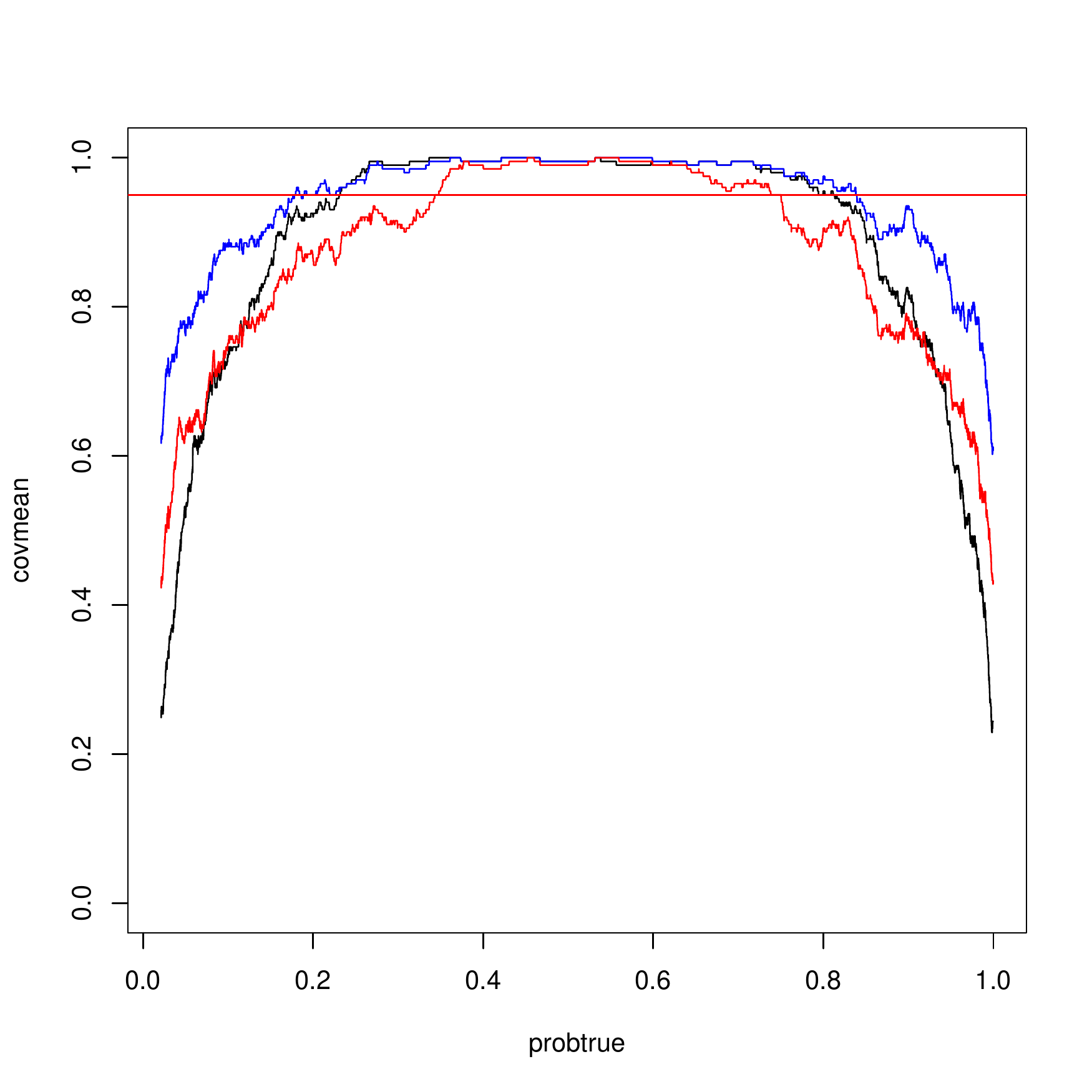}
  		}
  \end{minipage}\hfill
  \begin{minipage}[c]{0.50\linewidth}
    \centering
  		\subfigure[][HPD interval]{
  			\includegraphics[width=1.00\textwidth]{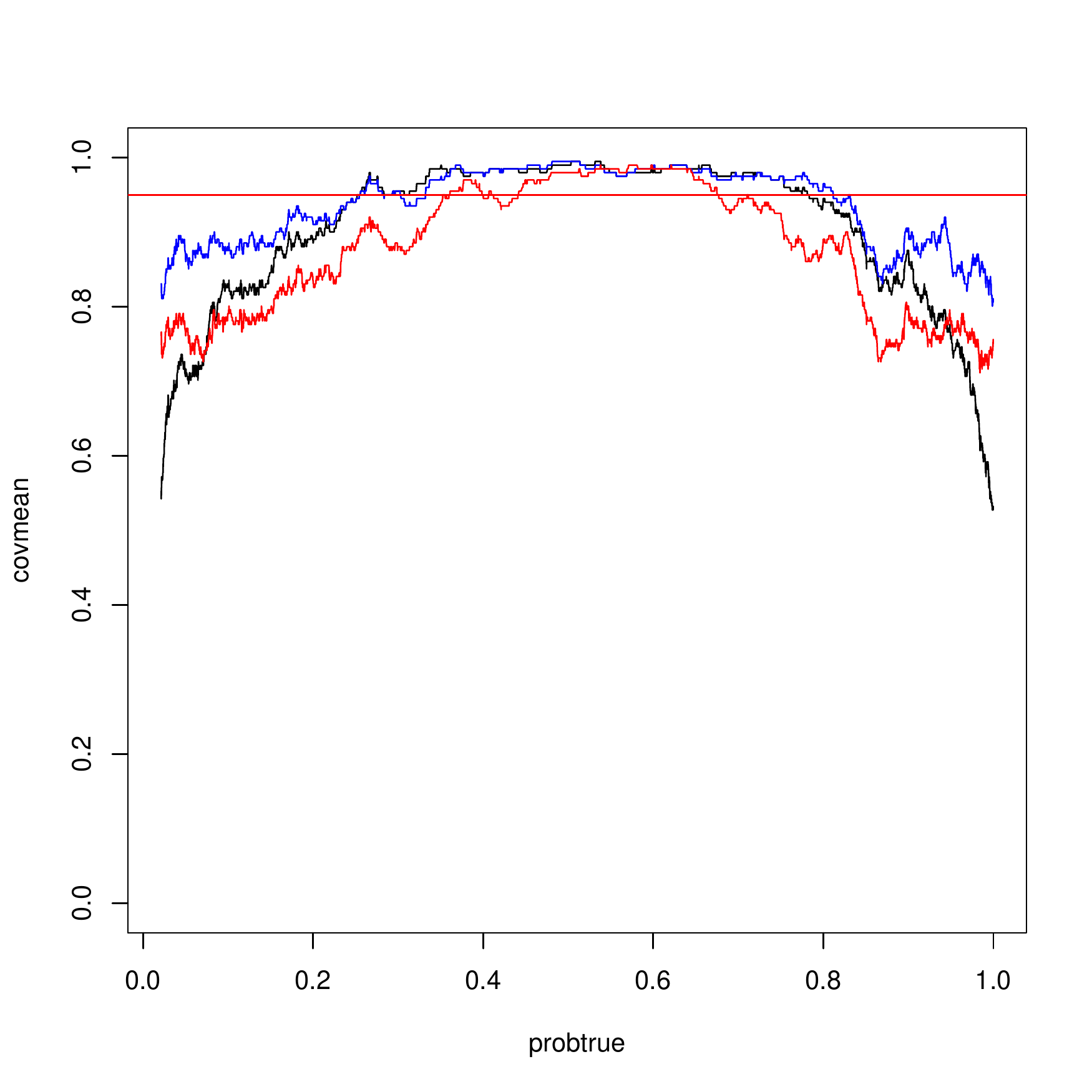}

  		}
  \end{minipage}
\caption{X-axis: True event probability $q_j$; Y-axis: mean coverage of 95\% posterior intervals for event probability. Mean is estimated by moving average. Case: $G=2, p_g=30, p = G*p_G = 60, n_{\text{train}} = 100$. Methods:
\textcolor{blue}{Hyb}, \textcolor{black}{EB}, \textcolor{red}{FB}}\label{hpdeq}
\end{figure}

\begin{figure}[h]
	\begin{minipage}[c]{0.50\linewidth}
    \centering
  		\subfigure[][$p_g = 10, G=5$]{
  			\includegraphics[width=1.00\textwidth]{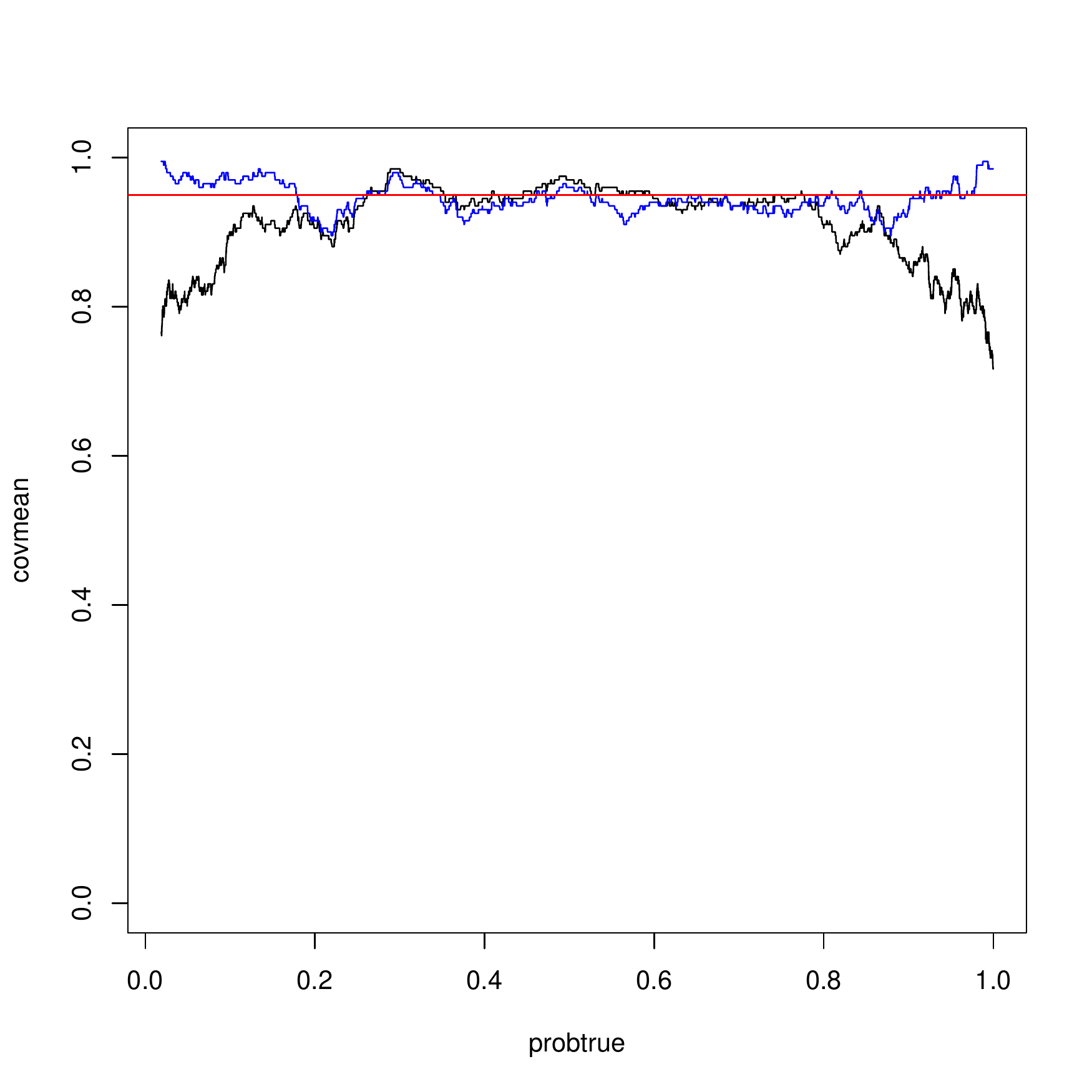}
  		}
  \end{minipage}\hfill
  \begin{minipage}[c]{0.50\linewidth}
    \centering
  		\subfigure[][$p_g = 40, G =5$]{
  			\includegraphics[width=1.00\textwidth]{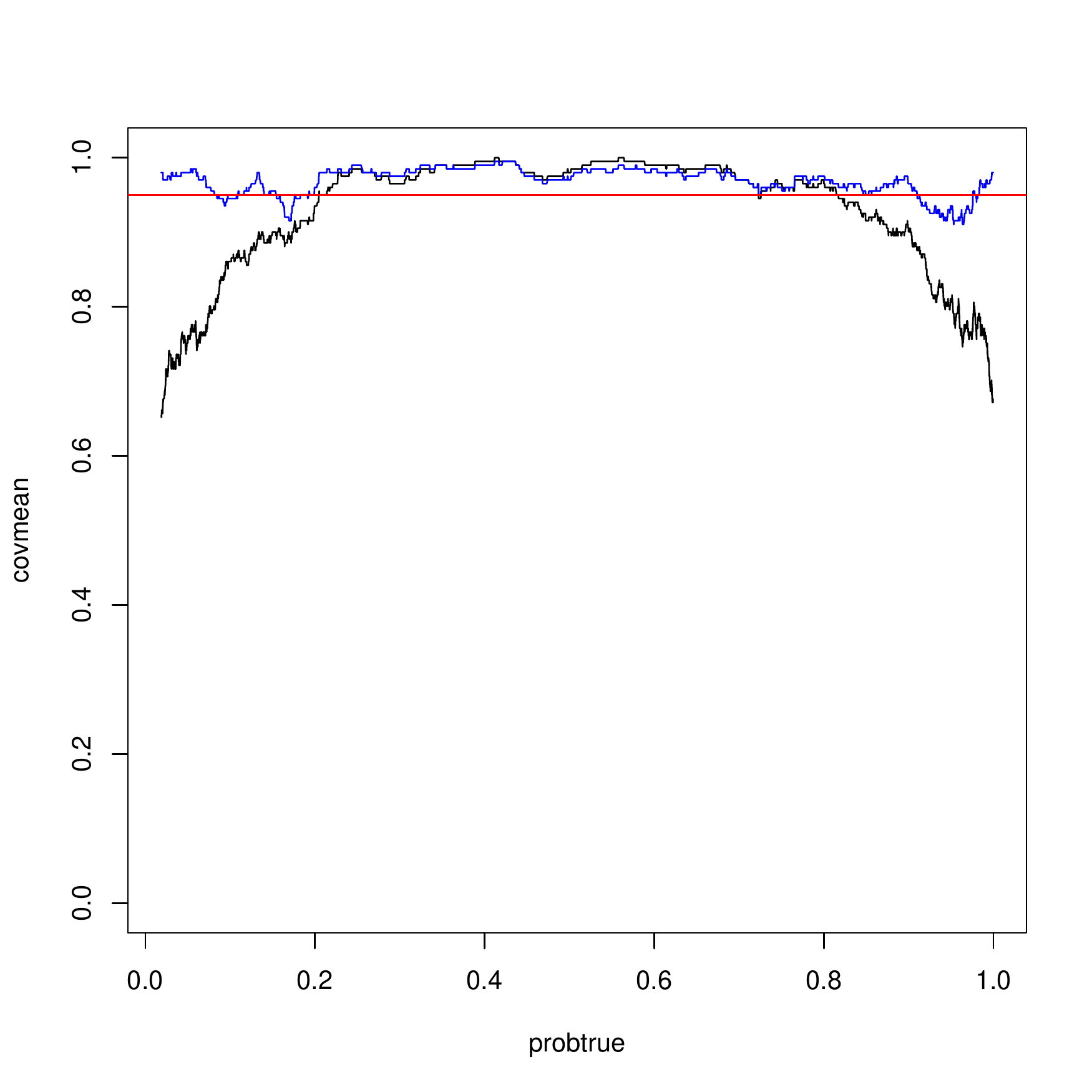}
  		}
  \end{minipage}
\caption{X-axis: True event probability $q_j$; Y-axis: mean coverage of 95\% posterior intervals for event probability. Mean is estimated by moving average. Cases: $G=5, p_g=10, 40; p = G*p_G = 50, 200; n_{\text{train}} = 200$. Methods:
\textcolor{blue}{Hyb}, \textcolor{black}{EB}}\label{comp}
\end{figure}

\clearpage
\section{APPENDIX}

\subsection{Baseball batting example, revisited}
We shortly revisit the famous baseball batting example \cite[]{Efron1975data}, often used as a scholarly example of EB. While this is an estimation problem instead of a prediction problem, we revisit it for several reasons:
i) it is a well-known example for which the true values are known; ii) the EB objective function is the same as for diagonal linear discriminant analysis; and iii) by casting the problem to a large $p$ setting it allows us to show the importance of $p$ being large.

For 18 baseball players, their batting averages over the first 45 bats are recorded and denoted by $B_i$. The batting averages over the remainder of the season are also known, and considered
to be the truth. We follow \cite{Houwelingen2014} by modeling $B_i \sim N(\theta_i,\sigma^2_i), $ where the aim is to estimate $\theta_i$. The variances are estimated by $\hat{\sigma}^2_i = B_i(1-B_i)/45$. Then, to effectuate shrinkage \cite{Houwelingen2014} applies a Gaussian prior $N(\mu,\tau^2)$ to $\theta_i$.
In the formulation of the marginal likelihood (see Main Document), this implies hyper-parameter $\balpha = (\mu,\tau^2)$, and estimation of $\balpha$  is straightforward due to the conjugacy of the likelihood and the prior:
$\hat{\tau}^2 = V(B_i) - \overline{\hat{\sigma}_i^2}$ and $\hat{\mu} = \sum_{i=1}^{n} w_iB_i/\sum_{i=1}^{n} w_i,$  with $w_i= (\hat{\tau}^2 + \hat{\sigma}^2_i)^{-1}$. Then, the posterior mean estimate equals
$\hat{\theta}_i = E(\theta_i|B_i; (\hat{\mu},\hat{\tau}^2)) = \hat{\mu} + \hat{\tau}^2(\hat{\tau}^2+\hat{\sigma}_i^2)^{-1}(B_i-\hat{\mu})$.
The conclusion in \cite{Houwelingen2014} is that the shrinkage prior slightly reduces the mean squared error, but enforces too strong shrinkage for the extremes. E.g. for the best player $\hat{\theta}_1 = 0.271,$ whereas $X_1=0.400$ and true $\theta_1 = 0.346$. Two possible explanations come to mind: the estimate of the prior parameters is not good due to $p=n$ being small and/or the prior does not accommodate the extremes well. We investigate this.

First, we simulate 10,000 additional true values from a density estimate with Gaussian kernel (using \texttt{R}'s \texttt{density} command) applied to $(\theta_1, \ldots, \theta_{18})$. To obtain $B_i$, $i = 19, \ldots, 10018$, Gaussian noise was added with variances $\theta_i(1-\theta_i)/45$. The estimates obtained in \cite{Houwelingen2014} were $\hat{\mu} = 0.256$ and $\hat{\tau}^2= 0.000623$. The latter seems to be a major cause of over-shrinkage: the true variance computed from the 18 known $\theta_i$'s equals $0.00143$. If we estimate $\tau^2$ from the large data set, a much better estimate is obtained: $\tilde{\tau}^2 = 0.00195$, as compared to the variance of the 18 known plus 10,000 generated true $\theta_i$'s, equaling $0.00166$. From this, we obtain posterior mean estimate $\tilde{\theta}_1 = 0.293,$ which is substantially closer to $\theta_1 = 0.346$ than $\hat{\theta}_1 = 0.271$.
Estimates for all 18 players are displayed in Figure \ref{baseball}(a).

In this example, it is natural to replace the Gaussian prior by a 3-component Gaussian mixture prior (bad, mediocre and good players): $\theta_i \sim \sum_{k=1}^3 p_k N(\mu_k,\tau_k^2).$ Then, $\balpha$ consists of 8 hyper-parameters given that $p_3 = 1-p_1-p_2$. We employed the EM-type algorithm of \cite{WielShrinkSeq} to maximize the marginal likelihood (see Main Document) in terms of $\balpha$. Here, we use that the likelihood is Gaussian, and the Gaussian mixture prior is conjugate to it. The latter also facilitates straightforward computation of the shrunken estimator $\hat{\theta}_i^{\text{Mixt}} = E(\theta | B_i; \balpha)$. In this setting, the mixture prior is fairly close to the estimated Gaussian prior, and so are the shrunken estimates, as displayed in Figure \ref{baseball}(b). Slightly less shrinkage for the extremes is observed, though. For example,
$\tilde{\theta}_1^{\text{Mixt}} = 0.298$.

\subsection{Bayesian elastic net}\label{app}

The Bayesian linear elastic net model, as used in the Main Document, is \cite[]{Li2010bayesian}:

\begin{align*}
\Y | \X, \bbeta, \sigma^2 & \sim \N (\X \bbeta, \sigma^2 \I) \\
\bbeta | \sigma^2 & \sim \prod_{j=1}^p g(\lambda_1, \lambda_2, \sigma^2) \cdot \exp \[ -\frac{1}{2 \sigma^2} (\lambda_1 |\beta_j| + \lambda_2 \beta_j^2) \] \\
\sigma^2 & \sim f(\sigma^2),
\end{align*}
with some arbitrary (possibly improper) density $f(\sigma^2)$. The normalizing constant $g(\lambda_1, \lambda_2, \sigma^2)$ is given by:

$$
g(\lambda_1, \lambda_2, \sigma^2) = \sqrt{\frac{\lambda_2}{4\sigma^2}} \phi \(\frac{\lambda_1}{\sqrt{4 \lambda_2 \sigma^2}}\) \Phi \(\frac{-\lambda_1}{\sqrt{4 \lambda_2 \sigma^2}} \)^{-1}
$$

\para
Since the simulations are for illustrative purposes only, the error variance was kept fixed at its true value ($\sigma^2 = 1$) throughout the simulations. Then, after introducing the latent variables $\btau = \begin{bmatrix} \tau_1 \cdots \tau_p \end{bmatrix} \tr$, we have the following conditional distributions for $\bbeta$ and $\btau$:

\begin{align*}
\bbeta | \Y, \sigma^2, \btau & \sim \N (\A^{-1} \X \tr \Y, \sigma^2 \A^{-1})\\
(\btau - \ones)| \Y, \sigma^2, \bbeta & \sim \prod_{j=1}^p \GIG \(\frac{1}{2}, \psi=\frac{\lambda_1^2}{4 \lambda_2 \sigma^2}, \chi_j=\frac{\lambda_2 \beta_j^2}{\sigma^2}\),
\end{align*}
where $\A=\X \tr \X + \lambda_2 \diag[\tau_j/(\tau_j - 1)]$, and $\GIG$ denotes the generalized
inverse Gaussian distribution.

\subsubsection{Marginal likelihood from Gibbs samples}
According to \cite{chib95}, the log marginal likelihood of a Bayesian model may be calculated from the converged Gibbs samples as:

\begin{equation}
\log m (\Y) = \log \[ \frac{f(\Y | \bbeta^*) \pi(\bbeta^*)}{p(\bbeta^* | \Y)} \] \approx \[\log \frac{f(\Y | \bbeta^*) \pi(\bbeta^*)}{K^{-1} \sum_{k=1}^K p(\bbeta^* | \Y, \btau^{(k)})}\],
\end{equation}
where $\bbeta^*$ is some high posterior density point of $p(\bbeta | \Y)$ and $\btau^{(k)}$ are Gibbs samples indexed by $k=1 , \dots, K$. In principle, any point $\bbeta^*$ may be used, but for the sake of efficiency a high-density point of $\bbeta$ is preferred, such as the posterior mode.
Then, for fixed $\sigma^2$, the log marginal likelihood is approximated by:

\begin{align*}
\log m (\Y) & \approx \[\log \frac{f(\Y | \bbeta^*) \pi(\bbeta^*)}{K^{-1} \sum_{k=1}^K p(\bbeta^* | \Y, \btau^{(k)})}\] = \log K - \log \[ \sum_{k=1}^K \frac{p(\bbeta^* | \Y, \btau^{(k)})}{f(\Y | \bbeta^*) \pi(\bbeta^*)} \] \\
& = \log K - \log \Bigg\{ \sum_{k=1}^K \exp \bigg[ \frac{n - p}{2} \log(2 \pi) + \frac{n}{2} \log \sigma^2 + \frac{p}{2} \log 4 - \frac{n}{2} \log \lambda_2 \\
& - p \log \phi \(\frac{\lambda_1}{\sqrt{4 \lambda_2 \sigma^2}}\) + p \log \Phi \(\frac{-\lambda_1}{\sqrt{4 \lambda_2 \sigma^2}} \) + \frac{\lambda_1}{2 \sigma^2} \sum_{j=1}^p |\beta^*_j| - \frac{\lambda_2}{2 \sigma^2} \sum_{j=1}^p \frac{(\beta_j^*)^2}{\tau_j^{(k)} - 1} \\
& + \frac{1}{2} \sum_{j=1}^p \log \( \frac{\tau_j^{(k)}}{\tau_j^{(k)} - 1} \) + \frac{1}{2} \log \left| \lambda_2 \I_n + \X (\I_p - (\T^{(k)})^{-1}) \X \tr \right| \\
& + \frac{\lambda_2}{2 \sigma^2} \Y \tr \( \lambda_2 \I_n + \X (\I_p - (\T^{(k)})^{-1}) \X \tr \)^{-1} \Y \bigg] \Bigg\},
\end{align*}
where $\T^{(k)} = \diag(\tau_j^{(k)})$.

\para
Sampling from the multivariate normal is a costly operation in high dimensions. In \cite{bhattacharya_fast_2015} an efficient sampling scheme for $\bbeta$ is described: \\

\begin{algorithmic}
	\State{Set $\D = \lambda_2^{-1} \I_p + \lambda_2^{-1} \diag(\tau_j^{-1})$}
	\State{Generate $\mathbf{u} \sim \N (\zeros, \D)$}
	\State{Generate $\mathbf{v} \sim \N(\zeros,\I_n)$}
	\State{\Return{$\bbeta = \mathbf{u} + \D \X \tr (\X \D \X \tr + \I_n)^{-1} (\Y - \X \mathbf{u} + \mathbf{l})$}}.
\end{algorithmic}

\para
Furthermore, if $(\tau_j - 1) | \Y, \sigma^2, \beta_j \sim \GIG (1/2, \psi, \chi_j)$, then $1/(\tau_j - 1)| \Y, \sigma^2, \beta_j \sim \IGauss (\mu_j=\sqrt{\psi/\chi_j}, \lambda=\psi)$. Sampling from this inverse Gaussian is done by the following scheme: \\

\begin{algorithmic}
	\State{Generate $U \sim \Un (0,1)$}
	\State{Generate $Y \sim \N(0,1)$}
	\State{Set $z= \displaystyle\sqrt{\frac{\psi}{\chi_j}} + \displaystyle\frac{y^2}{2 \chi_j} - \sqrt{\frac{\sqrt{\psi} y^2}{\chi_j^{1 \frac{1}{2}}} + \frac{y^4}{4 \chi_j^2}}$}
	\If{$u \leq (1 + z \sqrt{\chi_j/\psi})^{-1}$}
	\State{\Return{$\tau_j = 1/z + 1$}}
	\Else
	\State{\Return{$\tau_j = \chi_j z/\psi + 1$} }.
	\EndIf
\end{algorithmic}

\subsection{Proof EMSE $\tau^2$ for linear regression}
First, write
\begin{equation}\label{emse}
\text{EMSE}(\hat{\tau}^2) = E_{\X}\biggl[E_{\bbeta}\bigl[\text{MSE}(\hat{\tau}^2 | \bbeta, \X)\bigr]\biggr] = E_{\X}\biggl[E_{\bbeta}\bigl[(E_{\Y}(\hat{\tau}^2 | \bbeta, \X)-\tau^2)^2 + V_{\Y}(\hat{\tau}^2 | \bbeta, \X)\bigr]\biggr].
\end{equation}

\noindent
Then, let us first compute the expected squared bias w.r.t.  $\bbeta$:
\begin{equation*}
\begin{split}
E_{\bbeta}(\text{bias}^2) &= E_{\bbeta}[(E_{\Y}(\hat{\tau}^2 | \bbeta, \X)-\tau^2)^2]  = p^{-2}E_{\bbeta}[(\sum_{j=1}^p (E_{\Y}(\hat{\beta}_j^2) - v_j) - p\tau^2)^2]\\
&= p^{-2}E_{\bbeta}[(\sum_{j=1}^p \beta_j^2 - p\tau^2)^2] = p^{-2}E_{\bbeta}[(\sum_{j=1}^p \beta_j^2)^2 - 2p\tau^2\sum_{j=1}^p\beta_j^2 + p^2\tau^4]\\
&= p^{-2}(\sum_{j=1}^p 3\tau^4 + 2\sum_{j,k \neq j}\tau^4 - p^2\tau^4) = p^{-2}( 3p\tau^4 + p(p-1)\tau^4 - p^2\tau^4)\\
&=  \frac{2\tau^4}{p},
\end{split}
\end{equation*}
where we used the central moments of Gaussian random variables, available from Isserlis' Theorem \citep{isserlis1918formula}:
$E(\beta_j^4) = 3\tau^4$ and $E(\beta_j^2\beta_k^2) = E(\beta_j^2)E(\beta_k^2) = \tau^4.$
The result is constant in $\X$, so

\begin{equation}\label{Ebias}
E_{\X}[E_{\bbeta}(\text{bias}^2)] = \frac{2\tau^4}{p}.
\end{equation}

\para Next, we compute $E_{\X}[E_{\bbeta}[V_{\Y}(\hat{\tau}^2 | \bbeta, \X)]].$ Denoting
$V(\hat{\tau}^2) = V_{\Y}(\hat{\tau}^2 | \bbeta, \X)$, we have:
\begin{equation}\label{vartau}
V(\hat{\tau}^2) = V((\tau')^2) = \frac{1}{p^2}\biggl(\sum_{j=1}^p V(\hat{\beta}_j^2) + \sum_{j,k \neq j}^p \text{Cov}(\hat{\beta}_j^2,\hat{\beta}_k^2)\biggr).
\end{equation}

\noindent
Hence, we need to compute $V(\hat{\beta}_j^2)$ and $\text{Cov}(\hat{\beta}_j^2,\hat{\beta}_k^2)$. These are again derived from expressions of the central moments of Gaussian random variables. Let us first express the non-central moments in $\text{Cov}_{\Y}(\hat{\beta}_j^2,\hat{\beta}_k^2) = \text{Cov}(\hat{\beta}_j^2,\hat{\beta}_k^2) = E(\hat{\beta}_j^2\hat{\beta}_k^2) - E(\hat{\beta}_j^2)E(\hat{\beta}_k^2)$ in terms of the central ones.
Denote the centralized value of $\hat{\beta}_j$ by $\tbeta_j = \hat{\beta}_j - \beta_j.$ Then,
\begin{equation*}
\begin{split}
E[\hat{\beta}_j^2\hat{\beta}_k^2] &= E[((\hat{\beta}_j -\beta_j) + \beta_j)^2((\hat{\beta}_k -\beta_k) + \beta_k)^2]
= E[(\tbeta_j + \beta_j)^2(\tbeta_k + \beta_k)^2]\\ &= T_1 + T_2 := E[\tbeta_j^2\tbeta_k^2 + 4\beta_j\beta_k\tbeta_j\tbeta_k + \beta_j^2\beta_k^2 + \tbeta_j^2\beta_k^2 + \beta_j^2\tbeta_k^2] \\
& + E[2\beta_j\tbeta_j\tbeta_k^2 + 2\beta_k\tbeta_k\tbeta_j^2  + 2\beta_j\tbeta_j\beta_k^2 + 2\beta_k\tbeta_k\beta_j^2]\\
& = T_1,
\end{split}
\end{equation*}
because $T_2=0$ due to the symmetry of the central Gaussian distribution.
Likewise, the second term of the covariance equals:
$$E(\hat{\beta}_j^2)E(\hat{\beta}_k^2) = E(\tbeta_j^2)E(\tbeta_k^2) + \beta_j^2E(\tbeta_k^2) + \beta_k^2E(\tbeta_j^2) + \beta_j^2\beta_k^2.$$
Subtracting the latter from $T_1$ cancels the latter 3 terms in both expressions, rendering

\begin{equation}\label{cov}
\begin{split}
\text{Cov}(\hat{\beta}_j^2,\hat{\beta}_k^2) &= E(\tbeta_j^2\tbeta_k^2) + 4\beta_j\beta_k E(\tbeta_j\tbeta_k) - E(\tbeta_j^2)E(\tbeta_k^2)\\
&= (v_jv_k + 2v_{jk}^2) + 4\beta_j\beta_k v_{jk} - v_jv_k =  2v_{jk}^2 + 4\beta_j\beta_k v_{jk},
\end{split}
\end{equation}
where we used the equations for the central moments of Gaussian random variables \citep{isserlis1918formula}.
Noting that $V(\hat{\beta}_j^2) = \text{Cov}(\hat{\beta}_j^2,\hat{\beta}_j^2)$ we directly obtain
\begin{equation}\label{var}
\begin{split}
V(\hat{\beta}_j^2) &= E(\tbeta_j^4) + 4\beta_j\beta_jE(\tbeta_j^2) - (E(\tbeta_j^2))^2\\
&= 3v_j^2  + 4\beta_j^2v_{j} - v_j^2 =  2v_{j}^2 + 4\beta_j^2 v_{j}.
\end{split}
\end{equation}
Note that the latter can also be obtained by writing $\hat{\beta}_j^2 = v_j (\hat{\beta}_j/\sqrt(v_j))^2 = v_j (\beta'_j)^2.$ Then $\beta'_j \sim N(\beta_j/\sqrt(v_j), 1),$ so
$(\beta'_j)^2 \sim \chi^2(\nu = \beta_j^2/v_j, k = 1)$ with $V((\beta'_j)^2) = 2(k + 2\nu) = 2(1+2\beta_j^2/v_j).$ Hence, indeed $V(\hat{\beta}_j^2) = v_j^2 V((\beta'_j)^2) =  2v_{j}^2 + 4\beta_j^2 v_{j}$.

\noindent
Substituting (\ref{cov}) and (\ref{var}) into (\ref{vartau}) renders:
$$V(\hat{\tau}^2) = \frac{2}{p^2}[\sum_{j=1}^p (v_{j}^2 + 2\beta_j^2 v_{j}) + \sum_{j,k \neq j} (v_{jk}^2 + 2\beta_j\beta_k v_{jk})].  $$

\noindent
Taking expectation w.r.t. $\bbeta$ gives:
$$E_{\bbeta}[V(\hat{\tau}^2)] = \frac{2}{p^2}[\sum_{j=1}^p (v_{j}^2 + 2\tau^2 v_{j}) + \sum_{j,k \neq j}^p v_{jk}^2],$$
because we assume i.i.d. central priors for $\beta_j$.
Now to compute
\begin{equation}\label{expvar}
E_{\X}[E_{\bbeta}[V(\hat{\tau}^2)]] = \frac{2}{p^2}[\sum_{j=1}^p (E_{\X} (v_{j}^2) + 2\tau^2 E_{\X} (v_{j})) + \sum_{j,k \neq j}^p E_{\X} (v_{jk}^2)]
\end{equation} we need to know $E_{\X}(v_j^2), E_{\X}(v_j)$ and $E_{\X}(v_{jk}^2)$, where $v_j=v_{jj}$ and $v_{jk} = V_{jk}, V = (\X^T\X)^{-1}, $ with $X_{i} \sim N(0,\Sigma = \Sigma_{p \times p})$ and $\Sigma_{jj}=1$. By definition, $V$ follows an inverse-Wishart distribution: $V \sim \mathcal{W}^{-1}(\Psi=\Sigma^{-1},n)$. Hence, the requested moments are known \citep{Press1982}:
\begin{equation}\label{invwish}
\begin{split}
E_{\X} (v_j) &= \psi_{jj}/(n-p-1)\\
E_{\X} (v_j^2) &= V_{\X}(v_j) + (E_{\X}(v_j))^2 = \frac{2\psi^2_{jj}}{(n-p-1)^2(n-p-3)} + \frac{\psi^2_{jj}}{(n-p-1)^2}\\
E_{\X} (v_{jk}^2) &= V_{\X}(v_{jk}) + (E_{\X}(v_{jk}))^2 = \frac{(n-p+1)\psi_{jk}^2 + (n-p-1)\psi_{jj}\psi_{kk}}{(n-p)(n-p-1)^2(n-p-3)} + \frac{\psi^2_{jk}}{(n-p-1)^2},
\end{split}
\end{equation}
where we assume $p<n-3$. Substituting (\ref{invwish}) into (\ref{expvar}) and aggregating with the expected squared bias (\ref{Ebias}) finalizes the result:
\begin{equation}\label{emseresult}
\begin{split}
\text{EMSE}(\hat{\tau}^2) &= \frac{2}{(n-p-1)p^2}\biggl[\sum_{j=1}^p \bigl(\frac{2\psi^2_{jj}}{(n-p-1)(n-p-3)} + \frac{\psi^2_{jj}}{(n-p-1)}) + 2\tau^2\sum_{j=1}^p \psi_{jj}\\
&+ \sum_{j,k \neq j}^p\bigl(\frac{(n-p+1)\psi_{jk}^2 + (n-p-1)\psi_{jj}\psi_{kk}}{(n-p)(n-p-1)(n-p-3)} + \frac{\psi^2_{jk}}{(n-p-1)}\bigr)\biggr] + \frac{2\tau^4}{p}.
\end{split}
\end{equation}

\noindent
This simplifies for independent $X_i$, because then $\psi_{jj} = 1$ and $\psi_{jk}=0$:
\begin{equation}\label{emseresultind}
\begin{split}
\text{EMSE}_{\perp}(\hat{\tau}^2) &= \frac{2}{(n-p-1)p}\biggl[\frac{2}{(n-p-1)(n-p-3)} + \frac{1}{n-p-1} + 2\tau^2\\
&+ \frac{p-1}{(n-p)(n-p-3)}\biggr] + \frac{2\tau^4}{p}.
\end{split}
\end{equation}

\section{Supplementary Figures}
\subsection{Baseball Example}
\begin{figure}[!h]
	\begin{minipage}[c]{0.50\linewidth}
    \centering
  		\subfigure[(a)][18 vs 10018]{
  			\includegraphics[width=1.00\textwidth]{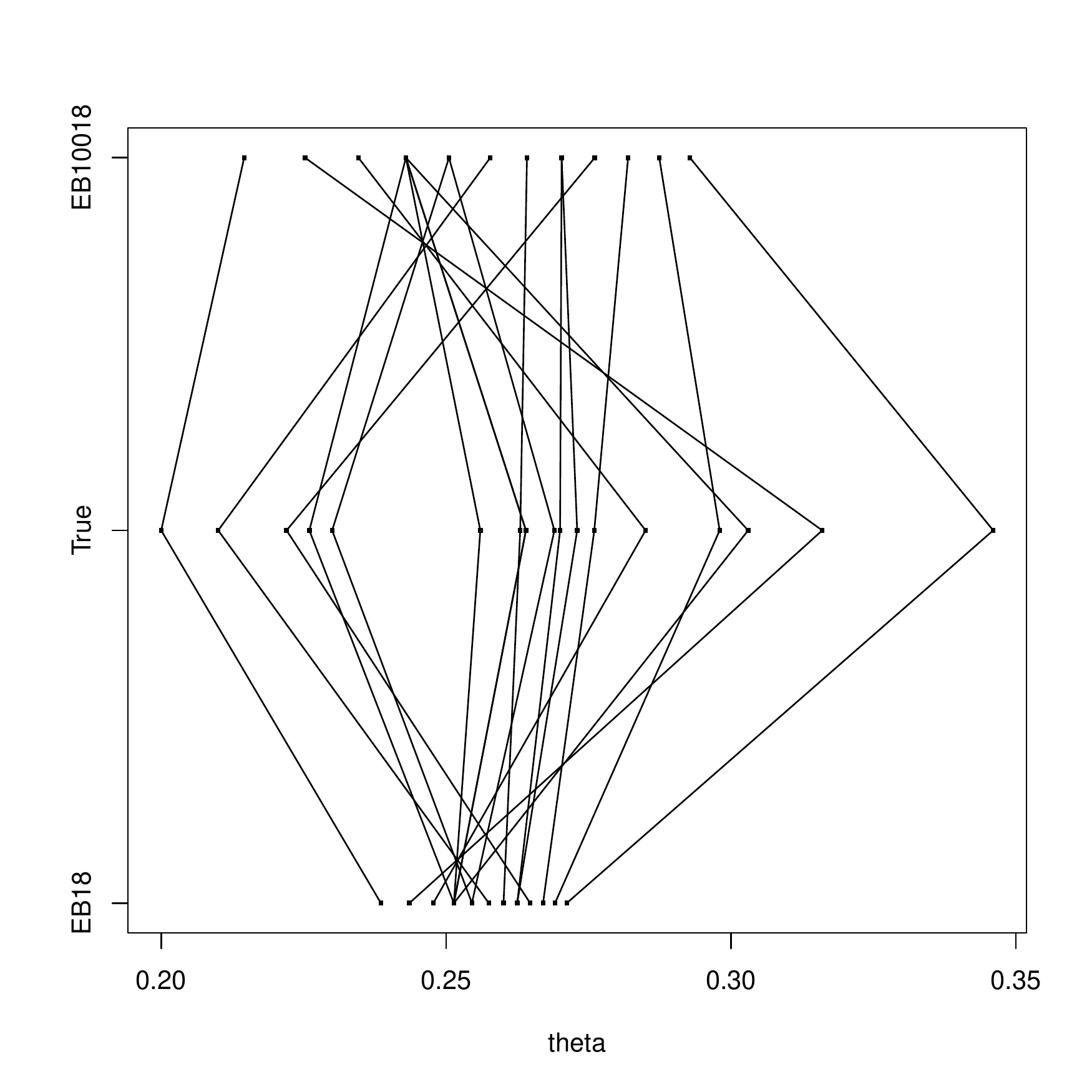}
  		}
  \end{minipage}\hfill
  \begin{minipage}[c]{0.50\linewidth}
    \centering
  		\subfigure[(b)][Gaussian vs Mixture]{
  			\includegraphics[width=1.00\textwidth]{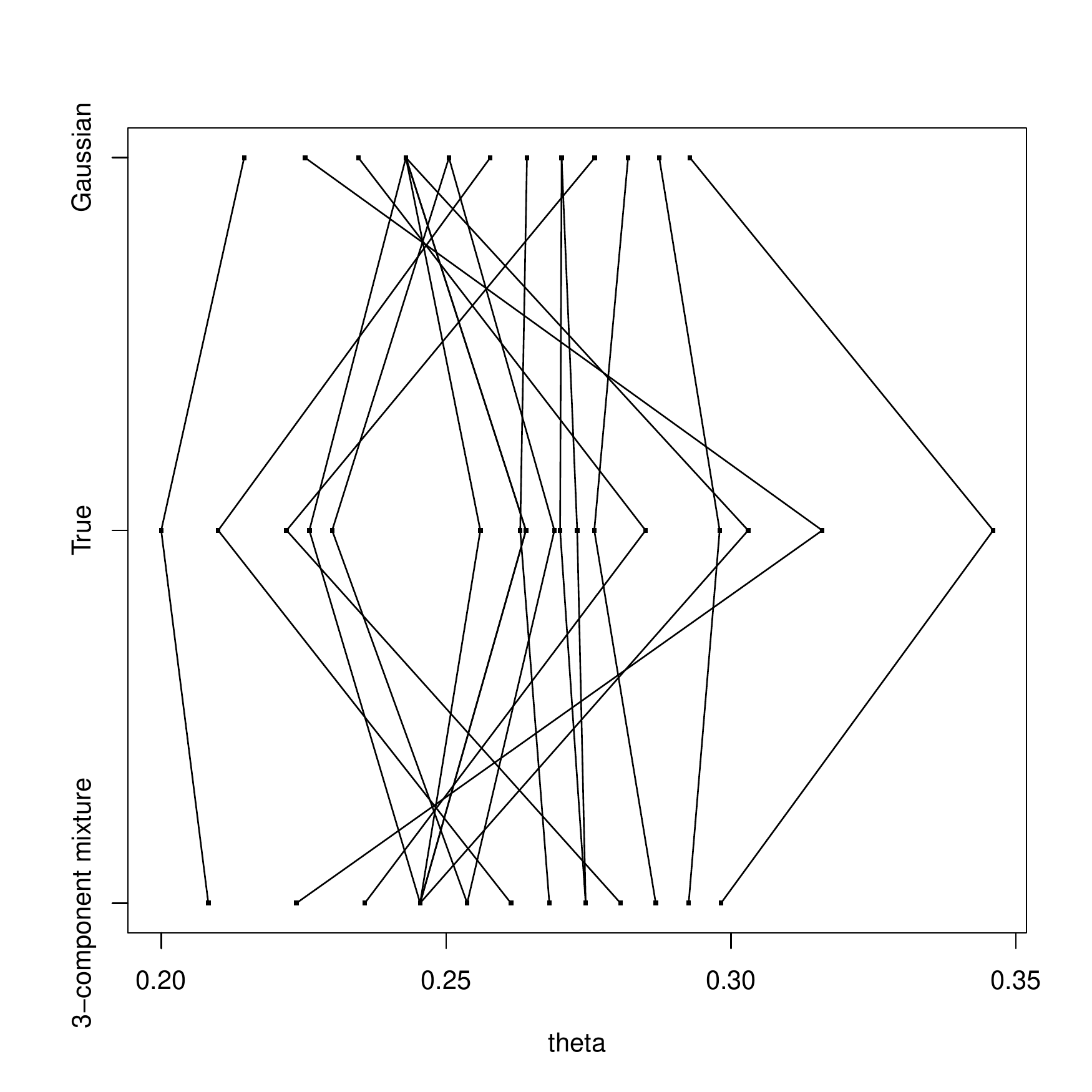}

  		}
  \end{minipage}
\caption{Shrinkage estimators vs true value. Left: Parameter estimates for 18 real players when EB estimation of hyper-parameters is based on either 18 (bottom) or 10018 (top) players. Right:
Parameter estimates for 18 real players when EB estimation of hyper-parameters is based on 10018 players, using as prior either a 3-component mixture of Gaussians (bottom) or a Gaussian (top).}\label{baseball}
\end{figure}


\subsection{Simulation Example}
Here, we show the results for all simulation settings presented in the Simulation example.

\begin{figure}[!h]
	\begin{minipage}[c]{0.50\linewidth}
    \centering
  		\subfigure[][Equal-tailed interval]{
  			\includegraphics[width=1.00\textwidth]{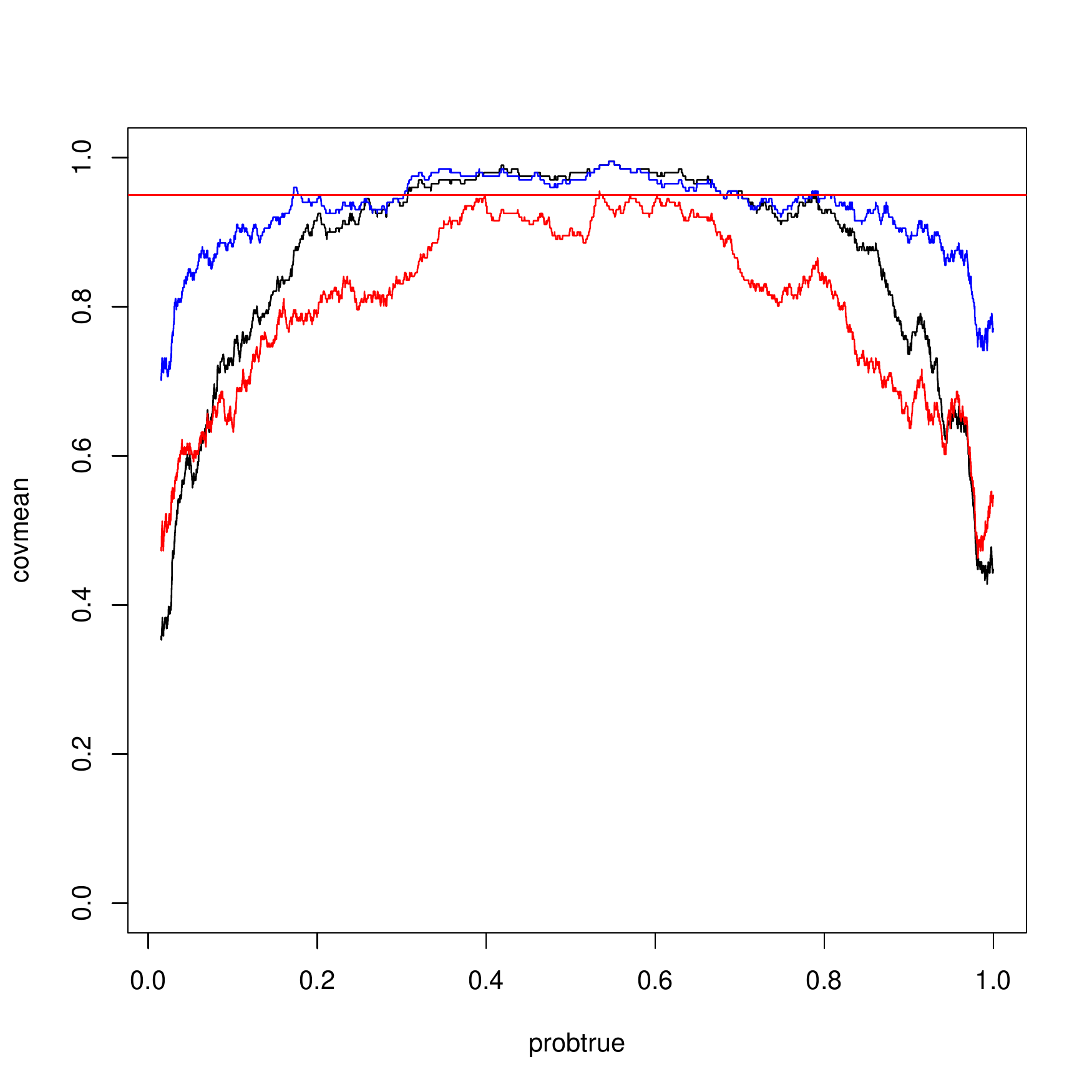}
  		}
  \end{minipage}\hfill
  \begin{minipage}[c]{0.50\linewidth}
    \centering
  		\subfigure[][HPD interval]{
  			\includegraphics[width=1.00\textwidth]{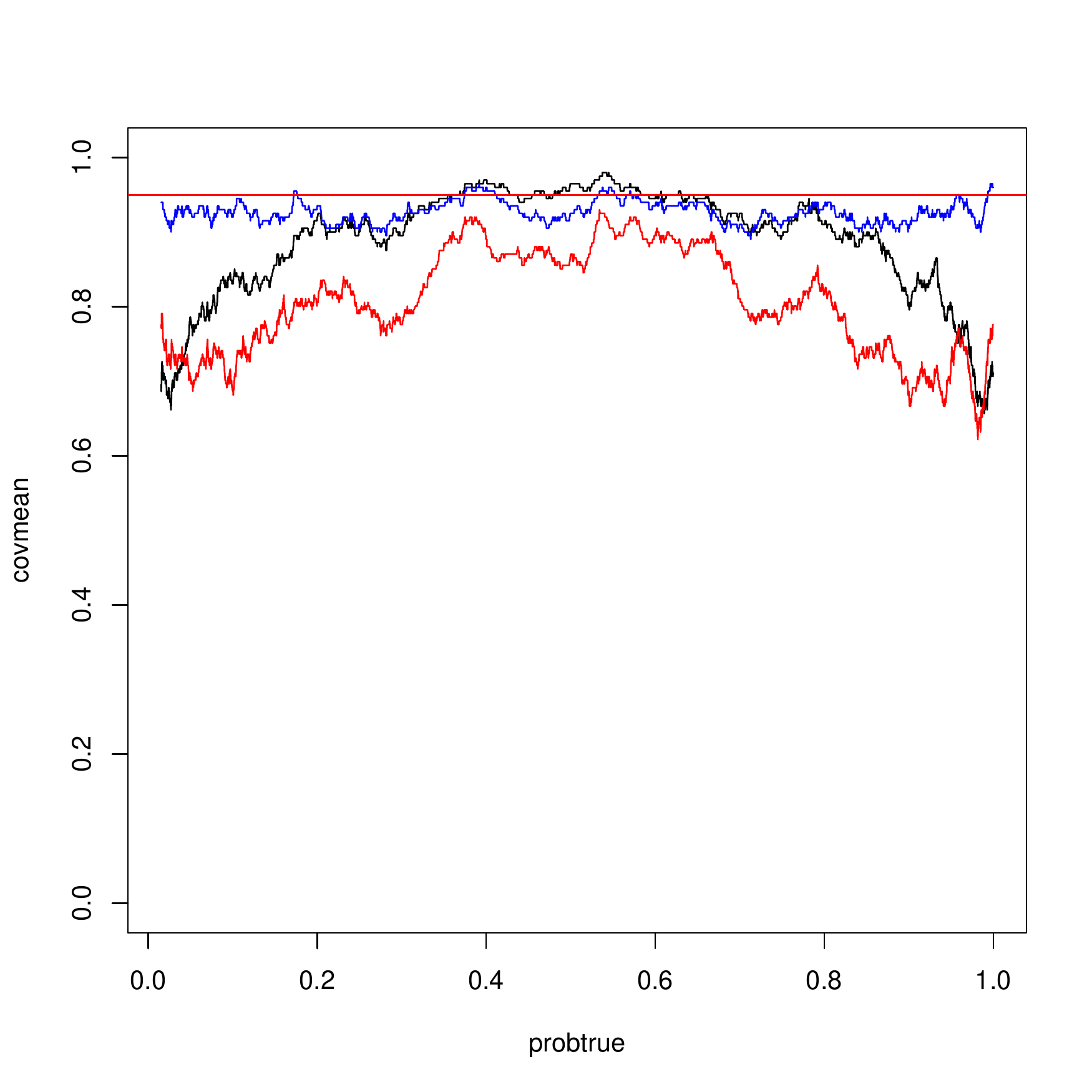}

  		}
  \end{minipage}
\caption{X-axis: True event probability $q_j$; Y-axis: mean coverage of 95\% posterior intervals for event probability. Mean is estimated by moving average. Case: $G=2, p_g=10, p = G*p_G = 20, n_{\text{train}} = 100$. Methods:
\textcolor{blue}{Hyb}, \textcolor{black}{EB}, \textcolor{red}{FB}}
\end{figure}

\begin{figure}[!h]
	\begin{minipage}[c]{0.50\linewidth}
    \centering
  		\subfigure[][Equal-tailed interval]{
  			\includegraphics[width=1.00\textwidth]{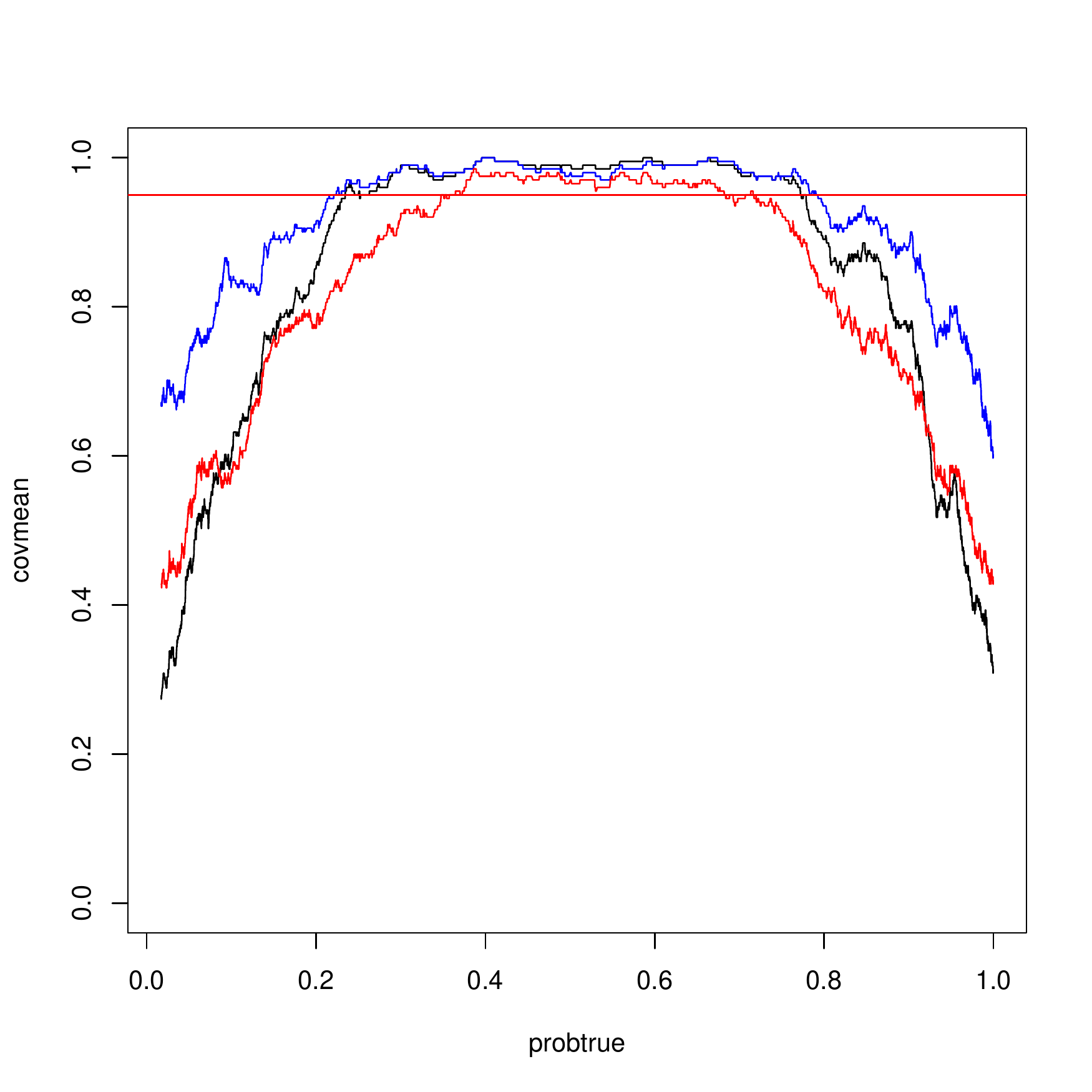}
  		}
  \end{minipage}\hfill
  \begin{minipage}[c]{0.50\linewidth}
    \centering
  		\subfigure[][HPD interval]{
  			\includegraphics[width=1.00\textwidth]{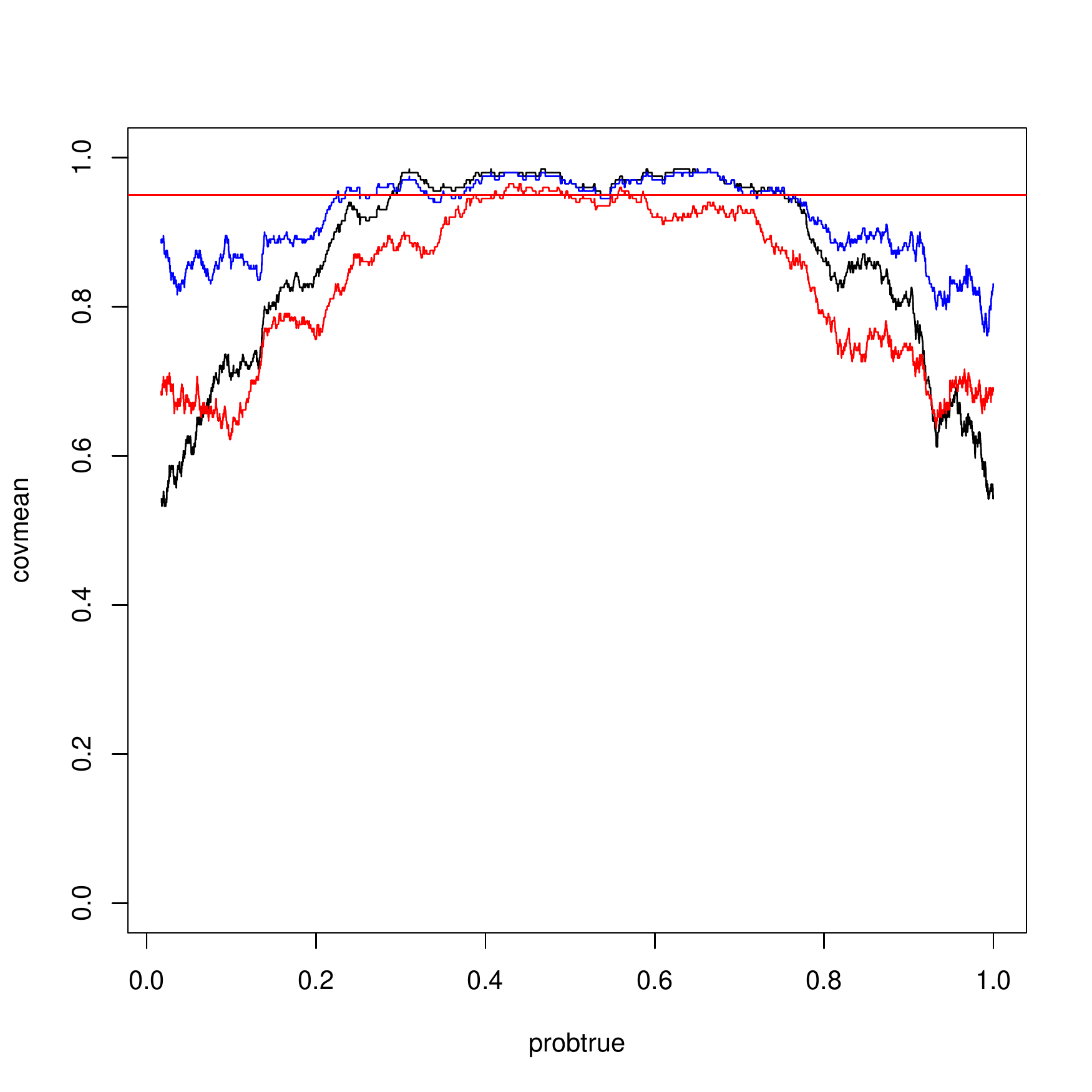}

  		}
  \end{minipage}
\caption{X-axis: True event probability $q_j$; Y-axis: mean coverage of 95\% posterior intervals for event probability. Mean is estimated by moving average. Case: $G=2, p_g=20, p = G*p_G = 40, n_{\text{train}} = 100$. Methods:
\textcolor{blue}{Hyb}, \textcolor{black}{EB}, \textcolor{red}{FB}}
\end{figure}

\begin{figure}[!h]
	\begin{minipage}[c]{0.50\linewidth}
    \centering
  		\subfigure[][Equal-tailed interval]{
  			\includegraphics[width=1.00\textwidth]{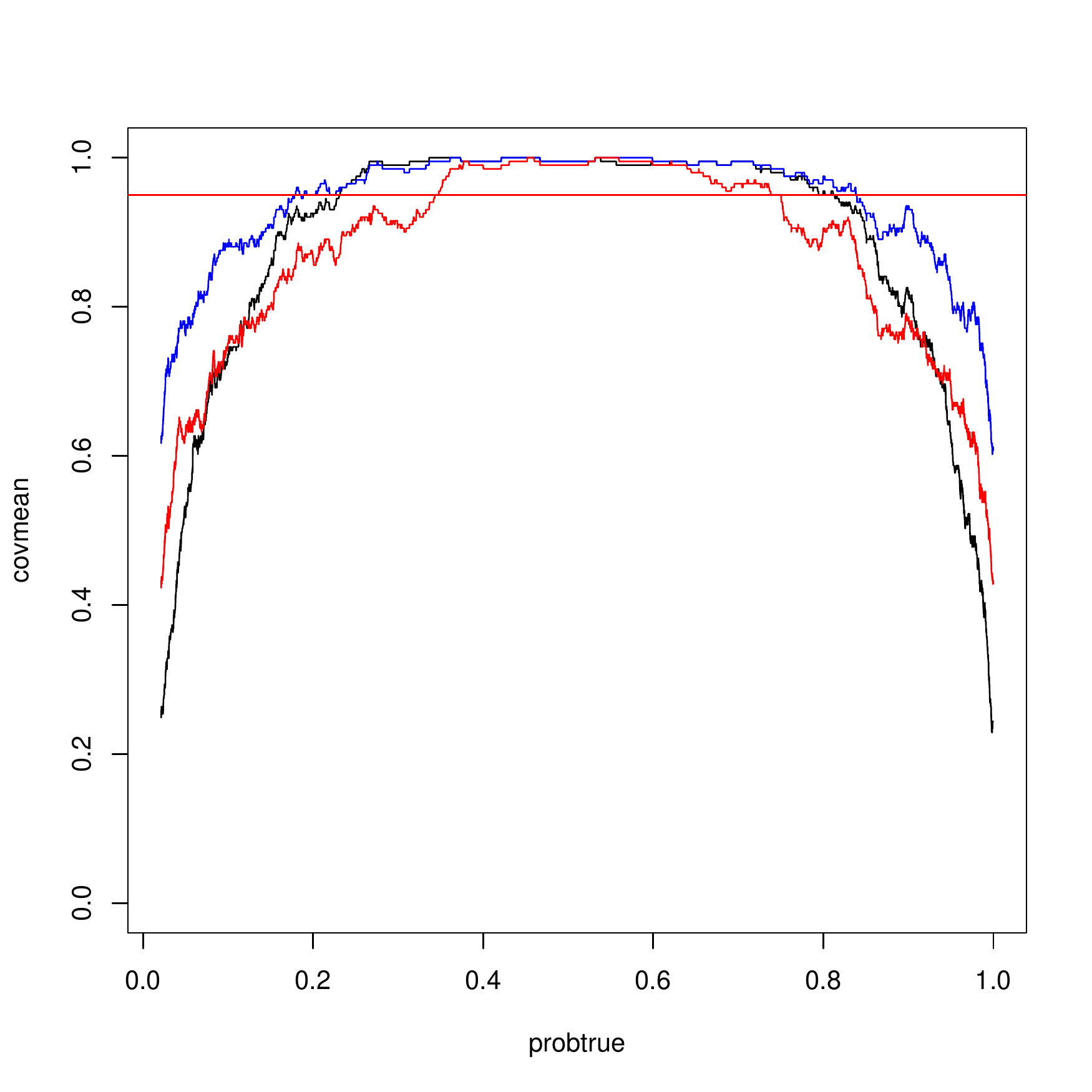}
  		}
  \end{minipage}\hfill
  \begin{minipage}[c]{0.50\linewidth}
    \centering
  		\subfigure[][HPD interval]{
  			\includegraphics[width=1.00\textwidth]{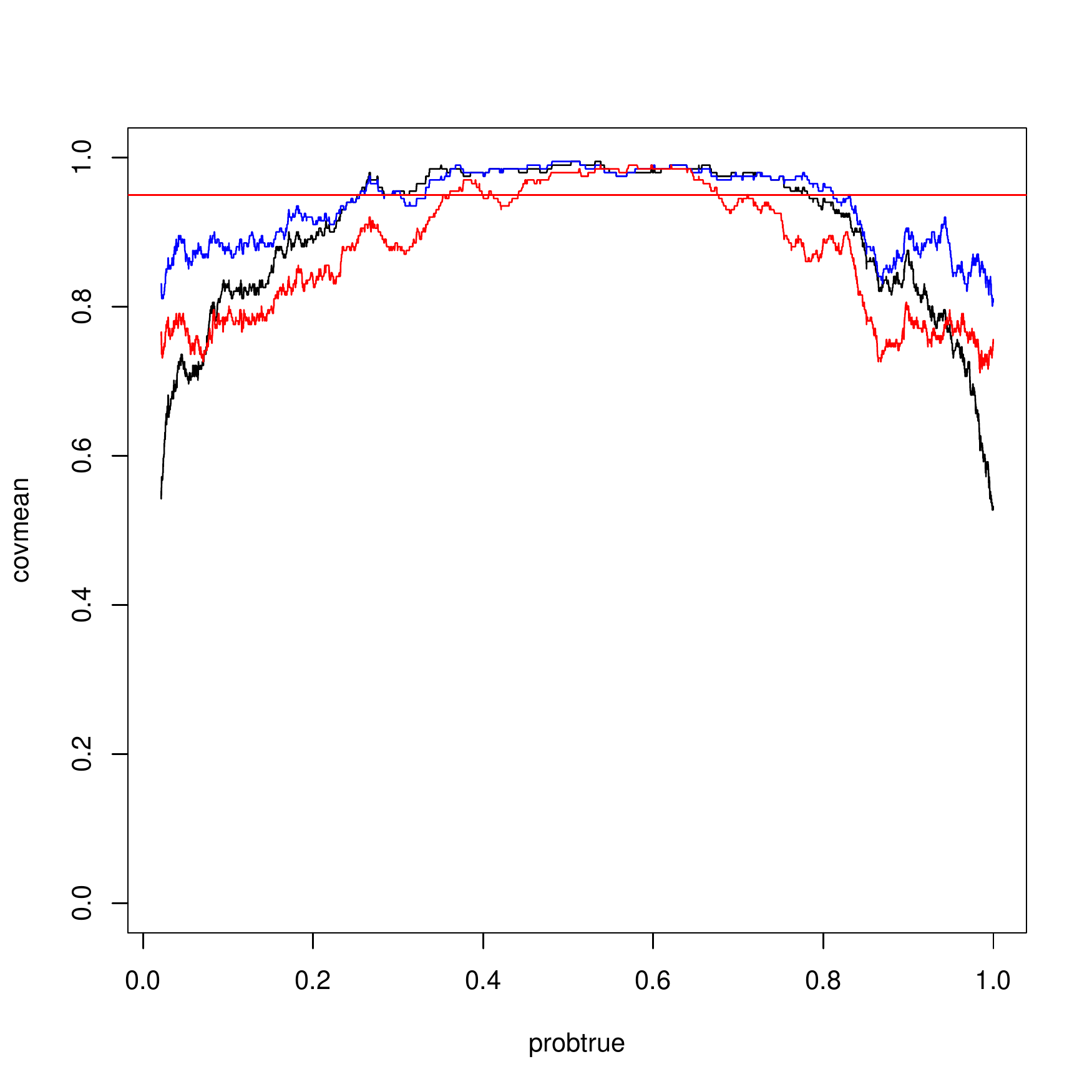}

  		}
  \end{minipage}
\caption{X-axis: True event probability $q_j$; Y-axis: mean coverage of 95\% posterior intervals for event probability. Mean is estimated by moving average. Case: $G=2, p_g=30, p = G*p_G = 60, n_{\text{train}} = 100$. Methods:
\textcolor{blue}{Hyb}, \textcolor{black}{EB}, \textcolor{red}{FB}}
\end{figure}

\begin{figure}[!h]
	\begin{minipage}[c]{0.50\linewidth}
    \centering
  		\subfigure[][Equal-tailed interval]{
  			\includegraphics[width=1.00\textwidth]{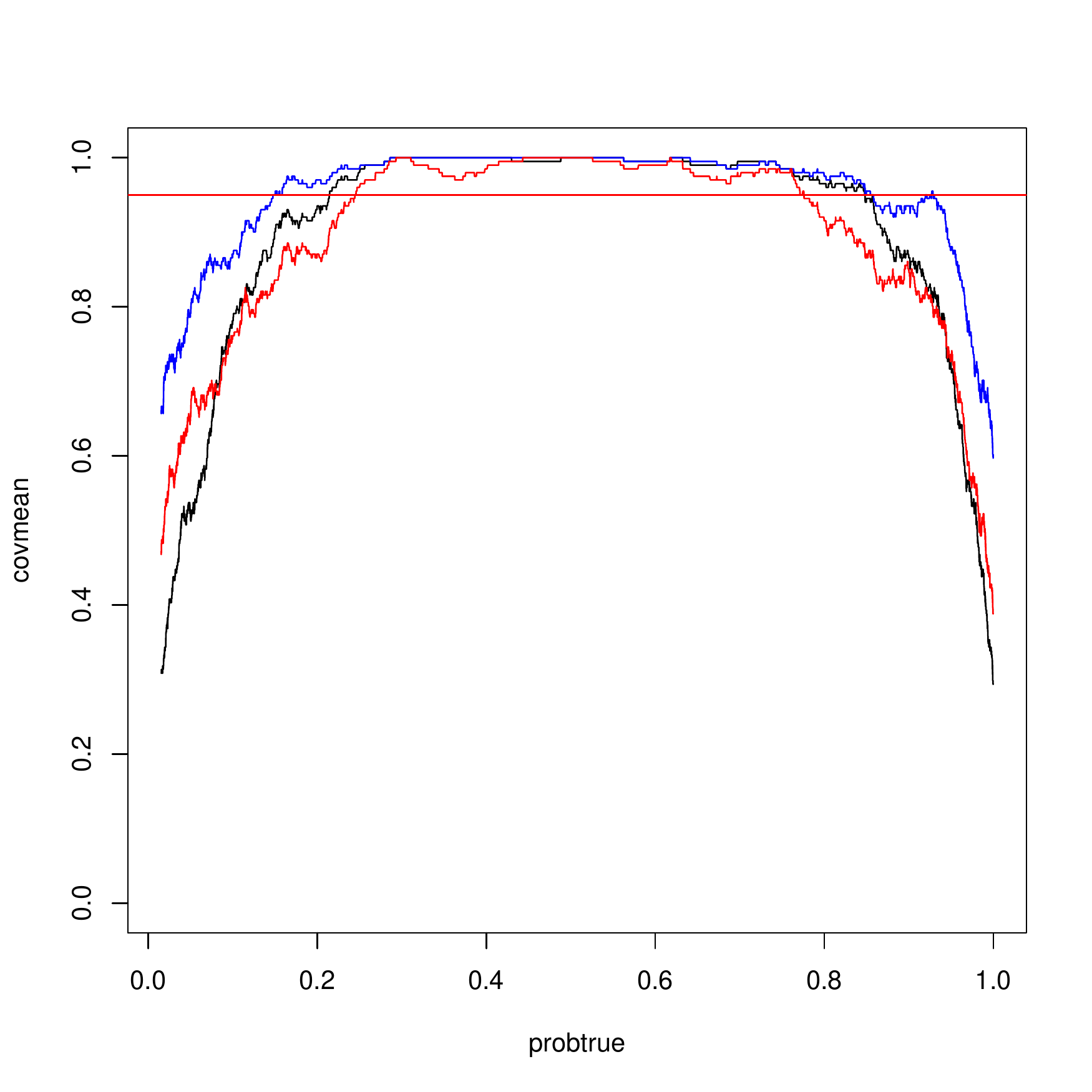}
  		}
  \end{minipage}\hfill
  \begin{minipage}[c]{0.50\linewidth}
    \centering
  		\subfigure[][HPD interval]{
  			\includegraphics[width=1.00\textwidth]{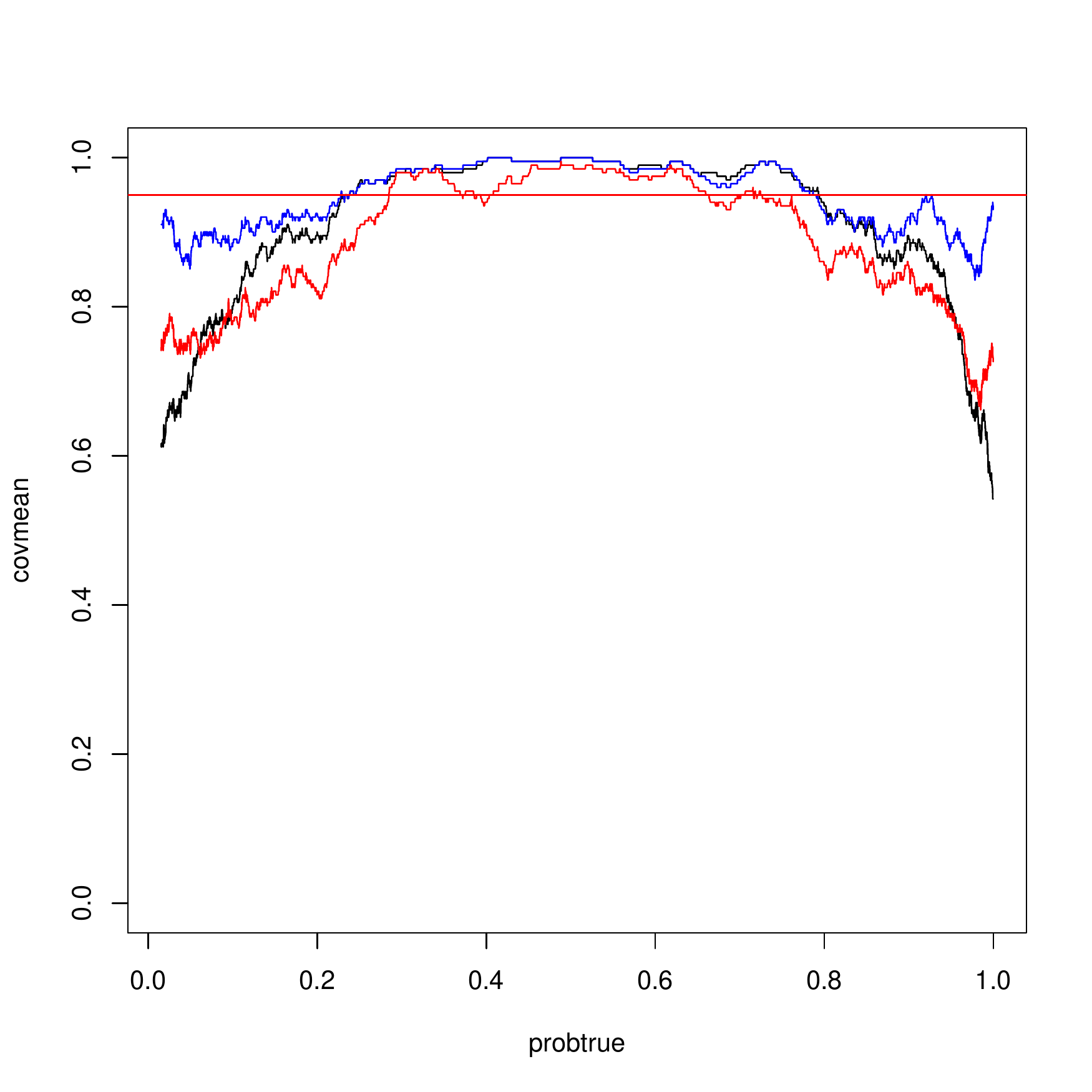}

  		}
  \end{minipage}
\caption{X-axis: True event probability $q_j$; Y-axis: mean coverage of 95\% posterior intervals for event probability. Mean is estimated by moving average. Case: $G=2, p_g=40, p = G*p_G = 80, n_{\text{train}} = 100$. Methods:
\textcolor{blue}{Hyb}, \textcolor{black}{EB}, \textcolor{red}{FB}}
\end{figure}

\begin{figure}[!h]
	\begin{minipage}[c]{0.50\linewidth}
    \centering
  		\subfigure[][Equal-tailed interval]{
  			\includegraphics[width=1.00\textwidth]{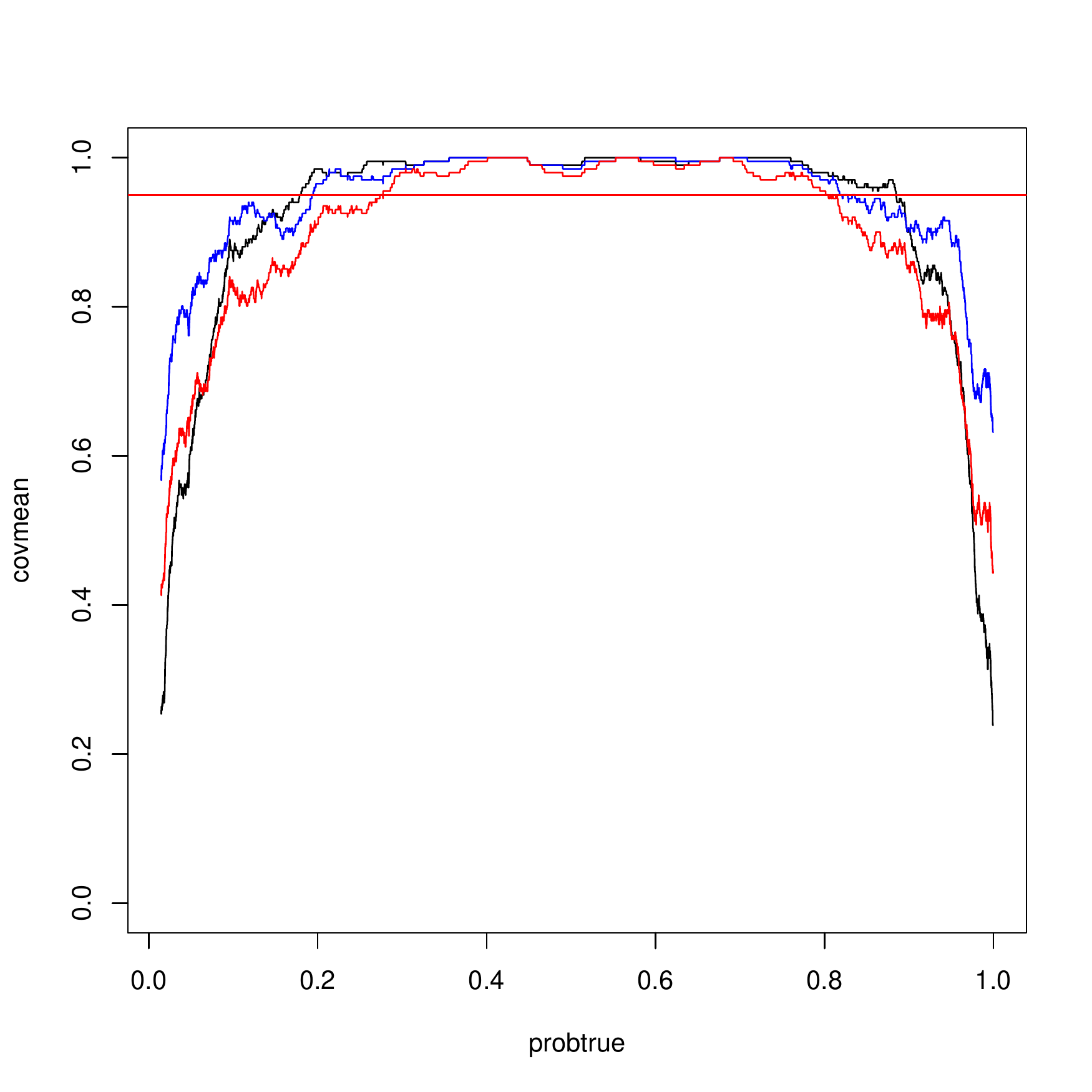}
  		}
  \end{minipage}\hfill
  \begin{minipage}[c]{0.50\linewidth}
    \centering
  		\subfigure[][HPD interval]{
  			\includegraphics[width=1.00\textwidth]{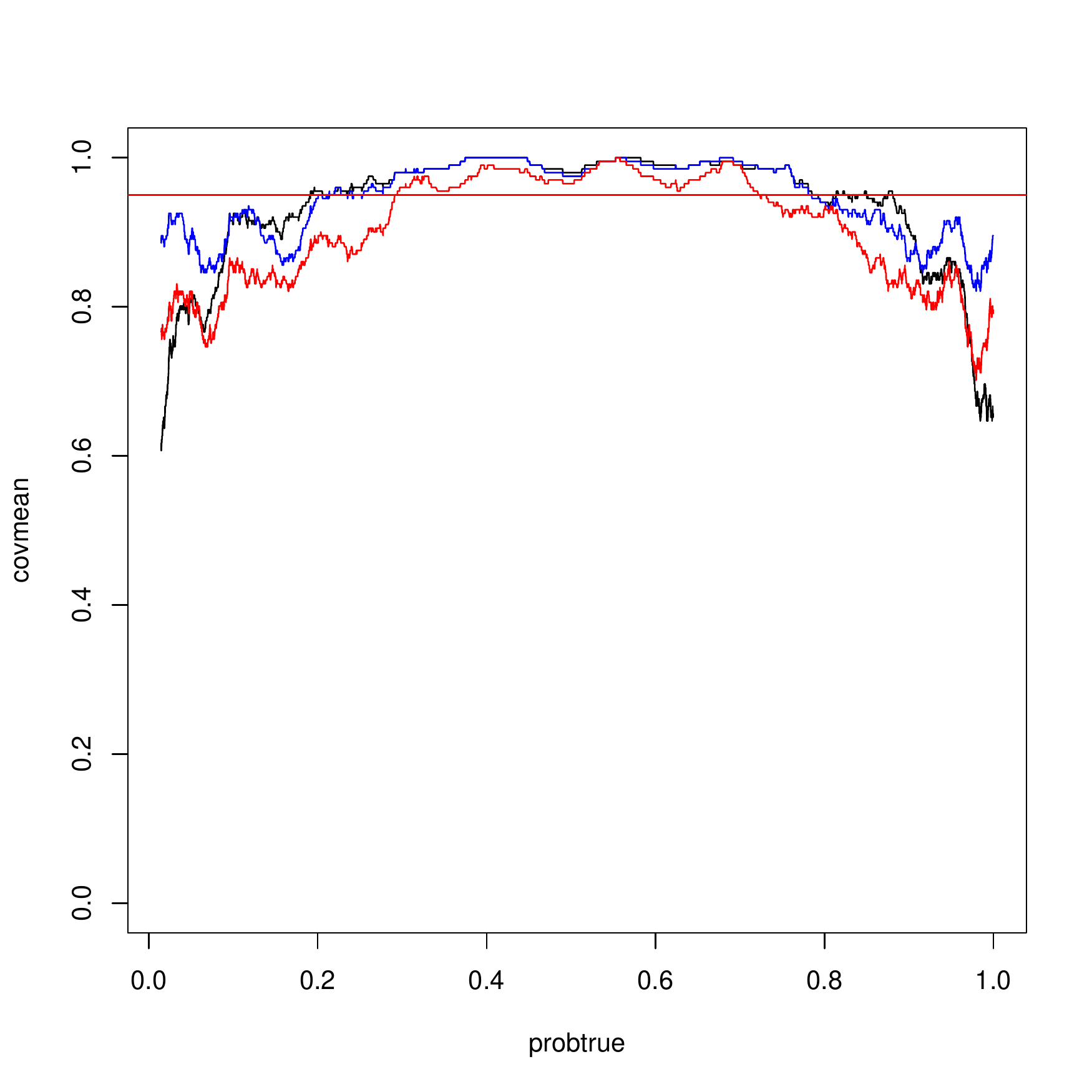}

  		}
  \end{minipage}
\caption{X-axis: True event probability $q_j$; Y-axis: mean coverage of 95\% posterior intervals for event probability. Mean is estimated by moving average. Case: $G=2, p_g=50, p = G*p_G = 50, n_{\text{train}} = 100$. Methods:
\textcolor{blue}{Hyb}, \textcolor{black}{EB}, \textcolor{red}{FB}}
\end{figure}

\begin{figure}[!h]
	\begin{minipage}[c]{0.50\linewidth}
    \centering
  		\subfigure[][Equal-tailed interval]{
  			\includegraphics[width=1.00\textwidth]{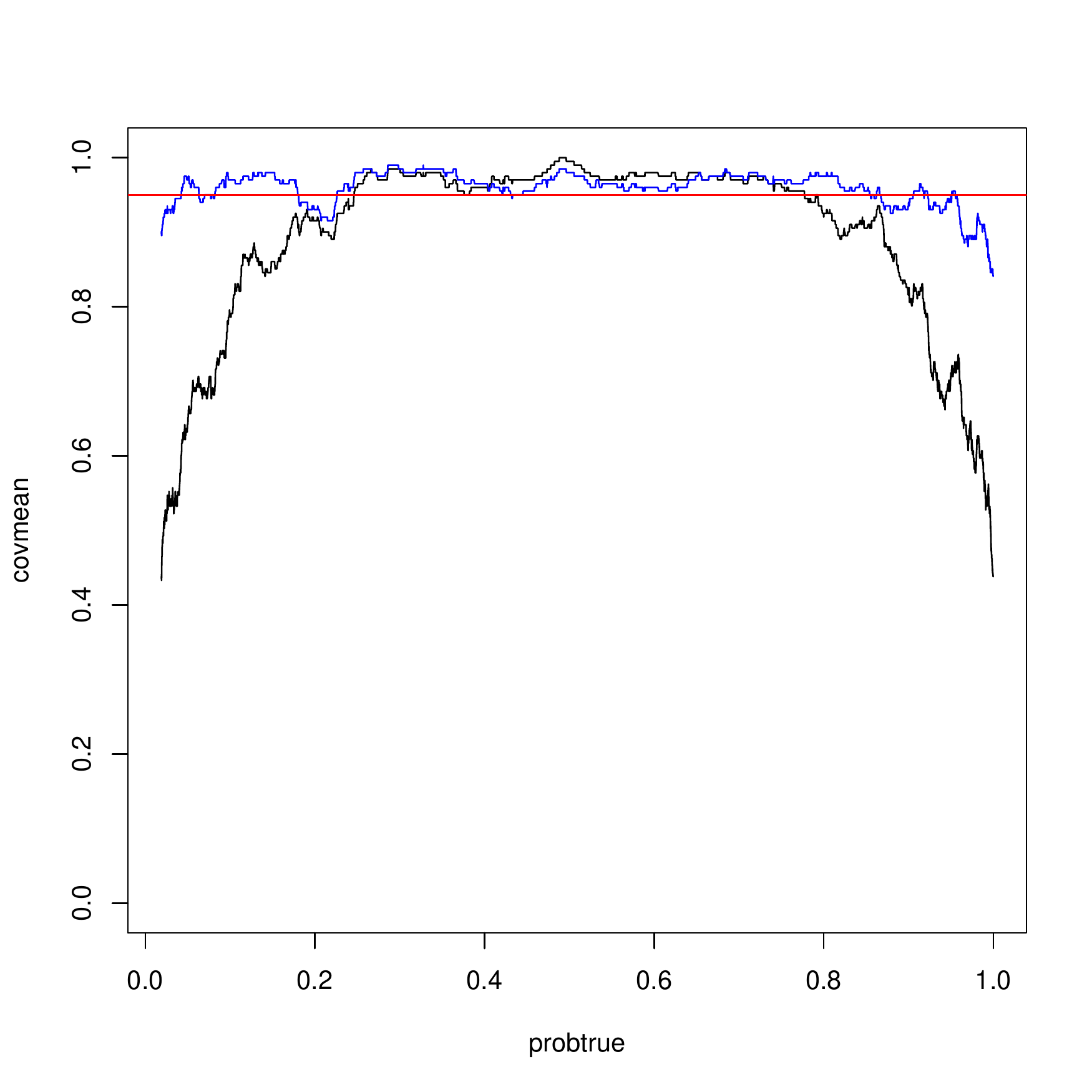}
  		}
  \end{minipage}\hfill
  \begin{minipage}[c]{0.50\linewidth}
    \centering
  		\subfigure[][HPD interval]{
  			\includegraphics[width=1.00\textwidth]{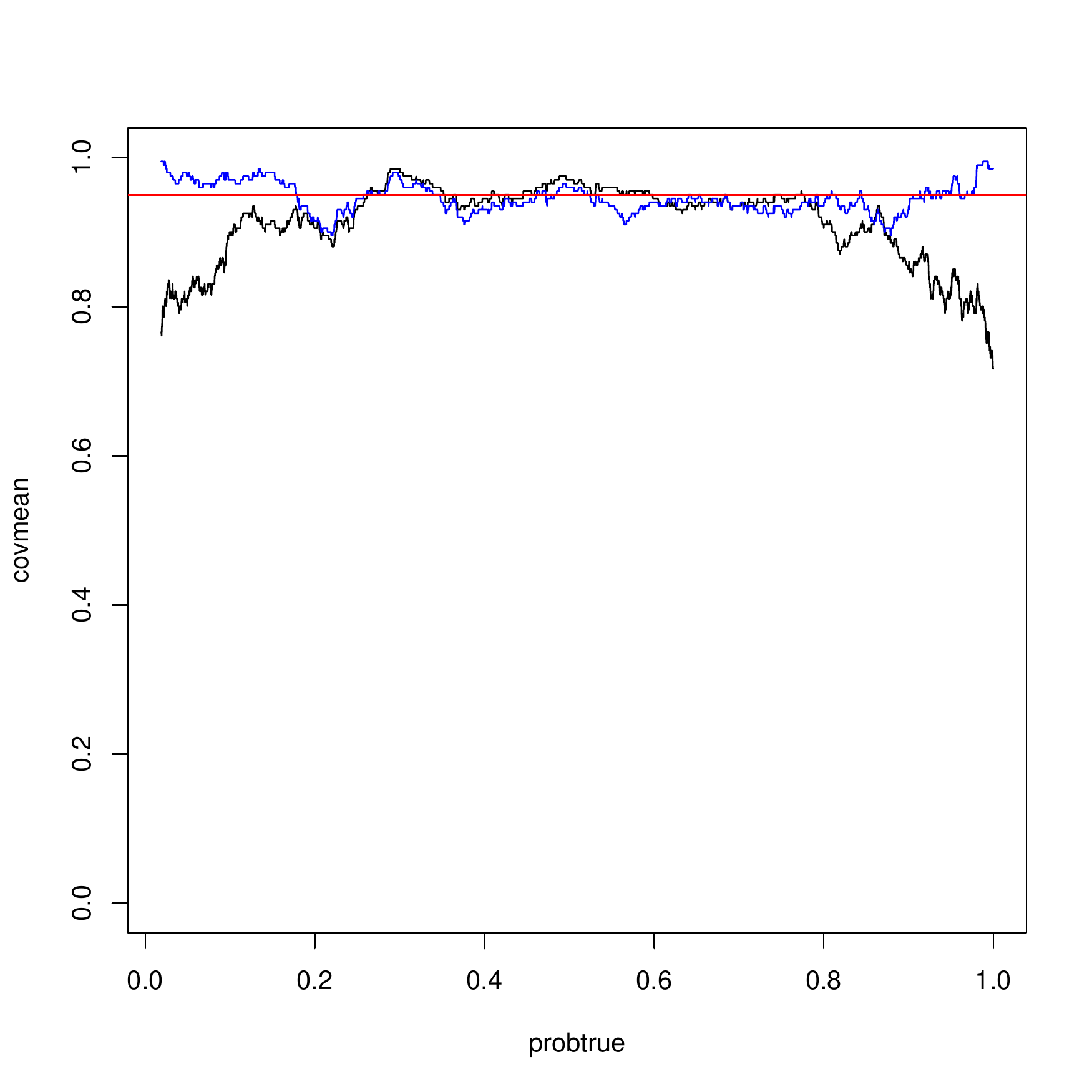}

  		}
  \end{minipage}
\caption{X-axis: True event probability $q_j$; Y-axis: mean coverage of 95\% posterior intervals for event probability. Mean is estimated by moving average. Case: $G=5, p_g=10, p = G*p_G = 50, n_{\text{train}} = 200$. Methods:
\textcolor{blue}{Hyb}, \textcolor{black}{EB}}
\end{figure}

\begin{figure}[h]
	\begin{minipage}[c]{0.50\linewidth}
    \centering
  		\subfigure[][Equal-tailed interval]{
  			\includegraphics[width=1.00\textwidth]{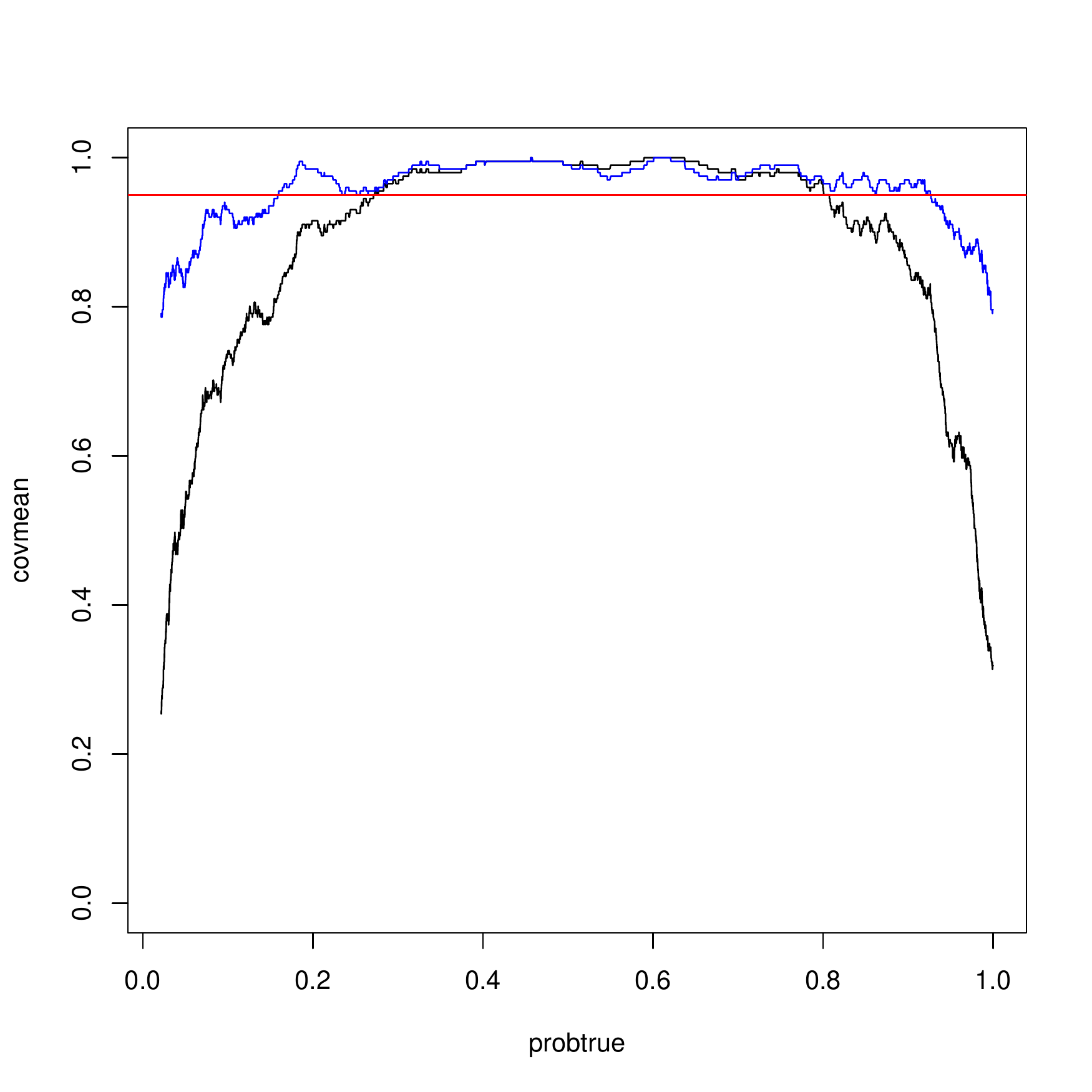}
  		}
  \end{minipage}\hfill
  \begin{minipage}[c]{0.50\linewidth}
    \centering
  		\subfigure[][HPD interval]{
  			\includegraphics[width=1.00\textwidth]{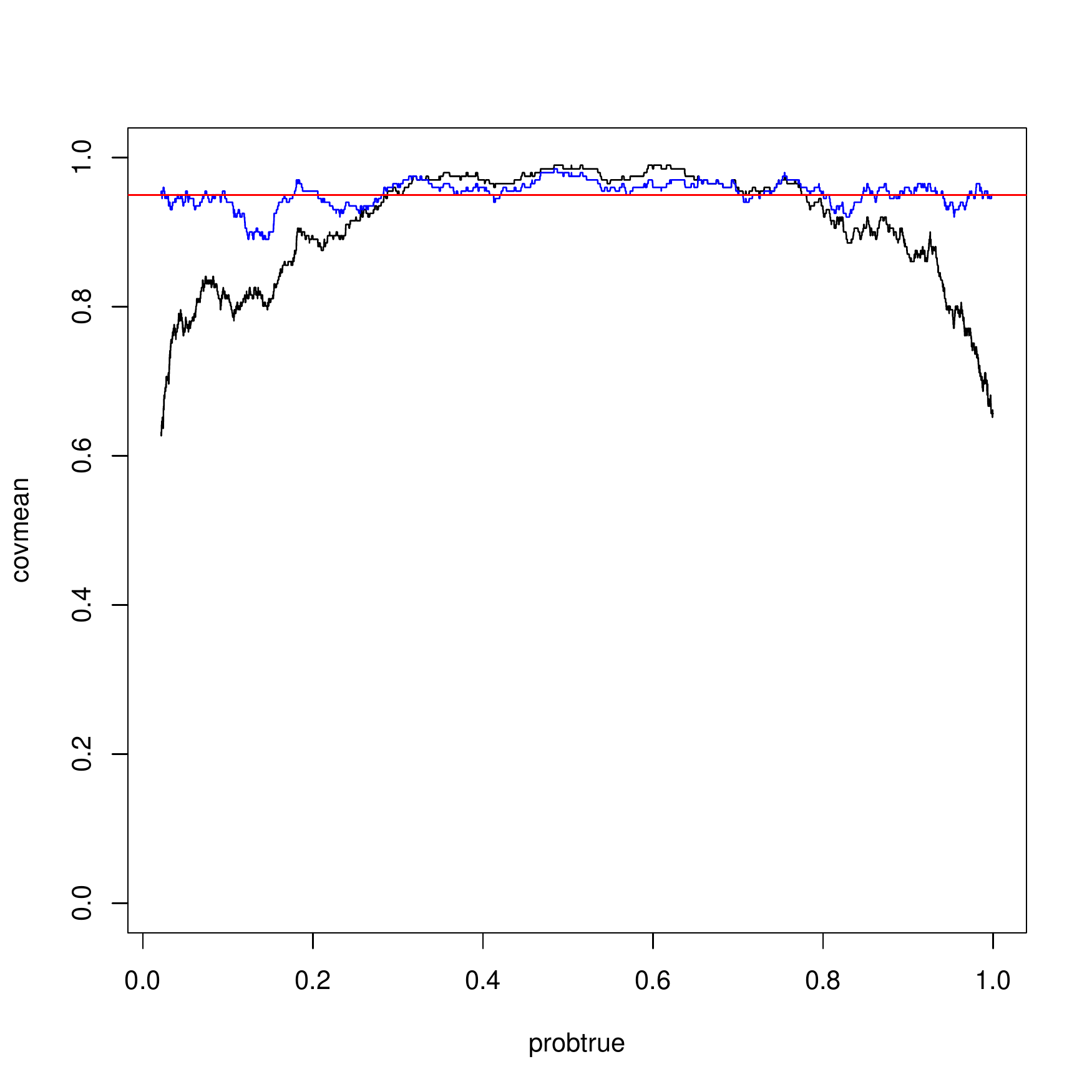}

  		}
  \end{minipage}
\caption{X-axis: True event probability $q_j$; Y-axis: mean coverage of 95\% posterior intervals for event probability. Mean is estimated by moving average. Case: $G=5, p_g=20, p = G*p_G = 100, n_{\text{train}} = 200$. Methods:
\textcolor{blue}{Hyb}, \textcolor{black}{EB}}
\end{figure}

\begin{figure}[h]
	\begin{minipage}[c]{0.50\linewidth}
    \centering
  		\subfigure[][Equal-tailed interval]{
  			\includegraphics[width=1.00\textwidth]{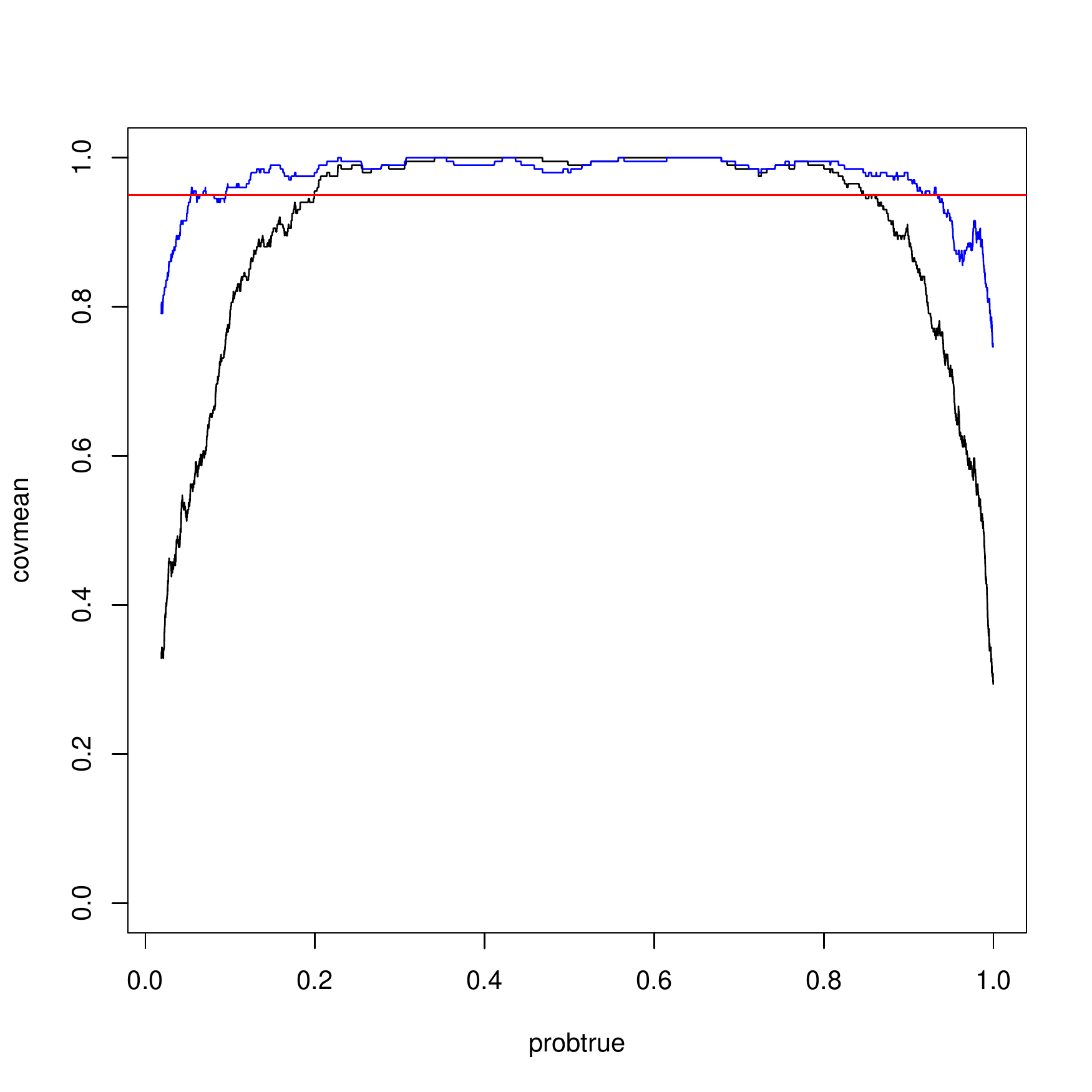}
  		}
  \end{minipage}\hfill
  \begin{minipage}[c]{0.50\linewidth}
    \centering
  		\subfigure[][HPD interval]{
  			\includegraphics[width=1.00\textwidth]{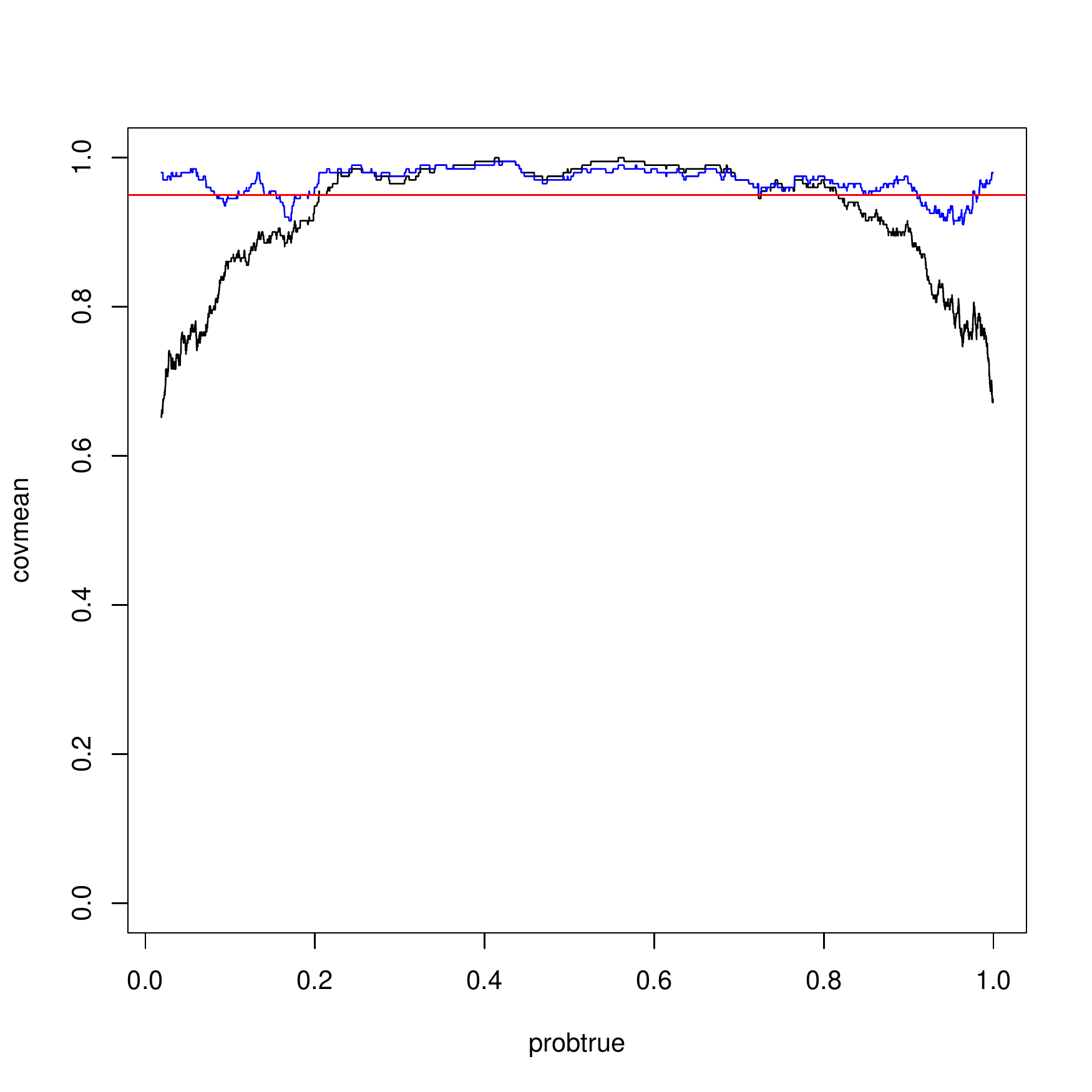}

  		}
  \end{minipage}
\caption{X-axis: True event probability $q_j$; Y-axis: mean coverage of 95\% posterior intervals for event probability. Mean is estimated by moving average. Case: $G=5, p_g=40, p = G*p_G = 200, n_{\text{train}} = 200$. Methods:
\textcolor{blue}{Hyb}, \textcolor{black}{EB}}
\end{figure}

\clearpage
\bibliographystyle{C://Synchr//Stylefiles//author_short3}
\bibliography{C://Synchr//Bibfiles//empbayes,C://Synchr//Bibfiles//bibarrays}

\begin{thebibliography}{66}
\providecommand{\natexlab}[1]{#1}
\providecommand{\url}[1]{\texttt{#1}}
\providecommand{\urlprefix}{URL }
\expandafter\ifx\csname urlstyle\endcsname\relax
  \providecommand{\doi}[1]{doi:\discretionary{}{}{}#1}\else
  \providecommand{\doi}{doi:\discretionary{}{}{}\begingroup
  \urlstyle{rm}\Url}\fi

\bibitem[{Bar and Schifano(2011)}]{Bar2011}
Bar, H.Y. and Schifano, E.D. (2011).
\newblock Empirical and fully {B}ayesian approaches for random effects models
  in microarray data analysis.
\newblock \emph{Stat. Model.}, \textbf{11}, 71--88.

\bibitem[{Barber et~al.(2016)}]{Barber2016}
Barber, R.F. et~al. (2016).
\newblock \emph{Statistical Analysis for High-Dimensional Data: The Abel
  Symposium 2014}, chapter Laplace Approximation in High-Dimensional {B}ayesian
  Regression, pages 15--36.
\newblock Springer International Publishing, Cham.

\bibitem[{Basu et~al.(2003)}]{basu2003empirical}
Basu, R. et~al. (2003).
\newblock Empirical {B}ayes prediction intervals in a normal regression model:
  higher order asymptotics.
\newblock \emph{Statist. Prob. Let.}, \textbf{63}, 197--203.

\bibitem[{Belitser and Nurushev(2015)}]{Belitser2015needles}
Belitser, E. and Nurushev, N. (2015).
\newblock Needles and straw in a haystack: empirical {B}ayes confidence for
  possibly sparse sequences.
\newblock Technical report, arXiv:1511.01803, https://arxiv.org/abs/1511.01803.

\bibitem[{Bergersen et~al.(2011)}]{Bergersen2011}
Bergersen, L.C. et~al. (2011).
\newblock {{W}eighted lasso with data integration}.
\newblock \emph{Stat. Appl. Genet. Mol. Biol.}, \textbf{10}, 1--29.

\bibitem[{Bernardo et~al.(2003)}]{bernardo2003variational}
Bernardo, J.M. et~al. (2003).
\newblock The variational {B}ayesian {EM} algorithm for incomplete data: with
  application to scoring graphical model structures.
\newblock \emph{Bayesian statistics}, \textbf{7}, 453--464.

\bibitem[{Bhattacharya et~al.(2015)}]{bhattacharya_fast_2015}
Bhattacharya, A. et~al. (2015).
\newblock Fast sampling with {Gaussian} scale-mixture priors in
  high-dimensional regression.
\newblock \emph{arXiv:1506.04778 [stat]}.
\newblock ArXiv: 1506.04778.
\newline\urlprefix\url{http://arxiv.org/abs/1506.04778}

\bibitem[{Bickel and Levina(2004)}]{bickel2004some}
Bickel, P.J. and Levina, E. (2004).
\newblock Some theory for {F}isher's linear discriminant function,'naive
  {B}ayes', and some alternatives when there are many more variables than
  observations.
\newblock \emph{Bernoulli}, \textbf{10}, 989--1010.

\bibitem[{Blei et~al.(2017)}]{blei2017variational}
Blei, D.M. et~al. (2017).
\newblock Variational inference: A review for statisticians.
\newblock \emph{J Amer Statist Assoc}, \textbf{XX}, To appear.

\bibitem[{Booth and Hobert(1999)}]{Booth1999}
Booth, J.G. and Hobert, J.P. (1999).
\newblock Maximizing generalized linear mixed model likelihoods with an
  automated {M}onte {C}arlo {EM} algorithm.
\newblock \emph{J. Roy. Statist. Soc. B}, \textbf{61}, 265--285.

\bibitem[{Boulesteix et~al.(2017)}]{boulesteix2017ipf}
Boulesteix, A.L. et~al. (2017).
\newblock {IPF-LASSO}: Integrative-penalized regression with penalty factors
  for prediction based on multi-omics data.
\newblock \emph{Comp. Math. Meth. Med.}, \textbf{2017}.

\bibitem[{Carbonetto and Stephens(2012)}]{carbonetto2012scalable}
Carbonetto, P. and Stephens, M. (2012).
\newblock Scalable variational inference for {B}ayesian variable selection in
  regression, and its accuracy in genetic association studies.
\newblock \emph{Bayesian analysis}, \textbf{7}, 73--108.

\bibitem[{Carlin and Louis(2000)}]{CarlinLouis}
Carlin, B. and Louis, T. (2000).
\newblock \emph{Bayes and Empirical Bayes Methods for Data Analysis}.
\newblock Chapman and Hall/CRC.

\bibitem[{Casella(2001)}]{Casella2001empirical}
Casella, G. (2001).
\newblock Empirical {B}ayes {G}ibbs sampling.
\newblock \emph{Biostatistics}, \textbf{2}, 485--500.

\bibitem[{Castillo and {Van der Vaart}(2012)}]{Castillo2012needles}
Castillo, I. and {Van der Vaart}, A.W. (2012).
\newblock Needles and straw in a haystack: {P}osterior concentration for
  possibly sparse sequences.
\newblock \emph{Ann. Statist.}, \textbf{40}, 2069--2101.

\bibitem[{Chib(1995)}]{chib95}
Chib, S. (1995).
\newblock Marginal likelihood from the {G}ibbs output.
\newblock \emph{J. Amer. Statist. Assoc.}, \textbf{90}, 1313--1321.

\bibitem[{Cule and De~Iorio(2013)}]{cule2013ridge}
Cule, E. and De~Iorio, M. (2013).
\newblock Ridge regression in prediction problems: automatic choice of the
  ridge parameter.
\newblock \emph{Genetic epidemiology}, \textbf{37}, 704--714.

\bibitem[{Dicker and Zhao(2016)}]{dicker2016high}
Dicker, L.H. and Zhao, S.D. (2016).
\newblock High-dimensional classification via nonparametric empirical {B}ayes
  and maximum likelihood inference.
\newblock \emph{Biometrika}, \textbf{103}, 21--34.

\bibitem[{Duin(2000)}]{duin2000classifiers}
Duin, R.P.W. (2000).
\newblock Classifiers in almost empty spaces.
\newblock In \emph{IEEE Proceedings of the 15th International Conference on
  Pattern Recognition}, volume~2, pages 1--7.

\bibitem[{Efron(2010)}]{EfronBook}
Efron, B. (2010).
\newblock \emph{Large-scale inference}.
\newblock Institute of {M}athematical {S}tatistics {M}onographs. Cambridge
  University Press, Cambridge.

\bibitem[{Efron and Morris(1975)}]{Efron1975data}
Efron, B. and Morris, C. (1975).
\newblock Data analysis using {S}tein's estimator and its generalizations.
\newblock \emph{J. Amer. Statist. Assoc.}, \textbf{70}, 311--319.

\bibitem[{Efron(2009)}]{Efron2009Prediction}
Efron, B. (2009).
\newblock Empirical {B}ayes estimates for large-scale prediction problems.
\newblock \emph{J. Amer. Statist. Assoc.}, \textbf{104}, 1015--1028.

\bibitem[{Gelfand and Smith(1990)}]{gelfand1990sampling}
Gelfand, A.E. and Smith, A.F.M. (1990).
\newblock Sampling-based approaches to calculating marginal densities.
\newblock \emph{J. Amer. Statist. Assoc.}, \textbf{85}, 398--409.

\bibitem[{George and Foster(2000)}]{george2000calibration}
George, E.I. and Foster, D.P. (2000).
\newblock Calibration and empirical {B}ayes variable selection.
\newblock \emph{Biometrika}, \textbf{87}, 731--747.

\bibitem[{Heisterkamp et~al.(1999)}]{Heisterkamp1999}
Heisterkamp, S. et~al. (1999).
\newblock Empirical bayesian estimators for a poisson process propagated in
  time.
\newblock \emph{Biom. J.}, \textbf{41}, 385--400.

\bibitem[{Hoerl et~al.(1975)}]{hoerl1975ridge}
Hoerl, A.E. et~al. (1975).
\newblock Ridge regression: some simulations.
\newblock \emph{Comm. Statist.-Th. Meth.}, \textbf{4}, 105--123.

\bibitem[{Ishwaran and Rao(2005)}]{Ishwaran2005spike}
Ishwaran, H. and Rao, J.S. (2005).
\newblock Spike and slab variable selection: frequentist and {B}ayesian
  strategies.
\newblock \emph{Ann. Statist.}, \textbf{33}, 730--773.

\bibitem[{Isserlis(1918)}]{isserlis1918formula}
Isserlis, L. (1918).
\newblock On a formula for the product-moment coefficient of any order of a
  normal frequency distribution in any number of variables.
\newblock \emph{Biometrika}, \textbf{12}, 134--139.

\bibitem[{Jiang et~al.(2016{\natexlab{a}})}]{Jiang2016}
Jiang, J. et~al. (2016{\natexlab{a}}).
\newblock On high-dimensional misspecified mixed model analysis in genome-wide
  association study.
\newblock \emph{Ann Statist}, \textbf{44}, 2127--2160.

\bibitem[{Jiang et~al.(2016{\natexlab{b}})}]{jiang2016variable}
Jiang, Y. et~al. (2016{\natexlab{b}}).
\newblock Variable selection with prior information for generalized linear
  models via the prior lasso method.
\newblock \emph{J. Amer. Statist. Assoc.}, \textbf{111}, 355--376.

\bibitem[{Johnstone and Silverman(2004)}]{johnstone2004needles}
Johnstone, I. and Silverman, B. (2004).
\newblock Needles and straw in haystacks: {E}mpirical bayes estimates of
  possibly sparse sequences.
\newblock \emph{Ann. Statist.}, \textbf{32}, 1594 -- 1649.

\bibitem[{Joo(2017)}]{joo2017bayesian}
Joo, L. (2017).
\newblock \emph{Bayesian lasso: An extension for genome-wide association
  study}.
\newblock Ph.D. thesis, New York University,
  https://search.proquest.com/openview/197063cc5f32e5e0aaac827bbfa7791a/.
\newline\urlprefix\url{https://search.proquest.com/openview/197063cc5f32e5e0aaac827bbfa7791a/}

\bibitem[{Karabatsos(2017)}]{karabatsos2017marginal}
Karabatsos, G. (2017).
\newblock Marginal maximum likelihood estimation methods for the tuning
  parameters of ridge, power ridge, and generalized ridge regression.
\newblock \emph{Comm Statist-Sim Comp}.

\bibitem[{Kuhn and Lavielle(2004)}]{kuhn2004coupling}
Kuhn, E. and Lavielle, M. (2004).
\newblock Coupling a stochastic approximation version of {EM} with an {MCMC}
  procedure.
\newblock \emph{ESAIM: Probability and Statistics}, \textbf{8}, 115--131.

\bibitem[{{Le Cessie} and van Houwelingen(1992)}]{leCessie1992}
{Le Cessie}, S. and van Houwelingen, J.C. (1992).
\newblock Ridge estimators in logistic regression.
\newblock \emph{Appl. Statist.}, \textbf{41}, 191--201.

\bibitem[{Leday et~al.(2017)}]{Leday2017}
Leday, G.G.R. et~al. (2017).
\newblock Gene network reconstruction using global-local shrinkage priors.
\newblock \emph{Ann Appl Statist}, \textbf{11}, 41--68.

\bibitem[{Levine and Casella(2001)}]{Levine2001}
Levine, R.A. and Casella, G. (2001).
\newblock Implementations of the {M}onte {C}arlo {EM} algorithm.
\newblock \emph{J. Comput. Graph. Statist.}, \textbf{10}, 422--439.

\bibitem[{Li and Lin(2010)}]{Li2010bayesian}
Li, Q. and Lin, N. (2010).
\newblock The {B}ayesian elastic net.
\newblock \emph{Bayesian Analysis}, \textbf{5}, 151--170.

\bibitem[{Meier et~al.(2008)}]{Meier2008}
Meier, L. et~al. (2008).
\newblock The group {L}asso for logistic regression.
\newblock \emph{J. R. Stat. Soc. Ser. B Stat. Methodol.}, \textbf{70}, 53--71.

\bibitem[{Morris(1983)}]{morris1983parametric}
Morris, C.N. (1983).
\newblock Parametric empirical {B}ayes inference: theory and applications.
\newblock \emph{J Amer Statist Assoc}, \textbf{78}, 47--55.

\bibitem[{Neuenschwander et~al.(2016)}]{neuenschwander2016use}
Neuenschwander, B. et~al. (2016).
\newblock On the use of co-data in clinical trials.
\newblock \emph{Statist. Biopharm. Res.}, \textbf{8}, 345--354.

\bibitem[{Newcombe et~al.(2014)}]{newcombe2014weibull}
Newcombe, P.J. et~al. (2014).
\newblock Weibull regression with {B}ayesian variable selection to identify
  prognostic tumour markers of breast cancer survival.
\newblock \emph{Stat. Meth. Med. Res.}, pages 1--23.

\bibitem[{Novianti et~al.(2017)}]{novianti2017better}
Novianti, P.W. et~al. (2017).
\newblock Better diagnostic signatures from {RNA}seq data through use of
  auxiliary co-data.
\newblock \emph{Bioinformatics}, \textbf{33}, 1572--1574.

\bibitem[{O'Hara and Sillanp{\"a}{\"a}(2009)}]{sillanpaa2009review}
O'Hara, R.B. and Sillanp{\"a}{\"a}, M.J. (2009).
\newblock A review of {B}ayesian variable selection methods: what, how and
  which.
\newblock \emph{Bayesian analysis}, \textbf{4}, 85--117.

\bibitem[{Park and Casella(2008)}]{ParkCasella2008}
Park, T. and Casella, G. (2008).
\newblock The {B}ayesian lasso.
\newblock \emph{J. Amer. Statist. Assoc.}, \textbf{103}, 681--686.

\bibitem[{Peltola et~al.(2012)}]{peltola2012finite}
Peltola, T. et~al. (2012).
\newblock Finite adaptation and multistep moves in the {M}etropolis-{H}astings
  algorithm for variable selection in genome-wide association analysis.
\newblock \emph{PloS one}, \textbf{7}, e49445.

\bibitem[{Polson et~al.(2013)}]{Polson2013}
Polson, N.G. et~al. (2013).
\newblock Bayesian inference for logistic models using {P}\'olya$-${G}amma
  latent variables.
\newblock \emph{J. Amer. Statist. Assoc.}, \textbf{108}, 1339--1349.

\bibitem[{Press(1982)}]{Press1982}
Press, S.J. (1982).
\newblock \emph{Applied {M}ultivariate {A}nalysis, 2nd ed.}
\newblock Dover Publications, New York.

\bibitem[{Quintana and Conti(2013)}]{quintana2013integrative}
Quintana, M.A. and Conti, D.V. (2013).
\newblock Integrative variable selection via {B}ayesian model uncertainty.
\newblock \emph{Statist. Med.}, \textbf{32}, 4938--4953.

\bibitem[{Ro{\v{c}}kov{\'a} and George(2014)}]{rovckova2014emvs}
Ro{\v{c}}kov{\'a}, V. and George, E.I. (2014).
\newblock {EMVS}: The {EM} approach to {B}ayesian variable selection.
\newblock \emph{J. Amer. Statist. Assoc.}, \textbf{109}, 828--846.

\bibitem[{Rousseau and Szabo(2017)}]{rousseau2017asymptotic}
Rousseau, J. and Szabo, B. (2017).
\newblock Asymptotic behaviour of the empirical {B}ayes posteriors associated
  to maximum marginal likelihood estimator.
\newblock \emph{Ann. Statist.}, \textbf{45}, 833--865.

\bibitem[{Rue et~al.(2009)}]{Rue2009}
Rue, H. et~al. (2009).
\newblock Approximate {B}ayesian inference for latent {G}aussian models by
  using integrated nested {L}aplace approximations (with discussion).
\newblock \emph{J. Roy. Statist. Soc. B}, \textbf{71}, 319--392.

\bibitem[{Scott and Berger(2010)}]{scott_bayes_2010}
Scott, J.G. and Berger, J.O. (2010).
\newblock Bayes and empirical-{Bayes} multiplicity adjustment in the
  variable-selection problem.
\newblock \emph{Ann. Statist.}, \textbf{38}, 2587--2619.

\bibitem[{Shun and McCullagh(1995)}]{shun1995laplace}
Shun, Z. and McCullagh, P. (1995).
\newblock Laplace approximation of high dimensional integrals.
\newblock \emph{J. Roy. Statist. Soc., B}, pages 749--760.

\bibitem[{Simon et~al.(2013)}]{simon2013sparse}
Simon, N. et~al. (2013).
\newblock A sparse-group lasso.
\newblock \emph{J Comput Graph Stat}, \textbf{22}, 231--245.

\bibitem[{Stingo et~al.(2011)}]{stingo2011incorporating}
Stingo, F.C. et~al. (2011).
\newblock Incorporating biological information into linear models: A bayesian
  approach to the selection of pathways and genes.
\newblock \emph{Ann. Appl. Statist.}, \textbf{5}, 1202--1214.

\bibitem[{Taddy et~al.(2015)}]{taddy2015bayesian}
Taddy, M. et~al. (2015).
\newblock Bayesian and empirical {B}ayesian forests.
\newblock Technical report, arXiv:1502.02312, https://arxiv.org/abs/1502.02312.

\bibitem[{Tai and Pan(2007)}]{Tai2007}
Tai, F. and Pan, W. (2007).
\newblock Incorporating prior knowledge of predictors into penalized
  classifiers with multiple penalty terms.
\newblock \emph{Bioinformatics}, \textbf{23}, 1775--1782.

\bibitem[{{Te Beest} et~al.(2017)}]{Beest2017}
{Te Beest}, D.E. et~al. (2017).
\newblock Improved high-dimensional prediction with {R}andom {F}orests by the
  use of co-data.
\newblock \emph{BMC Bioinformatics}, \textbf{18}, 584.

\bibitem[{Tibshirani et~al.(2002)}]{Tibshirani2002}
Tibshirani, R. et~al. (2002).
\newblock Diagnosis of multiple cancer types by shrunken centroids of gene
  expression.
\newblock \emph{Proc. Natl. Acad. Sci.}, \textbf{99}, 6567--6572.

\bibitem[{{Van de Wiel} et~al.(2012)}]{WielShrinkSeq}
{Van de Wiel}, M.A. et~al. (2012).
\newblock Bayesian analysis of {RNA} sequencing data by estimating multiple
  shrinkage priors.
\newblock \emph{Biostatistics}, \textbf{14}, 113--128.

\bibitem[{{Van de Wiel} et~al.(2016)}]{WielGRridge}
{Van de Wiel}, M.A. et~al. (2016).
\newblock Better prediction by use of co-data: adaptive group-regularized ridge
  regression.
\newblock \emph{Statist. Med.}, \textbf{35}, 368--381.

\bibitem[{{Van Houwelingen}(2014)}]{Houwelingen2014}
{Van Houwelingen}, H.C. (2014).
\newblock The role of empirical {B}ayes methodology as a leading principle in
  modern medical statistics.
\newblock \emph{Biom. J.}, \textbf{56}, 919--932.

\bibitem[{Waldron et~al.(2011)}]{Waldron2011}
Waldron, L. et~al. (2011).
\newblock {{O}ptimized application of penalized regression methods to diverse
  genomic data}.
\newblock \emph{Bioinformatics}, \textbf{27}, 3399--3406.

\bibitem[{Wei and Tanner(1990)}]{wei1990monte}
Wei, G.C.G. and Tanner, M.A. (1990).
\newblock A {M}onte {C}arlo implementation of the {EM} algorithm and the poor
  man's data augmentation algorithms.
\newblock \emph{J. Amer. Statist. Assoc.}, \textbf{85}, 699--704.

\bibitem[{Zou and Hastie(2005)}]{ZouHastie2005}
Zou, H. and Hastie, T. (2005).
\newblock Regularization and variable selection via the elastic net.
\newblock \emph{J. R. Stat. Soc. Ser. B Stat. Methodol.}, \textbf{67},
  301--320.

\end{thebibliography}

\end{document}